\documentclass[12pt,preprint,twocolumn]{emulateapj}
\usepackage[utf8]{inputenc}
\usepackage{graphicx}
\usepackage{epstopdf}
\usepackage{amssymb}
\usepackage{subfigure}
\usepackage{bm}
\usepackage{natbib}
\usepackage{graphicx}
\usepackage{float}
\usepackage{rotating}
\usepackage{color,soul}
\usepackage{ulem}
\usepackage{multirow}

\newcommand{\ergcms}{erg cm$^{-2}$ s$^{-1}$}
\newcommand{\Msolar}{ M$_{\odot}$}
\newcommand{\Rsolar}{ R$_{\odot}$}
\newcommand{\ergs}{erg s$^{-1}$}
\def\arcsec{\hbox{$^{\prime\prime}$}}
\def\arcseck{\hbox{$^{\prime\prime}$ }}

\newcommand{\Lx}{$L_{\rm X}$}
\newcommand{\Fx}{$F_{\rm X}$}

\newcommand{\chandra}{{\it Chandra}}
\newcommand{\xmm}{{\it XMM-Newton}}
\newcommand{\hst}{{\it HST}}
\newcommand{\rosat}{{\it ROSAT}}

\shorttitle{ CG\,X-1: a WR-BH ULX}
\shortauthors{Qiu et al.}

\begin{document}
\title{CG\,X-1: an eclipsing Wolf-Rayet ULX in the Circinus galaxy}

\author{Yanli Qiu\altaffilmark{1,2}, Roberto Soria\altaffilmark{2,1,3,4}, Song Wang\altaffilmark{1}, Grzegorz Wiktorowicz\altaffilmark{1,2}, Jifeng Liu\altaffilmark{1,2,5},  Yu Bai\altaffilmark{1}, \\
Alexey Bogomazov\altaffilmark{6}, Rosanne Di Stefano\altaffilmark{7}, Dominic J. Walton\altaffilmark{8},  Xiaojie Xu\altaffilmark{9,10}}

\altaffiltext{1}{Key Laboratory of Optical Astronomy, National Astronomical Observatories, Chinese Academy of Sciences, Beijing 100101, China; ylqiu@bao.ac.cn;rsoria@nao.cas.cn; jfliu@bao.ac.cn}
\altaffiltext{2} {School of Astronomy and Space Sciences, University of Chinese Academy of Sciences, Beijing 100049, China}
\altaffiltext{3}{International Centre for Radio Astronomy Research, Curtin University, GPO Box U1987, Perth, WA 6845, Australia}
\altaffiltext{4}{Sydney Institute for Astronomy, School of Physics A28, The University of Sydney, Sydney, NSW 2006, Australia}
\altaffiltext{5}{ WHU-NAOC Joint Center for Astronomy, Wuhan University, Wuhan, Hubei 430072, China }
\altaffiltext{6}{ M. V. Lomonosov Moscow State University, P. K. Sternberg Astronomical Institute, 13, Universitetskij prospect, Moscow, 119991, Russia}
\altaffiltext{7}{Harvard-Smithsonian Center for Astrophysics, 60 Garden Street, Cambridge, MA 02138, USA}
\altaffiltext{8}{ Institute of Astronomy, University of Cambridge, Madingley Road, Cambridge CB3 0HA, UK}
\altaffiltext{9}{School of Astronomy and Space Science, Nanjing University, Nanjing 210023, People's Republic of China}
\altaffiltext{10}{Key Laboratory of Modern Astronomy and Astrophysics (Nanjing University), Ministry of Education, Nanjing 210023, People's Republic of China}
%\altaffiltext{}{}
%\altaffiltext{}{}

\begin{abstract}

We investigated the time-variability and spectral properties of the eclipsing X-ray source Circinus Galaxy X-1 (GG\,X-1), using {\it Chandra}, {\it XMM-Newton} and {\it ROSAT}. We phase-connected the lightcurves observed over 20 years, and obtained a best-fitting period $P = (25,970.0 \pm 0.1)$ s $\approx$7.2 hr, and a period derivative $\dot{P}/P = ( 10.2\pm4.6) \times 10^{-7}$ yr$^{-1}$. The X-ray lightcurve shows asymmetric eclipses, with sharp ingresses and slow, irregular egresses. The eclipse profile and duration vary substantially from cycle to cycle. We show that the X-ray spectra are consistent with a power-law-like component, absorbed by neutral and ionized Compton-thin material, and by a Compton-thick, partial-covering medium, responsible for the irregular dips. The high X-ray/optical flux ratio rules out the possibility that CG\,X-1 is a foreground Cataclysmic Variable; in agreement with previous studies, we conclude that it is the first example of a compact ultraluminous X-ray source fed by a Wolf-Rayet star or stripped Helium star. Its unocculted luminosity varies between $\approx$4 $\times 10^{39}$ erg s$^{-1}$ and $\approx$3 $\times 10^{40}$ erg s$^{-1}$. Both the donor star and the super-Eddington compact object drive powerful outflows: we suggest that the occulting clouds are produced in the wind-wind collision region and in the bow shock in front of the compact object. Among the rare sample of Wolf-Rayet X-ray binaries, CG\,X-1 is an exceptional target for studies of super-critical accretion and close binary evolution; it is also a likely progenitor of gravitational wave events.

\end{abstract}


\section{Introduction}
% general context
% ULX
Ultraluminous X-ray sources (ULXs) are non-nuclear point-like sources with X-ray luminosity $\gtrsim 3\times $10$^{39}$ \ergs\ \citep{Kaaret2017,Feng2011}. 
The luminosity distribution of this population is now fairly well determined
\citep{Wang2016, Mineo2012, Walton2011, Swartz2011, Liu2005}, and is consistent with the high-luminosity end of the high-mass X-ray binary (XRB) distribution; we also know that the number of ULXs in a star-forming galaxy is roughly proportional to its star formation rate,
and that the distribution may have a cut-off or downturn at an X-ray luminosity of $\approx$2 $\times$ 10$^{40}$ \ergs\ \citep{Mineo2012, Swartz2011}.
However, more specific properties of these sources are still poorly constrained. It is not known what fraction of them are powered by a BH and what fraction by a neutron star (NS)  (see {\it e.g.}, \citealt{Wiktorowicz2019}, and references therein); so far, an identification of the compact object has been possible only for a handful of ULXs with X-ray pulsations, signature of a 
magnetized NS \citep{Bachetti2014, Israel2017a, Israel2017b, Furst2016,Tsygankov2017, Carpano2018}. The relative distribution of stellar types and ages for the donor stars is also poorly known. In a few cases, the donor is 
identified as a blue supergiant \citep{Motch2014}, or a red supergiant \citep{Heida2016},
or a low-mass star \citep{Soria2012}, but in many other cases, it is hard to tell whether the observed optical counterpart corresponds to the donor star or the irradiated accretion disk \citep{Tao2011}. Likewise, the mechanism of mass transfer (Roche lobe overflow (RLOF) or wind accretion), the duty cycle, the duration of the super-Eddington phases and the total amount of mass that may be accreted by the compact object during its lifetime remain generally unknown. 

Population synthesis models %\citep[e.g.][]{Wiktorowicz2016,Artale2018}
predict super-Eddington mass transfer phases from several different types of donor stars at different ages, but our lack of empirical information on the fundamental system parameters for most ULXs makes it difficult to test such models. It also makes it hard to predict the future evolution of such binary systems, {\it e.g.}, what fraction of ULXs will evolve into double compact binaries (BH-BH, BH-NS or NS-NS), potential progenitors of gravitational wave mergers 
\citep{Marchant2017}. Perhaps the most important piece of information that is usually missing is the binary period; without a period, also the mass ratio and the binary separation remain unconstrained. Only in a few cases have periodic variations (in either the X-ray or the optical lightcurve) been detected and interpreted as a binary period; those periods vary between a few days to a few months \citep{Liu2013,Bachetti2014,Motch2014,Furst2018,Urquhart2016,Wang2018,Vinokurov2018}.

One rare ULX candidate with a well-determined period of 7.2 hr (as well as other intriguing X-ray properties) is the very bright X-ray source CG\,X-1 \citep{Bauer2001, Weisskopf2004, Esposito2015}. It is seen projected in the sky inside the inner region of the Circinus galaxy, $\approx$15$''$ to the east of its nucleus ($\approx$300 pc at the distance of Circinus), with the coordinates of $\alpha=14^{\rm h}13^{\rm m}12^{\rm s}.21$, $\delta = -65^{\circ}20^{\prime}13^{\prime\prime}.7$ (J2000). Circinus is a large Sb galaxy at a distance of 4.2 Mpc \citep{Tully2009} with a Seyfert nucleus and intense star formation, at a rate of $\approx$3--8 $M_{\odot}$ yr$^{-1}$ \citep{Freeman1977, For2012}. If CG\,X-1 belongs to the Circinus galaxy, its X-ray luminosity is $\approx$10$^{40}$ erg s$^{-1}$ (reaching $\approx$3 $\times 10^{40}$ erg s$^{-1}$ at some epochs), making it one of the most luminous ULXs in the local universe, near the potential break in the ULX luminosity function. By comparison, from the average X-ray luminosity function of \cite{Mineo2012}, we expect $\approx$0.2--0.6 sources at or above an X-ray luminosity of 10$^{40}$ erg s$^{-1}$ in a galaxy with the star formation rate of Circinus.
Thus, the presence of 1--2 luminous ULX in Circinus is not unexpected. Besides CG\,X-1, there is another ULX (known as ULX5) in the outskirts of this galaxy, with $L_{\rm X} \approx 2\times 10^{40}$ erg s$^{-1}$ \citep{Walton2013}. 
%With a likelihood of 0.6, it is not surprise to see one in the crowed nuclear region.} 

\begin{figure}
%\vspace{-1.2cm}
\center{
\includegraphics[width=8.5cm]{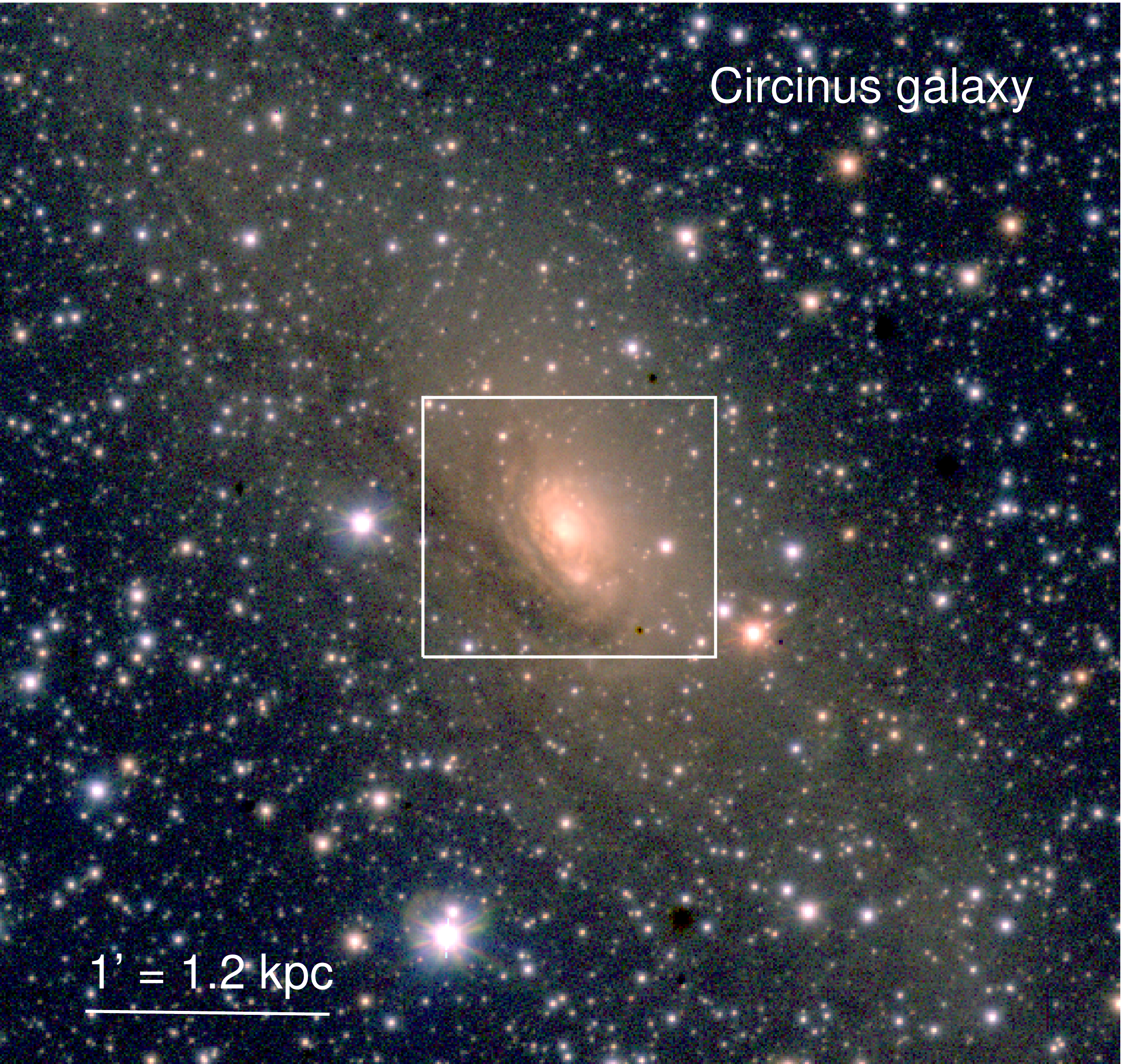}\\
\includegraphics[width=8.5cm]{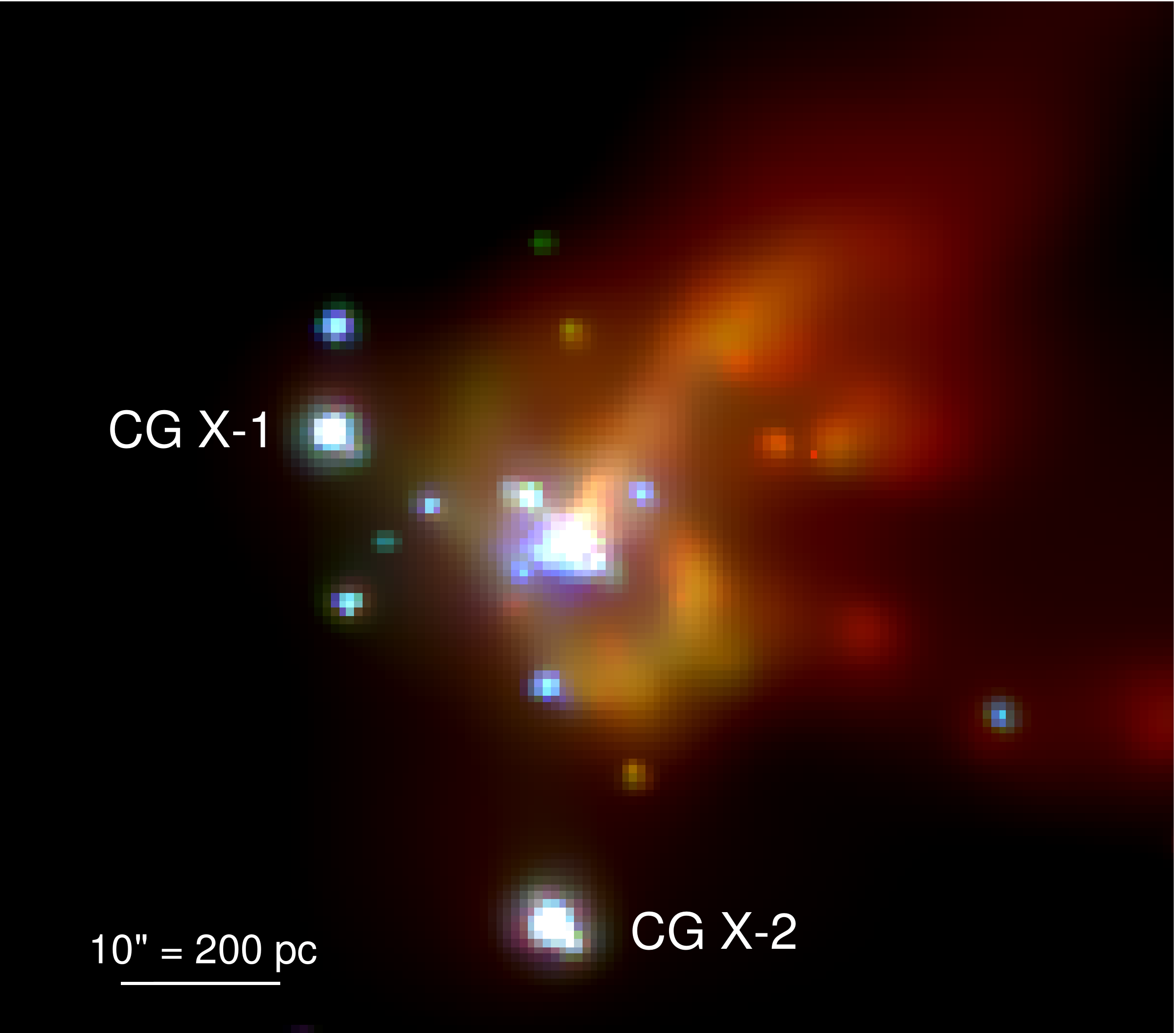}}
\caption{Top panel: archival true-colour optical image of the Circinus galaxy, from the 2.6-m ESO-VLT Survey Telescope ({\it g, r, i} bands). North is up and east to the left. Most of the bright point-like sources in the field are foreground stars in the Milky Way. The white box shows the region displayed in the bottom panel. Bottom panel: adaptively smoothed {\it Chandra}/ACIS image of the central region of Circinus (red $=$ 0.3--1.1 keV; green $=$ 1.1--2.0 keV; blue $=$ 2.0--7.0 keV). The two brightest non-nuclear point sources are known as CG\,X-1 (the ULX subject of our investigation) and CG X-2 (also known as SN 1996cr: \citealt{Bauer2008}). North is up and east to the left.\\}
\label{fig:galaxy}
\end{figure}

There has been some debate in the literature about whether CG\,X-1 really belongs to Circinus or is instead a foreground magnetic cataclysmic variable (mCV) in the Milky Way (as suggested by \citealt{Weisskopf2004}), projected by chance in front of Circinus. In the latter case, the 7.2-hr period would correspond to the spin period of the white dwarf rather than the binary period. There were several reasons behind the mCV suggestion.  Firstly, Circinus is located almost behind the Galactic plane, in a field crowded with foreground stars (Figure \ref{fig:galaxy}); in fact, two other X-ray sources projected in the field of Circinus (but not as close to its starburst region as CG\,X-1) turned out to be foreground mCVs \citep{Esposito2015}. Secondly, when first discovered two decades ago, empirical knowledge of ULXs was still scant, and an intrinsic luminosity in excess of $10^{40}$ erg s$^{-1}$ for a non-nuclear source was still regarded as suspiciously unlikely. Thirdly, the X-ray lightcurve of CG\,X-1, with its asymmetric eclipses, is very unusual for a ULX or more generally for a luminous X-ray binary and is  similar instead to the periodic eclipses of the accreting poles in an mCV. However, the mCV interpretation was substantially refuted by \cite{Esposito2015}, based on probability arguments. They also showed that the short orbital period and eclipsing lightcurve are typical of systems with a Wolf-Rayet (WR) donor stars (as already proposed by \citealt{Bauer2001}); a few such systems have been discovered in recent years \citep{Carpano2007,Esposito2013, Esposito2015, Maccarone2014}, although none of them has reached the super-Eddington luminosity of CG\,X-1. If CG\,X-1 survives the collapse of the WR donor and evolves into a double BH binary, the coalescence timescale via gravitational wave emission is only $\approx$50 Myr \citep{Esposito2015}.

Here, we present a detailed, multi-epoch analysis of CG\,X-1 to further probe the nature of this remarkable system. The main objectives of our paper are the following. In Section 2, we will describe the X-ray data we used. In Section 3,  we will illustrate the ever-changing lightcurve profiles observed from individual orbital cycles, using data from two \rosat,  24 \chandra\ and five \xmm\ observations of Circinus, obtained over a period of 21 years; we will attempt to distinguish between periodic features and stochastic dipping, and we will obtain a more precise measurement of the period and of the period derivative. In Section 4, we will model the X-ray spectral properties of the systems during peak-luminosity phases, during dipping phases, and during occultation phases (which still show faint, residual emission). 
%We substantially agree with the ULX interpretation 
We corroborate the ULX interpretation
of \cite{Esposito2015}, and in this paper we will only add a short discussion about the X-ray to optical flux ratio of CG\,X-1, which again disfavours the mCV scenario (Section 5). In Section 6, we will propose a physical scenario that may explain the puzzling timing and spectral properties, in terms of partial covering by optically thick clumps located mostly in front of the compact object, as it moves through the dense wind of the donor star (bow shock scenario). Then, we will briefly discuss the possible origin and evolutionary state of this system in the context of population synthesis models. Finally (Section 7), we present a summary of our results.

 % from /Users/qyl/work/cgx1/chandra/evt/reproj/
 % and /Users/qyl/work/cgx1/vst

%%%%%%%%%%%%%%%%%%%%%%%%%%%%%%%%%%%%%%%%%%%%%%%%%%%%%%
\section{         Observations and data analysis                   }

\subsection{         \chandra              }

CG\,X-1 was observed 24 times by \chandra's Advanced CCD Imaging Spectrometer (ACIS) from 2000 to 2010 (Table 1). Eight of the observations were centred on the back-illuminated S3 chip with no gratings; the other 16 were taken with the High Energy Transmission Grating (HETG) in front of the ACIS-S detectors.
We downloaded the data from the public archive and reprocessed them with the task {\it chandra\_repro} provided by the Chandra Interactive Analysis of Observations (\textsc{ciao}) software package, Version 4.9 \citep{Fruscione2006}.
We corrected photon arrival times to the solar system barycenter using the \textsc{ciao} task {\it{axbary}}. 

For the non-grating data, we used {\it dmcopy} to extract images in the 0.3--8 keV band, from the event files of the individual observations. 
We then applied the source-finding task {\it wavdetect} to the images of individual observations,
to determine the coordinates of CG\,X-1 and to identify an ellipse around CG\,X-1 that contains $\approx$99\% of the source counts ({\it wavdetect} parameter ellsigma $= 3$). The typical semi-major axis of the elliptical source region is $\approx 1\arcsec$.3.
We used that ellipse as the source extraction region for our timing and spectral analysis. For the local background regions, we used an elliptical annulus centered at the position of the source; the inner and outer sizes of the annulus were 2 and 4 times the size of the source ellipse, respectively. We extracted background-subtracted lightcurves of each observation, in the 0.3--8 keV band, with the task {\it{dmextract}}. We used {\it{specextract}} to create spectra and associated background and response files, for each observation; we used the {\it specextract} option ``correctpsf = yes'' for aperture correction, and we grouped the spectra to a minimum of 15 counts per bin, for subsequent $\chi^2$ fitting.

For the HETG observations, we only extracted lightcurves (also filtered to the 0.3--8 keV band), from the zeroth-order images. The source extraction region was a circle with a radius of 1$\arcsec$.3, centered at the averaged position determined from the individual non-grating images. The background region is a circle with radius of 16\arcseck in the northeast of CG\,X-1, free of point sources.
Subsequent analysis of spectra was done with standard tool {\sc xspec} from the  {\sc heasoft} package, Version 6.21 \citep{Blackburn1995}.

\subsection{         \xmm           }
\xmm\ has observed CG\,X-1 on six occasions before October 2018, with the European Photon Imaging Camera (EPIC) as prime instrument. For this work, we used the first five observations (Table 1), which are already available for download from the public archive.
%by the Principal Investigator D.J.~Walton
We reduced the data with the \xmm\ Science Analysis System ({\textsc{sas}}) version 16.7.0. In particular, we ran the standard tasks {\it cifbuild}, {\it odfingest}, {\it epproc} and {\it emproc} to obtain Processing Pipeline System (PPS) products from the Observation Data Files (ODFs), for EPIC-pn and EPIC-MOS. The {\sc sas} task {\it barycen} was applied for barycentric corrections.

For each observation, we used {\it xmmselect} to create preliminary lightcurves of the whole field of view above 10 keV, for the purpose of flagging and removing time intervals affected by background flares. Time intervals with count rate less than some certain criteria were remained to create the GTI intervals. The count rate criteria varies according to the observations and cameras, typically higher than 0.8 ct s$^{-1}$ for pn and 0.3 ct s$^{-1}$ for MOS, respectively. For the first four observations, we used the resulting GTI intervals for both lightcurve and spectral extractions. Instead, the most recent observation in our sample (ObsID 0780950201) was affected by moderate background flares throughout most of the exposure; for this reason, for this observation only, we decided not to remove the high background intervals, otherwise the GTI would have been too short for meaningful analysis.

Defining a source extraction region and a local background is much more complicated for \xmm\ than for \chandra, because of the larger point spread function (PSF) of the former. In fact, the full-width at half-maximum of the pn and MOS PSFs is approximately the same as the distance between CG\,X-1 and the (stronger) nuclear X-ray source of Circinus ($\approx$15$\arcsec$). Hence, we had to use a small extraction radius of only 5\arcsec\ for CG\,X-1, to reduce the nuclear source contamination from the wings of its PSF. The local background was extracted from three circles with radius of 5\arcsec, located  around the Circinus nucleus, and centred at approximately the same distance of 15\arcsec\ from the central source. This is the best way to ensure that the nuclear PSF-wing contamination to the source extraction region is as similar as possible to its contribution to the background regions, and can be effectively subtracted out. We used the same source and background regions for both the lightcurve and the spectral extraction.

We used ``{\sc \#xmmea\_em \&\&  (pattern $<=$ 12)}'' to filter MOS lightcurves and spectra; used ``{\sc \#xmmea\_ep \&\& (pattern $<=$ 4 \&\& flag==0)}'' to filter the  PN data. We extracted lightcurves from MOS and pn  with the {\sc sas} tasks {\it{evselect}}, followed by {\it{epiclccorr}} to correct for vignetting and bad pixels\footnote{http://www.cosmos.esa.int/web/xmm-newton/sas-threads}. The photon energy was filtered to 0.3--8 keV.  

For each observation, we extracted individual PN, MOS1 and MOS2 spectra with {\it{evselect}}, in 0,3--8 keV band.  Response and ancillary response files were generated with {\it{rmfgen}} and {\it{arfgen}}, respectively. We then combined the MOS and pn spectra of each observation with the {\sc{sas}} task {\it{epicspeccombine}}, to improve the signal-to-noise ratio of possible narrow features. We grouped the merged spectra to at least 20 counts per bin for $\chi^2$ fitting. Finally, as for the {\it Chandra} data, we used 
%{\sc{xronos}} tools for timing analysis, and 
{\sc{xspec}} for spectral modelling. 
%\qyl{I used matlab instead of {\sc{xronos}} for timing analysis.}

\subsection{ \rosat}

CG\,X-1 has been observed by the Roengten Satellite High Resolution Imager (\rosat/HRI) five times. Two of those observations have exposure times longer than 25 ks (Table \ref{tab:obs}), while the other three are shorter than 5 ks. We only used the two observations with the longest exposure time, {\it i.e.}, RH702058A02 and RH702058A03, taken at various intervals between 1997 March 3--11, and between 1997 August 17--September 9, respectively. We used the {\sc {heasoft}} tool {\it{rosbary}} to do the barycenter correction of the event lists. The source region used for extracting the lightcurve is a circle with a radius of 6\arcsec. The background region is an annulus with inner radius of 6\arcsec\ and outer radius of 10\arcsec\ around the source. Due to the low count rate and sparse snapshots in these two observations, we folded their lightcurves using the best-fitting period (determined from {\it Chandra} and {\it XMM-Newton} observations), for a better statistics. 
The epoch of this folded lightcurve was chosen at the orbital cycle,  where the two observations accumulated half of the total observed counts, {\it i.e.}, MJD 50681.437036 for phase $\phi$ = 1 (the phase will be defined in Section 3.1).  
The main purpose for using the folded \rosat\ lightcurve is to study the variation of the periodicity of CG\,X-1  over a long-term duration of more than 20 years.

% fig
% from cgx1/hst
% HST
% 

%----------------------------------------

% table1
% from /Users/qyl/work/cgx1/table/tab1.tex
%\clearpage
\begin{deluxetable*}{lccccc}
%\rotate
\centering
\tabletypesize{\footnotesize}
\tablewidth{0pt}
\tablecolumns{6}
\tablecaption{ Log of the X-ray observations considered in this work \label{tab:obs}}
%\tablewidth{0pt}
\tablehead{
% &  &  &  & & \\[5pt]
\colhead{ObsID} & \colhead{Instrument} & \colhead{Start date} & \colhead{Exp.~Time (ks) } & \colhead{Off-Axis Angle ($'$)}   &\colhead{counts}    }
\startdata
rh702058a02 (0)& \rosat/HRI & 1997-03-03 21:50:52 & 26.38 & 0.30 & 75 \\
rh702058a03 (0)& \rosat/HRI & 1997-08-17 10:44:29 & 45.89 & 0.30 & 169\\
355 & ACIS-S     & 2000-01-16 10:18:17 & 1.32 & 1.56 & 43 \\
365$^\ast$ & ACIS-S & 2000-03-14 06:01:26 & 4.97 & 1.09 & 1639 \\
356$^\ast$ & ACIS-S     & 2000-03-14 07:46:18 & 24.72 & 1.09 & 3579 \\
374 & HETG       & 2000-06-15 22:01:09 & 7.12 & 1.06 & 85 \\
62877 & HETG     & 2000-06-16 00:38:28 & 60.22 & 1.06 & 923 \\
2454 & ACIS-S& 2001-05-02 16:02:48 & 4.40  & 0.73 & 301 \\
0111240101$^*$ (1)& MOS+PN & 2001-08-06 08:54:51 & 109.85 & 0.27  & 15791 \\
4770 & HETG  & 2004-06-02 12:40:42 & 55.03 & 1.32 & 591 \\
4771 & HETG  & 2004-11-28 18:26:32 & 58.97 & 1.11 & 603 \\
9140$^*$ & ACIS-S& 2008-10-26 10:24:46 & 48.76 & 4.28 & 1176 \\
10226 & HETG & 2008-12-08 17:57:06 & 19.67 & 1.35 & 201 \\
10223 & HETG & 2008-12-15 15:46:15 & 102.93 & 1.34 & 642 \\
10832 & HETG & 2008-12-19 18:15:08 & 20.61 & 1.34 & 187 \\
10833 & HETG & 2008-12-22 07:29:35 & 28.36 & 1.34 & 211 \\
10224 & HETG & 2008-12-23 11:25:12 & 77.10 & 1.33 & 483 \\
10844 & HETG & 2008-12-24 23:17:37 & 27.17 & 1.33 & 290 \\
10225 & HETG & 2008-12-26 04:02:06 & 67.89 & 1.33 & 651 \\
10842 & HETG & 2008-12-27 12:03:26 & 36.74 & 1.33 & 332 \\
10843 & HETG & 2008-12-29 10:10:49 & 57.01 & 1.32 & 448 \\
10873 & HETG & 2009-03-01 23:28:35 & 18.10 & 1.12 & 151 \\
10850 & HETG & 2009-03-03 00:43:18 & 13.85 & 1.12 & 77 \\
10872 & HETG & 2009-03-04 15:29:52 & 16.53 & 1.12 & 83 \\
10937$^*$ & ACIS-S & 2009-12-28 21:10:27 & 18.31 & 2.96 & 464 \\
12823$^*$ & ACIS-S & 2010-12-17 18:10:27 & 152.36 & 1.52 & 10464 \\
12824$^*$ & ACIS-S & 2010-12-24 03:38:54 & 38.89 & 1.52 & 2368 \\
0701981001$^*$ (2) & MOS+PN & 2013-02-03 07:24:11 & 58.91 & 4.80 & 3177 \\
0656580601$^*$ (3) & MOS+PN & 2014-03-01 09:55:41 & 45.90& 0.27  & 1833 \\
0792382701$^*$ (4)& MOS+PN & 2016-08-23 16:53:33 &37.00 & 4.70 & 5939 \\
0780950201$^*$ (5) & MOS+PN & 2018-02-07 12:11:49 & 44.36 & 4.70 & 4302\\
\enddata
\tablenotetext{}{\hskip-2.5pt {\it Notes}: For spectral analysis in this paper, we used all observations marked with an asterisk. Numbers in brackets are a short-hand notation for the corresponding \xmm\ ObsIDs, used for convenience in later sections of this paper.\\
}
\end{deluxetable*}

%%%%%%%%%%%%%%%%%%%%%%%%%%%%%%%%%%%%%%%%%%%%%%%%%%%%%%
%\section{         RESULTS           }

\section{   X-ray Timing Results  }
\subsection{ Binary period   }

%The periodic modulation of CG\,X-1 is apparent. 
A $(27 \pm0.7)$ ks X-ray periodicity was first discovered by \citet{Bauer2001} from {\it Chandra}/HETG zeroth-order data, and it was subsequently confirmed with {\it BeppoSAX} and \chandra/ACIS-S observations \citep{Bianchi2002,Weisskopf2004}. More recently, \citet{Esposito2015} analyzed two long \chandra/ACIS-S observations from 2010 (ObsID 12823 and ObsID 12824) and obtained a more precise value of $(26.1 \pm 0.2)$ ks. To reduce the error even further, for this work we re-analyzed the whole archival set of twenty \chandra\ observations and five \xmm\ observations longer than about half of the period ($\approx$13 ks): they cover a time span of 19 years (Table 1). (We did not use an additional four archival \chandra\ observations because their duration was only $\lesssim$0.25 times the period).

\begin{figure*}[htp!]
\begin{center}
    \hspace*{-0.8cm}\includegraphics[width=20cm]{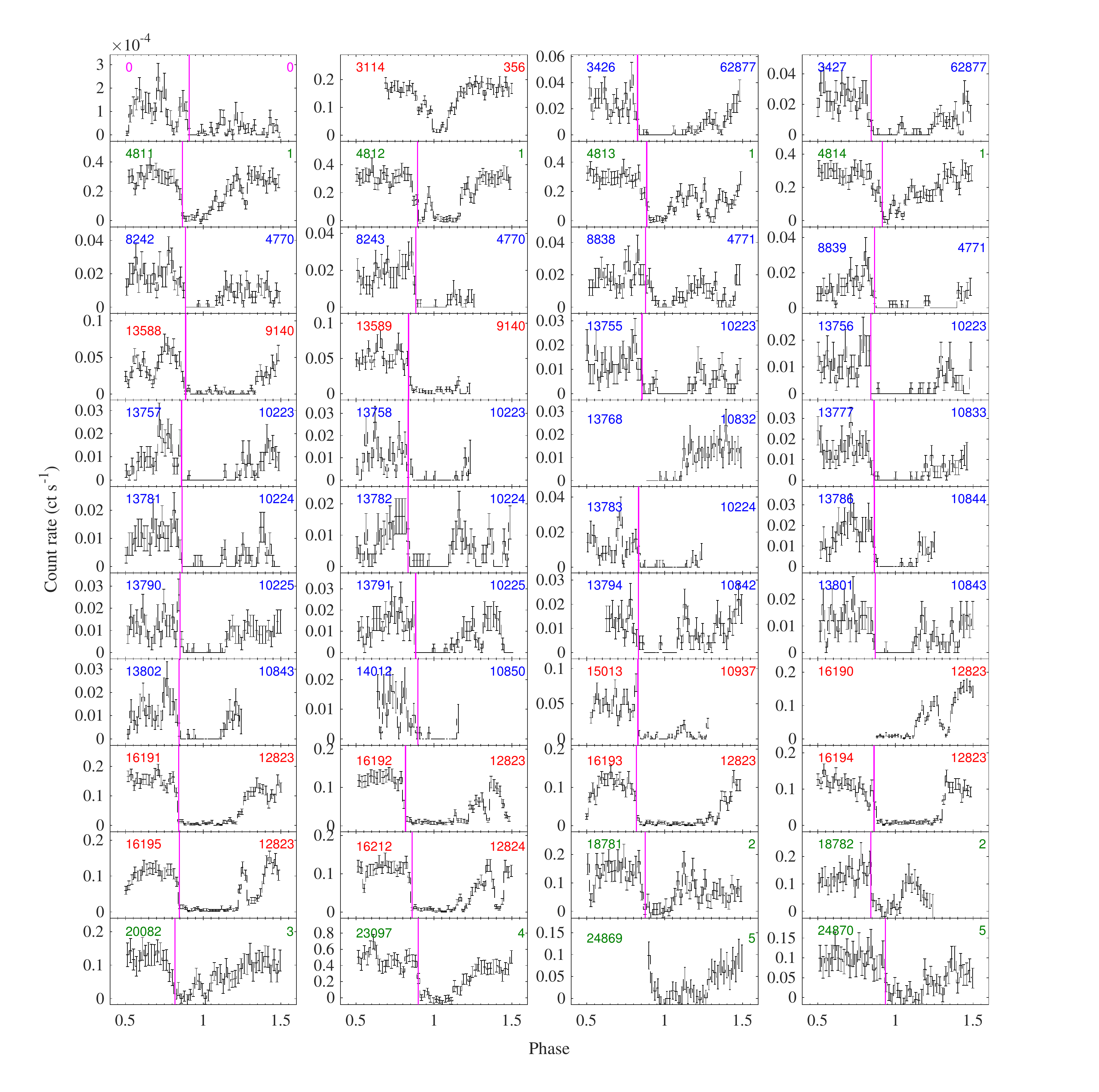}
    \caption{ X-ray lightcurves of CG\,X-1 sampled over twenty years. All lightcurves (except for the first one, from \rosat/HRI) are in the 0.3--8 keV band and are displayed with 500-s bins. Error bars represent the statistical combination of the error in the source counts and the error in the area-scaled background counts. They are folded on our best-fitting ephemeris; phase $\phi = 1$ is defined to occur at
    MJD $ 50681.437036 + 0.300579 \times N$ (d). $N$
    %MJD $ 50681.444551 + 0.300579 \times N$ (d). $N$ old
    is the orbital cycle, and is shown at the top left of each panel. We chose MJD 50681.437036 as the reference time (corresponding to $N=0$) because it is the count-weighted average between the times of the two {\it ROSAT} observations (rh702058a03 and rh702058a02) in 1997; the top left panel (labelled ``N=0'') is a folded \rosat/HRI lightcurve from those two 1997 observations. The colored numbers on the top right of all the other  panels are short forms of the corresponding observations IDs from which those lightcurves were extracted. Red numbers: \chandra/ACIS-S3 observations; blue numbers: \chandra/HETG observations; green numbers: \xmm/EPIC observations. ObsID 1, 2, 3, 4, and 5 are short for \xmm\ ObsID 0111240101, 0701981001, 0656580601, 0792382701, and 0780950201, respectively. The vertical magenta lines show the estimated mid-time of the eclipse ingress phase for that orbital cycle (See Section 3.2 for an explanation of how it was determined).}
%    Only observations covering more than one orbital cycle are shown here.   }
    \label{fig:alc}
\end{center}
\end{figure*}

For each \chandra\ or \xmm\ observation, we generated a 0.3--8 keV background-subtracted lightcurve, and rebinned each lightcurve to 500-s bins (Figure \ref{fig:alc}). 
%the error is the statistical combination of the error in the source count and the error in the background counts normalized with the area.
We then applied the phase dispersion minimization (PDM) method to compute the best-fitting period \citep{Stellingwerf1978, Schwarzenberg-Czerny1997}. 
%Since the PDM method is sensitive to the amplitude of the modulation, we normalized the count rates of all the lightcurves to 0--1. 
%{\bf {NEED TO EXPLAIN THIS SENTENCE MORE CLEARLY]}} 
The PDM method is more accurate if the amplitude of the modulation is constant. But the count rates measured by different instruments vary a lot, because of their different instrumental responses (Figure \ref{fig:alc}). Thus, we normalized the count rate of each observation by dividing the original count rate by the maximum count rate in each  observation. The value of the normalized count rate is in the range of 0--1.  
We managed to phase-connect all archival observations (Figure \ref{fig:alc}), with a best-fitting  period $P_{\rm best} = (25970.0\pm 0.1)$ s: an improvement in precision by a factor of 2000. 

\begin{figure}
\center{
\includegraphics[width=8.5cm]{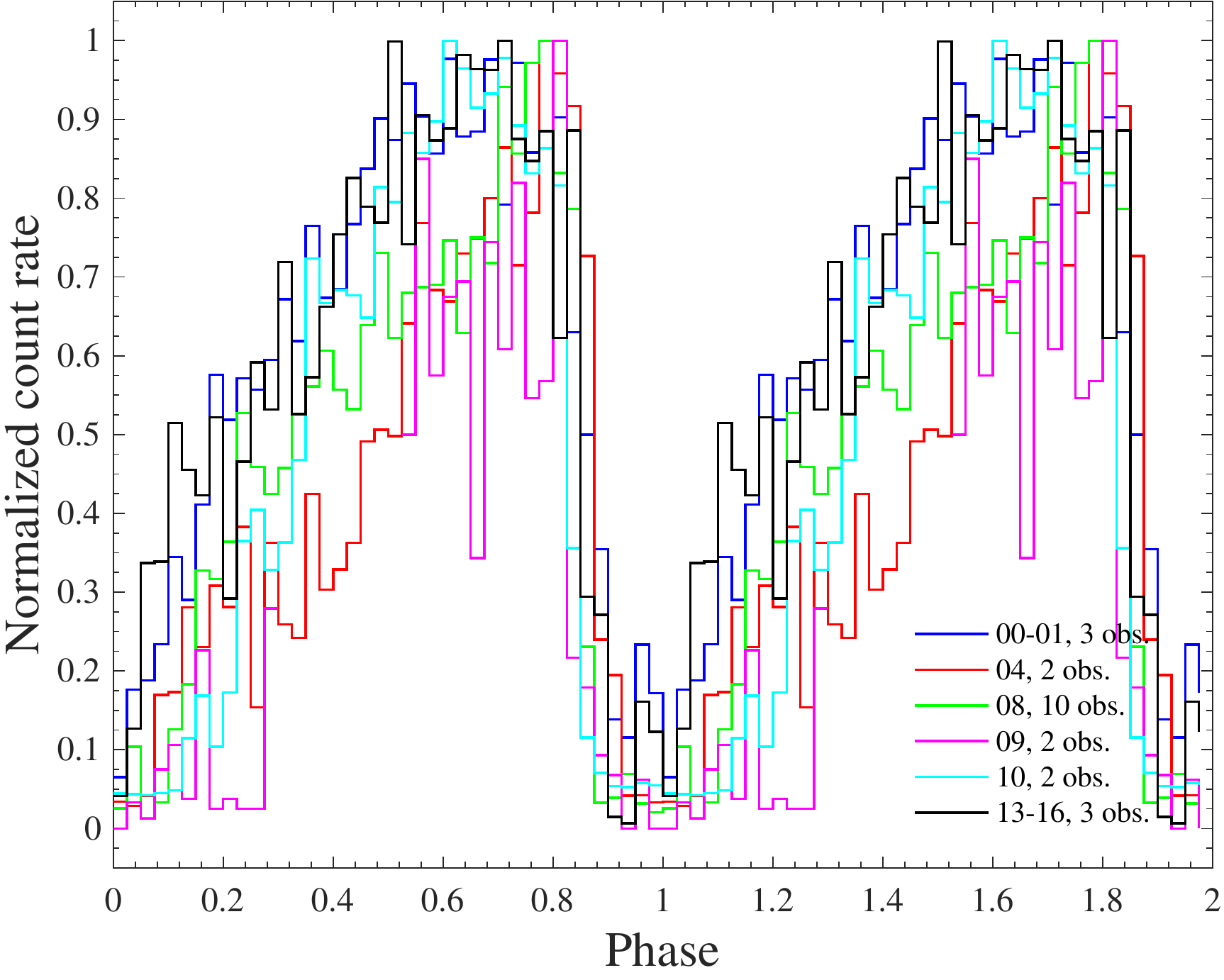}}
\caption{Folded profiles of the X-ray lightcurve, grouped into six observational epochs. The calendar years and number of observations for each lightcurve are shown in the inset box. In total, the six folded lightcurves include 22 observations, some of which cover more than one orbital cycle; see Figure \ref{fig:alc} for the lightcurves of the individual cycles. \\}
%They are repeated for two cycles for clarity. }
\label{fig:flc}
\end{figure}

We computed six folded lightcurves for different sub-groups of \chandra\ and \xmm\ observations (Figure \ref{fig:flc}): all those taken in 2000--2001; in 2004; in 2006; in 2008; in 2010; and in 2013--2016. Each of the six folded lightcurves confirms an approximate description of the {\it {average}} profile as a sharp ingress, a short, deep eclipse, and a slow egress. Using a fourth-order Fourier model, we fitted the eclipse section (defined as the bins with a count rate less than 30\% of the maximum count rate, in the phase interval between two consecutive peaks) of all the folded lightcurves simultaneously. We re-defined phase $\phi \equiv 1$ as the deepest point of the eclipse in the best-fitting model. As a first approximation, the phase of the eclipse is the same in each of the six folded lightcurves: we estimate that the deepest points of the six folded eclipses have only a small scatter of $\delta \phi \approx 0.034$ around the global best-fitting phase $\phi = 1$. (In Section 3.2, we will investigate this scatter further, and look for possible small changes in the period over the last 20 years.) This confirms that the X-ray eclipse is related to stable properties of the binary system: the most plausible explanation is that $\phi = 1$ corresponds to superior conjunction of the accreting compact object ({\it{i.e.}}, when it passes behind the star). However, the profile of each individual cycle is much more irregular and variable from cycle to cycle, as we have shown (Figure \ref{fig:alc}). We will attempt to explain the irregular profiles in Section 6.1.
 
In summary, assuming a constant period $P_{\rm best}$, our best-fitting ephemeris is $\phi = 1$ at MJD $50681.437036 + 0.300579 \times N$ (d).
%MJD $ 50681.444551 + 0.300579 \times N$ (d). old 
We have chosen to set the reference time ($N=0$) at an epoch covered by the \rosat\ observations\footnote{The \rosat\ observations span several time intervals between 1997 March and September. We arbitrarily defined the cycle $N=0$ as the one in which we reached 50\% of the counts in the stacked dataset.}. The cycle number $N$ of subsequent observations is plotted as a label in the top left corner of each frame of Figure \ref{fig:alc}. Choosing a different zeropoint (for example defining $N=0$ as the first \chandra\ observation) would obviously not change any physical interpretation.

%fig3
% from  cal_ingress_time.m  % pl_Figure_lc.m

%fig4 folded lc
% from pl_Figure_flc.m

\subsection{  Period derivative   }

%The ingress  of CG\,X-1  is  rather sharp and punctual, which can be used as an indicator of the changes of the orbital periodicity.
The next step of our analysis was to search for possible small changes in the binary period over the two decades of observations. This presents a practical challenge. Although the time of mid-eclipse (used to define the reference phase $\phi = 1$) is relatively easy to determine in all the lightcurves averaged over many cycles or several years (Figure \ref{fig:flc}), it is not obvious how to identify this point in any of the individual lightcurves (Figure \ref{fig:alc}), with eclipse durations and egress behaviours that differ markedly from cycle to cycle. Instead, we used the eclipse ingress as a phase marker for the individual lightcurves; the sharp flux drop leading to the eclipse is a feature that can be unequivocally identified in most of the individual lightcurves. It is already clear from a cursory inspection of Figure \ref{fig:alc} that the phase of mid-ingress differs from cycle to cycle; what we want to determine is whether this is a random scatter around a mean value, or represents a systematic drift, possible evidence of the period derivative $\dot{P} \neq 0$. 

First, we determined the average mid-time of the ingress from the stacked \chandra\ and \xmm\ lightcurves. We used the method developed by \citep{Hu2008}, which we adapted for our purposes. 
%is original used to parameterize  the arrival times of dippers and eclipse. But here we modified this method to detect the center time of ingress. 
We defined the count rate $I_0$ of the un-eclipsed high state as the average count rate from $\phi = 0.5$ to $\phi = 0.8$. We then introduced a parameter $\gamma_i \equiv \left(I_i-I_{i+1}\right)/(500\rm{s})$, to characterize the local slope of the lightcurve between two successive bins at times $t_{i}$ and $t_{i+1} = t_i + 500$s. 
%The boundary of the ingress is chosen from phase 0.7 to 1 by roughly guessing. 
Finally, we defined a mid-time $t_{\rm c}$ of the ingress section of a lightcurve as
\begin{equation}
t_{\rm c} =\frac{\sum_{i=1}^{N} \gamma_i \left(I_0-I_i\right)t_i}{\sum_{i=1}^{N} \gamma_i \left(I_0-I_i\right)},
\end{equation}
where $i$ is an integer index running from the left boundary ($i$=1) to the right boundary ($i$=N) of the ingress interval (approximately defined as the phase interval $0.7 < \phi < 1$), and $t_i$ and $I_i$ are the time and count rate of the $i$th bin, respectively. Eq.(1) is essentially a weighted average of the ingress times; the weights are the steepness of the decline times the total drop at a certain point. The reason why the average is weighted also by $\left(I_0-I_i\right)$ is because we want to reduce the uncertainty in the estimate of the time at which the lightcurve rolls over and the ingress starts; near the beginning of the ingress, $\left(I_0-I_i\right) \approx 0$. In any case, Eq.(1) provides a practical definition of a lightcurve feature that can be used to search for first-order period changes. Similarly, the physical width of the ingress is calculated as:
\begin{equation}
W=\left|\frac{\sum_{i=1}^{N} \gamma_i\left(I_0-I_i\right)\left(t_i-t_{\rm c}\right)^2}{\sum_{i=1}^{N}\gamma_i\left(I_0-I_i\right)}\right|^{1/2}.
\end{equation}

The mid-ingress phase for the stacked \chandra\ and \xmm\ observations occurs at phase $\phi = 0.85 \pm 0.03$, about 3896 s before mid-eclipse. Therefore, the predicted time of mid-ingress during an arbitrary cycle number $N$ is 
MJD $50681.391949  + 0.300579 \times N$ (d) $\equiv T_1 + P_{\rm best} \times N$. 
%MJD $ 50681.444551 + 0.300579 \times N$ (d).  old
%reference MJD $50681.43704 + 0.300579 \times N$ (d). new
The empirically determined location of $t_{\rm c}$ in each of the individual lightcurves plotted in Figure \ref{fig:alc} is marked with a magenta line.

%The ingress is sharp and  punctual relative to the egress, but the  arrival time of the ingress is slightly shifting over orbital cycle to cycle. For instance, the ingress time is at $\phi \approx0.8$ for \chandra\ ObsID 10937 and at $\phi \approx 0.9$ for \xmm\ ObsID 1, respectively. 
In order to investigate whether the period is changing with time over a 20-year timescale, we computed an "Observed minus Calculated" ($O-C$) diagram. 
%We used five {\it Chandra}/ACIS, eleven {\it Chandra}/HETG, and five {\it XMM-Newton}/EPIC observations. For the {\it Chandra}/ACIS, {\it XMM-Newton}/EPIC, and three of the {\it Chandra}/HETG observations, we had sufficient.
%We stacked seven of the {\it Chandra}/HETG observations (separated only by two weeks), to increase the signal to noise ratio; we used the observed mid-ingress phase from the corresponding   
%\sout{ \chandra\ ObsID 356 was not used {\bf {BECAUSE...?}}}
%
The predicted C values of the ingress times come from the ephemeris given above ($C = T_1 + P_{\rm best} \times N$), and are assumed to have no error. We calculated the O values from the data, using Eq.(1). In particular, we used one \rosat, four {\it Chandra}/ACIS, eleven {\it Chandra}/HETG, and five {\it XMM-Newton}/EPIC observations. For the {\it Chandra}/ACIS, {\it XMM-Newton}/EPIC, and three of the {\it Chandra}/HETG observations, we had sufficient counts to measure ingress times directly for individual cycles. We stacked the remaining eight {\it Chandra}/HETG observations (separated only by two weeks), to increase the signal to noise ratio; in that case, we measured the time difference between the average ingress phase in the stacked lightcurve (folded on the default ephemeris), and the predicted ingress phase. We used a similar method for those observations that covered multiple orbital cycles and therefore contained more than one ingress. In those cases, a single O-C datapoint was used for each observation, defined as the average of the individual time differences for all the ingresses covered during that  observation. We estimate the 1-$\sigma$ uncertainty (standard deviation) for each individual ingress measurement as $\approx$700s; smaller errors are associated to datapoints that are the average of multiple ingress times.

In an $O-C$ diagram, if the period remains constant with time, datapoints are scattered along a line:
   \begin{equation}
   O-C= (P_0-P_{\rm{best}}) \times N ,
   \end{equation}
where $P_0$ is the true period and $P_{\rm{best}}$ is the best-fitting period used to determine the values of C. If the fitted period is exactly equal to the true period, the $O-C$ diagrams follows a horizontal line around zero. Instead, if the orbital period has a small linear change, the datapoints follow a quadratic function: 
\begin{equation}
 O-C=c_1 + c_2 ~N+(1/2)P_{\rm{best}}\dot{P}\times N^2, 
\end{equation}
where $c_1$ and $c_2$ are normalization coefficients, and $\dot{P}$ is the period derivative.  

\begin{figure}
\center{
\includegraphics[width=8cm]{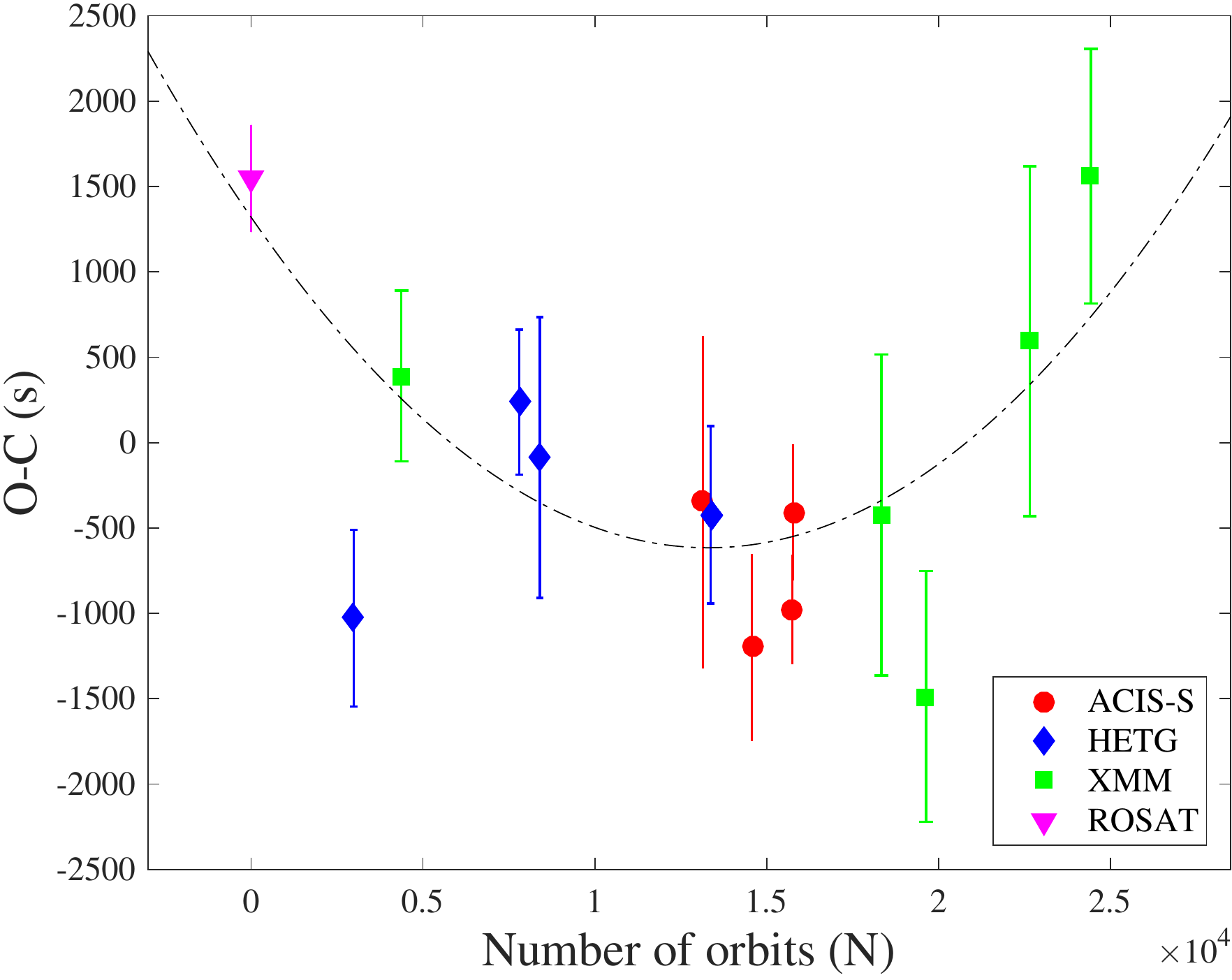}}
\caption{$O-C$ diagram for CG\,X-1, based on the empirically determined ingress times. The dashed line represents the best quadratic fit of the $O-C$ data. Datapoints have been color-coded according to the detectors used: magenta for \rosat, red for {\it Chandra}/ACIS, blue for {\it Chandra}/HETG, green for {\it XMM-Newton}/EPIC. For better statistics, we stacked eight \chandra/HETG observations separated by two weeks, {\it i.e.}, ObsIDs 10223, 10833, 10224, 10844, 10225, 10842, 10843, and 10850. \\}
\label{fig:oc}
\end{figure}

The resulting $O-C$ diagram is shown in Figure \ref{fig:oc}; it visually suggests an upwards curvature. We fitted the $O-C$ datapoints with a quadratic function. The best fitting curve (dashed-dotted line in Figure \ref{fig:oc}) 
is $O-C = 1320.6 - 0.29\times N + 1.09\times10^{-5}\times N^{2}$.
%9.0\times10^{-6} - 0.2\times N + 1465.3\times N^{2}$
%1465.3- 0.2\times N + 9.0\times10^{-6}\times N^{2}$. 
%P_dot/P = 10.2e-7 yr^-1, corresponding to the O-C solution listed above. The 90% error range (2.7 sigma) is P_dot/P =  (10.2 +/- 4.6)e-7 yr^-1. The 95% error range (3.8 sigma) is P_dot/P = (10.2 +/- 5.5)e-7 yr^-1. The value of P_dot/P is significantly > 0 to about 10 sigma (99.9% confidence level).
The orbital period derivative is $\dot{P} =  (8.4 \pm 3.8) \times 10^{-10}$ s s$^{-1}$, where the error range is the 90\% confidence level (2.7$\sigma$); this corresponds to $\dot{P}/P = (10.2\pm4.6) \times 10^{-7}$ yr$^{-1}$ at the 90\% confidence level.
%This quadratic fit gives a $\chi^2 (d.o.f)= (20.344/18)$.
The value of $\dot{P}/P$ is significantly $>$ 0 to approximately 10$\sigma$ (99.9\% confidence level).
As a further check, we used the F-test to evaluate the improvement from a linear fit to a quadratic fit. The low F-test probability ($p \approx 1.1 \times 10^{-3}$) confirms that the O-C datapoints follow a quadratic fit (curving upwards) better than a linear fit (constant period).
%The low F-test probability ($p=10^{-3}$) shows that the $O-C$ datapoints follow a quadratic fit better than a linear fit. }
This supports our conclusion that we are detecting a systematic increase of the binary period over 20 years.

%fig 4 O-C diagram
% from  cal_ingress_time.m   % old pl_Figure_o_c_all.m

\iffalse
\begin{deluxetable}{lccccc}
%\rotate
\centering
\tabletypesize{\footnotesize}
\tablewidth{0pt}
\tablecolumns{6}
\tablecaption{ Fitting of O-C. \label{tab:o_c}}
%\tablewidt
\tablehead{
 &  &  &  & & \\[5pt]
\colhead{Model} & \colhead{Reduced } & \colhead{ dof.} & \colhead{$\chi^2$} & \colhead{ftest} 
&\colhead{ftest}    \\
\colhead{} & \colhead{$\chi^2$ } & \colhead{ } & \colhead{} & \colhead{line} 
&\colhead{quadratic+sin}    \\[1pt]}
\startdata
line          & 1.292 & 19 & 31.716 &     - & 0.079 \\
quadratic     & 1.216 & 18 & 26.616 & 0.080 & 0.143 \\
sin           & 1.236 & 17 & 25.971 & 0.183 & 0.169 \\
quadratic+sin & 1.118 & 15 & 20.604 & 0.079 &     - \\
\enddata
\end{deluxetable}

\fi

%--------------------------------------------------------------------------------

\subsection{Duration of eclipse and egress phases}\label{setc:dphi_Lx}

Lightcurves folded over several cycles (Figure \ref{fig:flc}) show an eclipse lasting from $\phi \approx 0.9$ to $\phi \approx 1.1$, followed by a slow egress. In fact, as we have already mentioned, the profiles of the individual cycles tell a more complex story (Figure \ref{fig:alc}). In some cycles, such as those observed during \chandra\ ObsID 356 and the first three \xmm\ observations, the faintest phases last $\delta \phi \approx$ 0.15. At other epochs, such as \chandra\ ObsID 9140 and ObsID 12823, the observed flux remains close to zero for a longer time, $\delta \phi \approx 0.4$. In yet other cases, for example cycles $N=13,781$ and $N=13,782$ during \chandra\ ObsID 10224, we see a sequence of irregular dips and flares rather than a well-defined, single eclipse. This irregular behaviour cannot be explained with a simple stellar occultation of a point-like X-ray source; other factors must (also) be contributing to the duration of the periodic occultation and the variable dipping profile, as we shall discuss in Section 5.1.

We investigated whether there is a relation between the properties of the eclipsing phase in a cycle, and the unobscured luminosity before the ingress. However, because of the irregular dipping, in many orbital cycles it is difficult to determine when the eclipse ends (if indeed it is a true stellar eclipse) and the egress begins. Instead, we defined a different, empirical quantity that parameterizes the total duration of the most occulted phases during each orbital cycle. First, we determined the maximum count rate of a cycle as the average of the rates in the five 500-s bins with the highest count rates, in the phase range  $\phi=0.8$--$1.8$. Then, we defined a duration of the occultations as the number of 500-s phase bins ($N_{\rm {occ}}$) in which the count rate is lower than  40\% of the maximum count rate defined above. These bins typically include the candidate stellar eclipse around phase 1, as well as deep dips during the egress, often seen around phase 1.1--1.5; we used only observations with a complete phase coverage between 0.85 and 1.65. In terms of orbital phase, the duration of the occultations is $\Delta \phi_{\rm{occ}} = N_{\rm {occ}} \times 500/P_{\rm{best}} \approx 0.0193 N_{\rm {occ}}$. The uncertainty on the half-width was calculated as a Poisson error on the number of phase bins ({\it{i.e.}}, $0.0193 \sqrt{N_{\rm occ}}$).

Finally, in order to test whether the duration of the occultations is correlated with the peak luminosity, we used the online Portable, Interactive Multi-Mission Simulator\footnote{http://cxc.harvard.edu/toolkit/pimms.jsp.} ({\sc {pimms}}) version 4.9 to convert all count rates from different epochs and instruments to the effective count rate and unabsorbed luminosity for Cycle 3 \chandra\ ACIS-S3 (no grating) in 0.3--8 keV band. The spectral model used for the conversion is a power-law with photon index $\Gamma = 1.7$ and absorbing column density $N_{\rm H} = 1 \times 10^{22}\ cm^{-2}$. 
%Then we can compare the normalized count rates from \chandra\ and \xmm\ directly.

\begin{figure}[tp!]
\center
\includegraphics[width=8.5cm]{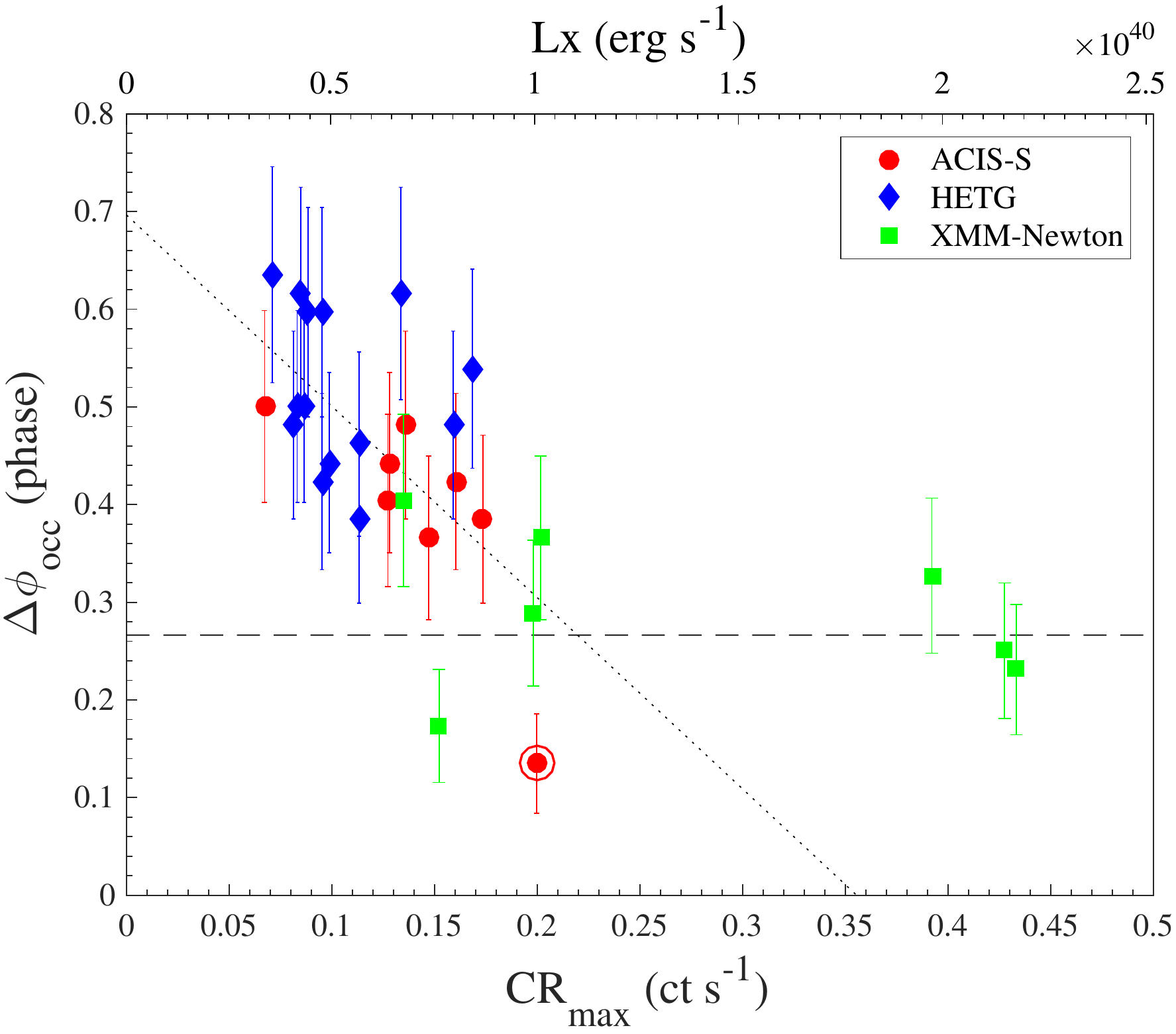}
\caption{ Duration of the occultation phase $\Delta \phi_{\rm {occ}}$ versus maximum count rate of the bright phase CR$_{\rm{max}}$; see Section 3.3 for the definition of occultation phase (eclipse and dips). Red dots, blue diamonds, and green squares represent \chandra/ACIS-S3, HETG, and \xmm\ observations, respectively; one datapoint (\chandra\ ObsID 356) is flagged with a red circle, to signal that its count rate is underestimated because of pileup. We converted all measured \xmm\ and \chandra\ count rates to an equivalent Cycle-3 \chandra/ACIS-S3 count rate, using a power-law model in {\sc {pimms}}. The dashed lines are the best linear fit for count rates $\lesssim$0.2 and $\gtrsim$0.2, respectively. We also converted count rates to peak unabsorbed 0.3--8 keV luminosities (labelled on the upper X axis) using {\sc {pimms}}, and assuming a distance at 4.2 Mpc.\\}
\label{fig:dphi}
\end{figure}

We found a correlation (Figure \ref{fig:dphi}) between the duration of the occultations and the peak luminosity, represented by the normalized maximum count rate $CR_{\rm max}$. At low luminosities ($L_{\rm X} \lesssim 10^{40}$ erg s$^{-1}$), the duration of the occultations is linearly anticorrelated with the count rate, as
%This anti-correlation of these two variables ($CR_{\rm max}$) is obviously, and can be fitted well by a linear line: 
$\Delta \phi_{\rm {occ}} = (-1.96 \pm 0.97) \times CR_{\rm max} + (0.70 \pm0.13)$.
%The number in the bracket is the uncertainty. 
To quantify the statistical significance of the correlation, we calculated the Spearman correlation coefficient $\rho$: we obtained $\rho \approx-0.73$, with a p value of 5$\times10^{-6}$.
At higher luminosities ($L_{\rm X} > 10^{40}$ erg s$^{-1}$), the duration of the occultations saturates around $\Delta \phi_{\rm {occ}} \approx 0.2$--0.3, independent of luminosity. The precise slope of the correlation depends on how the peak count rate and the occultation bins are defined, but the existence of a significant trend is a robust result.

% figure egress width
% from /Users/qyl/work/cgx1/matlab/pl_Figure_degress_cr.m

%--------------------------------------------------------------------------------

\section{   X-ray Spectral Results   }

\subsection{Outline of our spectral analysis}

The next question we addressed is whether/how the spectrum of the observed photons changes between fainter and brighter sections of an orbital cycle, and between orbital cycles over the years. 
%or instead only the normalization varies.  
We did this in three steps, as explained below. 

First (Section 4.2), we split a selected sample of high signal-to-noise lightcurves (from \chandra/ACIS observations) into phenomenological sub-structures (ingress, eclipse, dips, egress, bright phase) and compared the cumulative energy distribution of the observed photons in the various sub-structures. The objective of this part of our analysis is to detect spectral changes in a model-independent way. 

Second, we did a phase-resolved spectral modelling of the four highest-quality \chandra/ACIS-S3 observations (Section 4.3) and all five \xmm/EPIC observations (Section 4.4). The objective of this part of our analysis is to model how the fit parameters change between the brighter and fainter sections of an orbital cycle. Thus, we split each of the selected observations into three phase groups: a bright phase, an intermediate phase, and a faint phase, drawing on the results of our lightcurve analysis. We defined the bright phase as the time intervals when the count rates are higher than 70\% of the maximum count rate for that orbital cycle; the faint phase as the intervals when the count rates are less than 15\% of the maximum count rate for \xmm, and 10\% for \chandra;
%\footnote{ \qyl{The threshold chosen to define the faint phase in the  \xmm\ and \chandra\ lightcurves is different because their eclipse profiles are very different (see \qyl{Figure \ref{fig:alc}}). \chandra\ eclipses are long and flat, while \xmm\ eclipse are narrow and fluctuating.}}; 
the intermediate phase as the time bins in between the faint and the bright phase. To increase the number of counts in each of the three sub-intervals, we combined their spectra from the four \chandra\ datasets; instead, we had enough photons to analyze the five \xmm\ observations individually. For \xmm, we combined pn and MOS spectra together with the {\sc sas} task {\it epicspeccombine} to increase the signal-to-noise ratio. 

Finally, in the third step of our analysis (Section 4.5), we extracted and modelled spectra integrated over entire orbital cycles, from six \chandra\ observations and five \xmm\ ones, between 2000 and 2018; the datasets used for this modelling are marked by asterisks in Table \ref{tab:obs}). We determined average and peak fluxes and luminosities during those observations.

\subsection{   Model-independent photon energy distribution    }\label{sect:ks}

%fig7 dip k-s
% from pl_Figure_kstest_tc.m
\begin{figure*}%[p]
\center{
\includegraphics[width=19cm]{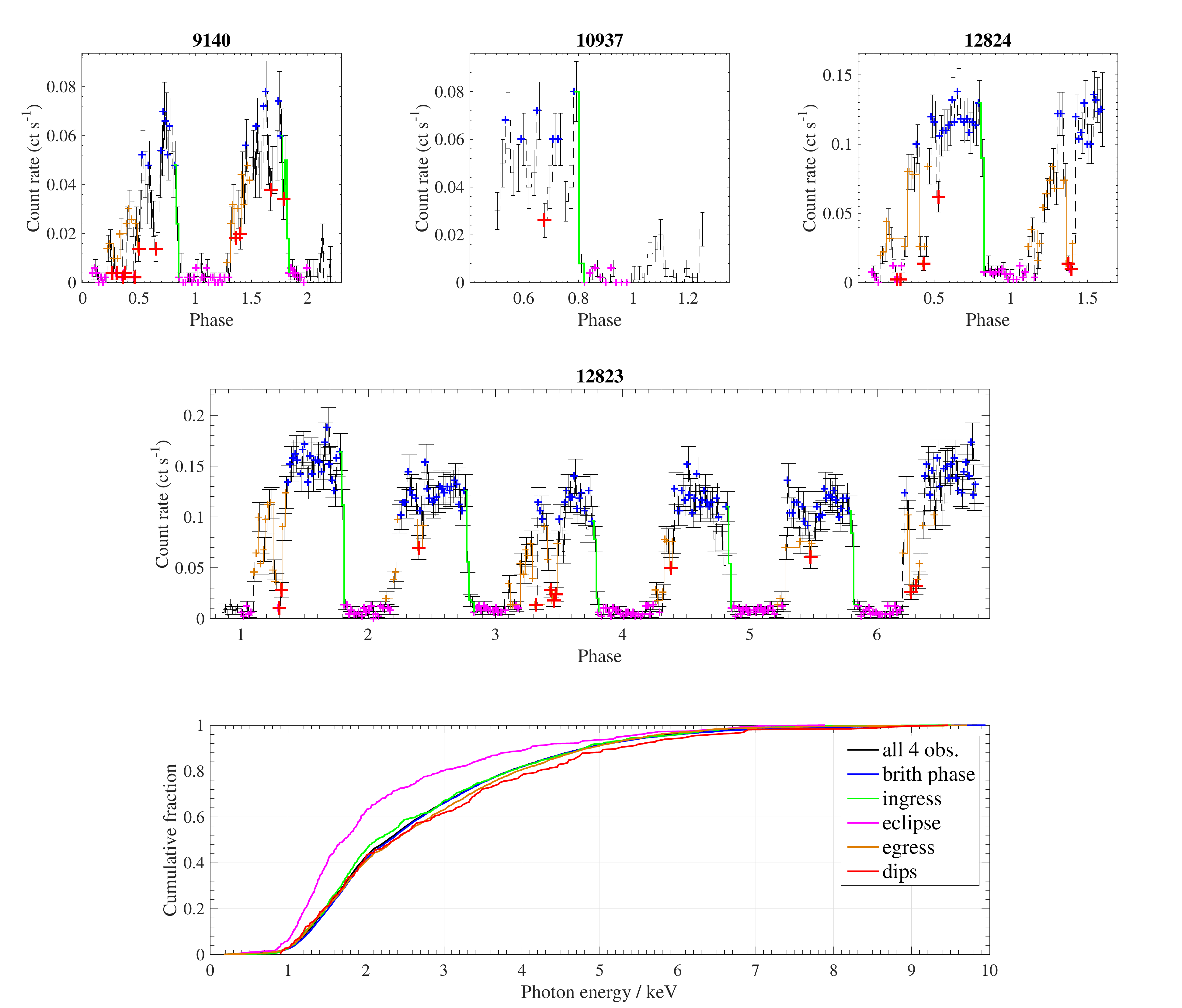}}
\caption{
%Cumulative distribution of the detected photon energy at different orbital phases. 
Top four panels: X-ray lightcurves extracted from four \chandra\ observations, which we then used for a K-S test to compare the cumulative energy distribution of the detected photons at different phases. We empirically divided the orbital cycle into different structures, identified with different colours: bright, ingress, faint, and egress phases are plotted in blue, green, magenta, and ocher, respectively; the time bins defined as dips are marked with red crosses. Bottom panel: cumulative fraction of photon numbers versus photon energy, normalized to 1 for each lightcurve structure; the colours correspond to those in the lightcurve panels. The comparison shows that the residual emission in the faint phase is softer than the emission at all other times in the orbital cycle. \\}
\label{fig:ks}
\end{figure*}
%\clearpage

%We didn’t use Chandra ObsID 365 and 356, because the exposure time of ObsID 365 is too short to define sub-structures and  it  has  discrepant  spectral  index  from  all  the  others.   ObsID 365 is suffering from high fraction of pileup (40\%) (Weisskopf et al. 2004).  But we used these two observations for obtaining long-time lightcurve in Section 4.4

The four \chandra\ observations chosen for our model-independent study of the photon energy distribution are ObsIDs 9140, 10937, 12823 and 12824, covering a total of about 11 orbital cycles\footnote{We did not use \chandra\ ObsID 365, because its exposure time is too short to enable a meaningful definitions of phase substructures. Also, we did not use \chandra\ ObsID 356 because of its  high pile-up fraction, $\approx$40\%. However, we did model the individual spectra of those two observations among the others in Section 4.5.}. The advantage of those particular observations is that it is relatively straightforward to identify five main sub-structures in the background-subtracted lightcurves: faint phase, bright phase, ingress, egress and dips; the five sub-structures are colour-coded in Figure \ref{fig:ks}.  More specifically, for this part of the analysis we defined a maximum count rate as the average value of the ten 500-s bins with the highest count rate during each observation; we then defined the faint phase as the time intervals when the count rates are lower than 10\% of the maximum count rate; the bright phase as the time intervals when the count rates are higher than 70\% of the maximum count rate; the ingress and egress phases as the intervals of decreasing and increasing count rates, respectively, between the bright and faint phases. The definition of a dip is somewhat more arbitrary, but corresponds to an approximate flux drop of at least a factor of 2 followed by an immediate recovery within $\lesssim 2000$ s; they are colour-coded as red datapoints in Figure \ref{fig:ks}. To extract the observed net counts from the dips, we considered the time of the local minimum count rate as the mid-time of a dip, and took the photons falling within $\pm$200 s of that mid time, in the unbinned event file. 
%starting and ending time of a dip to be the time 200 seconds prior or after the mid-time. There are 263 photons in total detected from all the dips.

We performed the Kolmogorov-Smirnov (K-S) statistical test for the null hypothesis that the cumulative photon energy distributions of faint, bright, ingress, egress and dip phases are the same. We find (Figure \ref{fig:ks}, bottom panel) that photons from the last four (bright, ingress, egress and dips) of those five structures do indeed follow the same energy distribution. Instead,  photons from the faint intervals are significantly softer. The K-S test rejects the null hypothesis that the faint intervals follow the same distribution as the other intervals, with a p value of $7\times10^{-8}$.  This suggests the presence of at least two emission components: a bright, harder one, and a faint, softer one that is seen as a residual component when the other component is occulted.

%--------------------------------------------------------------------------
\subsection{ \chandra\ spectra} \label{sect:spec_fit_chandra}

% from /Users/qyl/work/cgx1/chandra/evt/bright/4obs
\begin{deluxetable*}{lccccccc}
\centering
\tablewidth{0pt}
\tablecolumns{8}
\tablecaption{Best-fitting parameters of phase-resolved spectra from combined \chandra\ ObsIDs 9140, 10937, 12823, and 12824 \label{tab:spec_fit_one}}
\tablehead{
 %&  &  &  & \\[5pt]
 \colhead{Phase} & \colhead{ $N_{\rm H, intr}$} & \colhead{ $\Gamma$/ kT (keV)} &   \colhead{ p} &  \colhead{$N_{\rm H,absori}$} & \colhead{\Fx} & \colhead{\Lx}  &\colhead{$\chi^2_{\nu}$ (dof)}   \\[3pt]
 & $(10^{22}$ cm$^{-2})$ & & & $(10^{22}$ cm$^{-2})$ &$(10^{-13}$ \ergcms) & $(10^{38}$ \ergs) & \\[-3pt]}
\startdata
\\[-3pt]
\multicolumn{3}{l}{{\it tbabs}$_1 \times${\it tbabs}$_2 \times${\it po}}\\[3pt]
Bright  &  0.50$^{+0.06}_{-0.05}$   &  1.76$^{+0.06}_{-0.06}$  & -  & - & 17.1$^{+0.4}_{-0.4}$  &66.5$^{+3.1}_{-2.7}$  & 0.87 (267)\\[3pt] % bright_po.xcm
Intermediate    & 0.14$^{+0.11}_{-0.10}$ &  1.27$^{+0.14}_{-0.13}$ & -  &  - & 6.4$^{+0.4}_{-0.4}$  &18.2$^{+1.1}_{-1.0}$ & 0.85 (100)\\[3pt]
Faint      & $< 0.25$   &  2.42$^{+0.38}_{-0.29}$ & - &  - & 0.57$^{+0.06}_{-0.07}$ & 3.3$^{+1.7}_{-0.6}$  &  0.76 (30)\\
% bremss
 & &  &  & & &  &   \\[0.01pt]
\hline   
  & &  &  & & &  &   \\[0.01pt]
  \multicolumn{3}{l}{{\it tbabs}$_1 \times${\it tbabs}$_2 \times${\it bremss} }\\[3pt]
Bright  &   0.38$^{+0.04}_{-0.04}$    &  8.39$^{+1.36}_{-1.09}$ &- &-& 16.7$^{+0.4}_{-0.4}$ & 56.5$^{+1.4}_{-0.3}$ &  0.85 (267)\\[3pt] %bright_bremss.xcm 
Intermediate   &  0.13$^{+0.09}_{-0.04}$      &  $>24.3$  &-  &-& 5.8$^{+0.3}_{-0.3}$  &  17.4$^{+1.0}_{-0.9}$ & 0.84 (100)\\[3pt]
Faint   & $< 0.82$   &  2.34$^{+0.45}_{-0.53}$  &- &-& 0.52$^{+0.04}_{-0.06}$ &  2.27$^{+0.17}_{-0.16}$ & 0.91 (30)\\
% diskbb
 & &  &  & & &  &   \\[0.01pt]
\hline   
  & &  &  & & &  &   \\[0.01pt]
\multicolumn{3}{l}{{\it tbabs}$_1 \times${\it tbabs}$_2 \times${\it diskbb} }  \\[3pt]
Bright  &0.19$^{+0.03}_{-0.03}$   &  1.75$^{+0.08}_{-0.08}$  &- &- & 15.8$^{+0.4}_{-0.4}$  & 45.9$^{+1.0}_{-1.0}$ & 0.95 (267)\\[3pt]%bright_diskbb.xcm 
Intermediate   & $< 0.05$     & 2.50$^{+0.04}_{-0.03}$  &- &- & 6.0$^{+0.4}_{-0.4}$  &  15.6$^{+0.8}_{-0.8}$ & 0.82 (100)\\[3pt]
Faint   &  $< 0.82$    & 0.80$^{+0.10}_{-0.09}$    &- &- & 0.46$^{+0.05}_{-0.05}$  & 1.76$^{+0.13}_{-0.13}$ & 1.36 (30)\\
% diskpbb
 & &  &  & & &  &   \\[0.01pt]
\hline   
  & &  &  & & &  &   \\[0.01pt]
\multicolumn{3}{l}{{\it tbabs}$_1 \times${\it tbabs}$_2 \times${\it diskpbb}}\\[3pt]
Bright & 0.49$^{+0.06}_{-0.05}$  & 7.1$^{+\ast}_{-1.2}$  & 0.54$^{+0.01}_{-0.01}$ & - &  17.0$^{+0.4}_{-0.4}$ & 60.1$^{+9.2}_{-3.0}$ & 0.86 (266)\\[3pt] % bright_diskpbb.xcm
Intermediate     &$< 0.17$   & 2.68$^{+6.77}_{-0.68}$   & 0.73$^{+0.06}_{-0.12}$ &  - &   6.0$^{+0.5}_{-0.5}$  & 15.8$^{+2.8}_{-1.2}$& 0.83 (99)\\[3pt]
Faint     & $< 0.06$   & 1.47$^{+0.69}_{-0.37}$   & 0.50$^{+0.03}_{-0.05}$ &  -  & 0.53$^{+0.08}_{-0.7}$    & 2.53$^{+0.19}_{-0.28}$ & 0.86 (29)\\
 & &  &  & & &  &   \\[0.01pt]
\hline   
  & &  &  & & &  &   \\[0.01pt]
\multicolumn{3}{l}{{\it tbabs}$_1 \times${\it absori}$\times${\it tbabs}$_2 \times${\it po} }  \\[3pt]
Bright  &  0.54$^{+0.08}_{-0.08}$   &  2.03$^{+0.15}_{-0.14}$  & -  & 1.75$^{+1.03}_{-0.89}$ &16.8$^{+0.4}_{-0.4}$  & 90.1$^{+17.8}_{-13.1}$  & 0.83 (265)\\[3pt] % bright_po_absori.xcm
Intermediate    & 0.31$^{+0.21}_{-0.18}$ &  1.89$^{+0.41}_{-0.43}$ & -  &  3.8$^{+3.1}_{-2.7}$ & 6.1$^{+0.4}_{-0.4}$  &30.4$^{+19.2}_{-9.7}$ & 0.80 (98)\\[3pt] %eclipse_po_absori.xcm
Faint      & 0.12$^{+0.09}_{-0.12}$   &  2.63$^{+0.25}_{-0.31}$ & - &  2.3$^{+4.1}_{-2.3}$ & 0.58$^{+0.07}_{-0.06}$ & 5.3$^{+1.1}_{-2.4}$  &  0.73 (28)\\
 & &  &  & & &  &   \\[0.01pt]
\hline   
  & &  &  & & &  &   \\[0.01pt]
\multicolumn{3}{l}{ {\it tbabs}$_1 \times${\it tbabs}$_2 \times${\it bbodyrad}}  \\[3pt]
Bright & $<0.01$  & 0.87$^{+0.01}_{-0.01}$  & - & - &  14.0$^{+0.3}_{-0.3}$ & 35.7$^{+0.7}_{-0.7}$ & 1.76 (267)\\[3pt] % bright_bbodyrad.xcm
Intermediate     & $< 0.01$   & 0.99$^{+0.04}_{-0.04}$   & -  &  - &   5.0$^{+0.3}_{-0.3}$  & 12.4$^{+0.7}_{-0.6}$& 1.46 (100)\\[3pt]
Faint    & $< 0.01$   & 0.49$^{+0.05}_{-0.05}$   & - &  -  & 0.40$^{+0.05}_{-0.04}$    & 1.4$^{+0.1}_{-0.1}$ & 2.25 (30)\\
 & &  &  & & &  &   \\[0.01pt]
\enddata
\tablenotetext{}{\hskip-2.5pt {\it Notes}:
{\it tbabs}$_1$ is the line-of-sight absorption in the direction of Circinus, fixed at $N_{\rm H} = 0.56 \times 10^{22}$ cm$^{-2}$; {\it tbabs}$_2$ is the intrinsic neutral absorption; {\it absori} is the ionized absorption. $F_{\rm X}$ is the absorbed flux in the 0.3--8 keV band; $L_{\rm X} \equiv 4 \pi d^2 F_{\rm X}$ is the emitted 0.3--8 keV luminosity. See Section 4.1 for the definition of "bright", "intermediate" and "faint" phase intervals.\\
}
\end{deluxetable*}
% 0.3-8keV flux bin with 15 counts for eclipse and egress, 20 for brightt

We used the same four \chandra\ observations selected in Section 4.2 (ObsIDs 9140, 10937, 12823 and 12824). We extracted three phase-resolved spectra, averaged over the four observations but distinguished by count-rate brackets; we shall refer to them as the bright-phase, intermediate-phase and faint-phase spectra (or, more simply, the bright, intermediate and faint spectra). 
First, we fitted the three phase-resolved spectra independently, in the 0.5--7 keV range, using standard one-component models suitable to X-ray binaries: {\it power-law}, {\it bremsstrahlung}, {\it diskbb}, {\it diskpbb}, and {\it bbodyrad}. The emission components were convolved with two neutral absorbers ({\it tbabs} model): one for the Galactic line-of-sight absorption (column density $N_{\rm H}$ fixed at $5.6 \times 10^{21}$ cm$^{-2}$, from \citealt{Kalberla2005}) and one left free, for the intrinsic absorption inside the Circinus galaxy and the binary system. Three of those models ({\it power-law}, {\it bremsstrahlung} and {\it diskpbb} with $p < 0.75$) give good fits for all three phases (Table 2); the {\it diskbb} model is significantly less good, both for the bright phase and for the faint phase; the {\it bbodyrad} model is the worst one, especially for the faint phase, which has an unacceptable $\chi^2_{\rm \nu} =2.25$. The phenomenological interpretation is that both the bright and the intermediate spectra have a low degree of curvature in the {\it Chandra} bandpass, so they are best represented either by a power-law or by the (relatively flat) low-frequency section of a thermal continuum component ({\it e.g.}, bremsstrahlung with a temperature $>$8 keV, or p-free disk with peak temperature $>$2.5 keV). The intermediate-phase spectrum has a harder (flatter) slope than the bright spectrum, at least in the $\approx$1--5 keV range, but a lower intrinsic $N_{\rm H}$. Conversely, the faint spectrum (dominated by residual emission in the eclipse) is significantly softer (steeper). 

Instead of a simple power law, we also tried Comptonization models (in particular, {\it comptt}, {\it simpl} $\times$ {\it diskbb}, and {\it diskir}). However, such models add a layer of complexity and additional free parameters, without providing any improvement to the fit. Below 1 keV, the thermal seed component of the Comptonization models is unconstrained because of moderately high absorption; the possible high-energy downturn above $\sim$5 keV (typical of ULXs, \citealt{Stobbart2006,Gladstone2009,Sutton2013,Walton2018}) is also unconstrained because of low signal-to-noise in the {\it Chandra} spectra. 
Thus, we stick to the simple power-law model for the {\it Chandra} analysis.

Before we can attempt a more physical interpretation of the spectra, consistent with the proposed super-Eddington HMXB scenario, we need to account for one additional source of absorption, from ionized gas. We do that by adding an {\it absori} component. Here, we take the power-law model as the benchmark to determine whether the addition of an ionized absorber improves the fit. 
%(but we obtain similar conclusions if we apply the ionized absorber to the bremsstrahlung model).
We find that an ionized absorber with column density $N_{\rm H} \sim$ a few $10^{22}$ cm$^{-2}$ and an ionization parameter $\sim$ a few 100 does provide a significant improvement for both the bright and the intermediate spectra (Table 2). For the bright spectrum, the goodness-of-fit improves from $\chi^2_{\nu} = 232.7/267$ to $\chi^2_{\nu} = 218.7/265$: this is significant to $>$99.9\% probability\footnote{ Model  {\it tbabs$\times$ absori $\times$ tbabs $\times$ po} is equivalent to model {\it tbabs$\times$ tbabs $\times$ po}, when $N_{\rm H}$ from the {\it absori} component is equal to zero. The significance level is calculated using a $\chi^{2}$ distribution with two degrees of freedom,  {\it i.e.} $\xi$ and $N_{\rm H}$ from the {\it absori} component.} . For the intermediate spectrum, the improvement is from $\chi^2_{\nu} = 85.0/100$ to $\chi^2_{\nu} = 78.2/98$, significant at the $>$95\% level. A similar amount of ionized absorption is also consistent with the faint spectrum; however, in that case, because of the lower signal-to-noise level, the ionized absorber only improves the fit at the 1$\sigma$ level ($N_{\rm H} = 2.3^{+4.1}_{-2.3}\times 10^{22}$ cm$^{-2}$).

Moreover, when the ionized absorber is included, we find that the intrinsic power-law slope of the bright and intermediate spectra is consistent with being the same ($\Gamma \approx 2.0$); the reason the intermediate spectrum looks flatter is because of a factor-of-two higher column density of the ionized absorber. The faint spectrum remains significantly steeper ($\Gamma \approx 2.6$) than the other two (Table 2).

We can now start to identify some physical properties of the spectral evolution. The main reason for the increase in observed count rates from the intermediate to the bright spectrum is neither a dramatic decrease in absorption column density, nor a state transition in the intrinsic emission properties. There are changes in the fitted column densities of ionized and neutral components, but they only affect the shape of the spectrum below 2 keV. Instead, to a first approximation, the main difference between bright and intermediate spectra is consistent with a reduced normalization of the broadband emission component during the egress and dipping phases. Considering that such evolution and happens regularly during each orbital cycle, we consider it unlikely that it is due to intrinsic changes in the source emission. A much less contrived explanation (also by analogy with other dipping X-ray binaries) is that the flux changes are due to variable partial covering of the emission region by clumps of optically thick material.

We tested this physical interpretation with a new spectral model. We fitted the bright and intermediate spectra simultaneously, with a {\it tbabs} $\times$ {\it tbabs} $\times$ {\it absori} $\times$ {\it pcfabs} $\times$ {\it power-law} model, keeping both the slope and the normalization of the power-law component locked for the two spectra, while allowing column densities and ionization parameters to vary independently. The partial-covering absorber modelled with {\it pcfabs} has to be Compton thick, with $N_{\rm H} \gtrsim 2 \times 10^{24}$ cm$^{-1}$, so that it blocks at least 99\% of the flux below 7 keV; we also assumed a covering fraction $f = 0$ for the spectrum in the bright phase. We find that this model provides the best fit ($\chi^2_{\nu} = 297.7/363 = 0.82)$. The photon index of the intrinsic emission component is $\Gamma \approx 2.1$ and the partial covering fraction of the intermediate spectrum is $f \approx 0.60$. The average intrinsic (de-absorbed) luminosity during the four {\it Chandra} observations used in our spectral modelling is $L_{\rm X} \approx 9 \times 10^{39}$ erg s$^{-1}$, in the 0.3--8 keV band. (Higher luminosities in excess of 10$^{40}$ erg s$^{-1}$ were found during some of the {\it XMM-Newton} observations.)

We also recovered the result that the faint spectrum (residual emission in eclipse) is softer and less absorbed than the emission out of eclipse. It cannot be explained with the same $\Gamma \approx 2$ model used for the intermediate and bright spectra. The simplest way to model this spectrum is to assume that the dominant emission component seen in the bright and intermediate spectra is completely blocked by the opaque screen ($f \equiv 1$ in {\it pcfabs}) and there is instead a different emission component significantly detectable only in eclipse. We model this component also with a simple power-law, for want of higher signal-to-noise spectra; we find a photon index $\Gamma \approx2.6$ and a luminosity \Lx$\approx 5\times 10^{38}$ \ergs\ in the 0.3--8.0 keV band (Table 3). 

It is plausible (in the framework of our interpretation of the system) that part or all of the softer component seen in the eclipse spectra is also present in the intermediate and bright spectra. To check that, we refitted all three spectra simultaneously, including the eclipse-phase component as an additional (fixed) component in the intermediate and bright spectra. However, this does not change their best-fitting parameters, because the residual component is lost in the noise of the much brighter out-of-eclipse spectra. Therefore, for simplicity, we ignore this additional term when fitting the bright and intermediate spectra, for {\it Chandra} and, in the next section, for {\it XMM-Newton}.

%------------------------------------------------
\subsection{ \xmm\ spectra}

%{\bf PAR STILL TO BE POLISHED}

% new: /Users/qyl/work/cgx1/matlab/pl_spectra_chandra.m,chandra_spec_4obs.pdf
% from pl_spectra_xmm
% best xmm spectra
\begin{figure*}%[ht!]
%\centering
\resizebox{ \hsize}{!}{
\includegraphics[width=8cm]{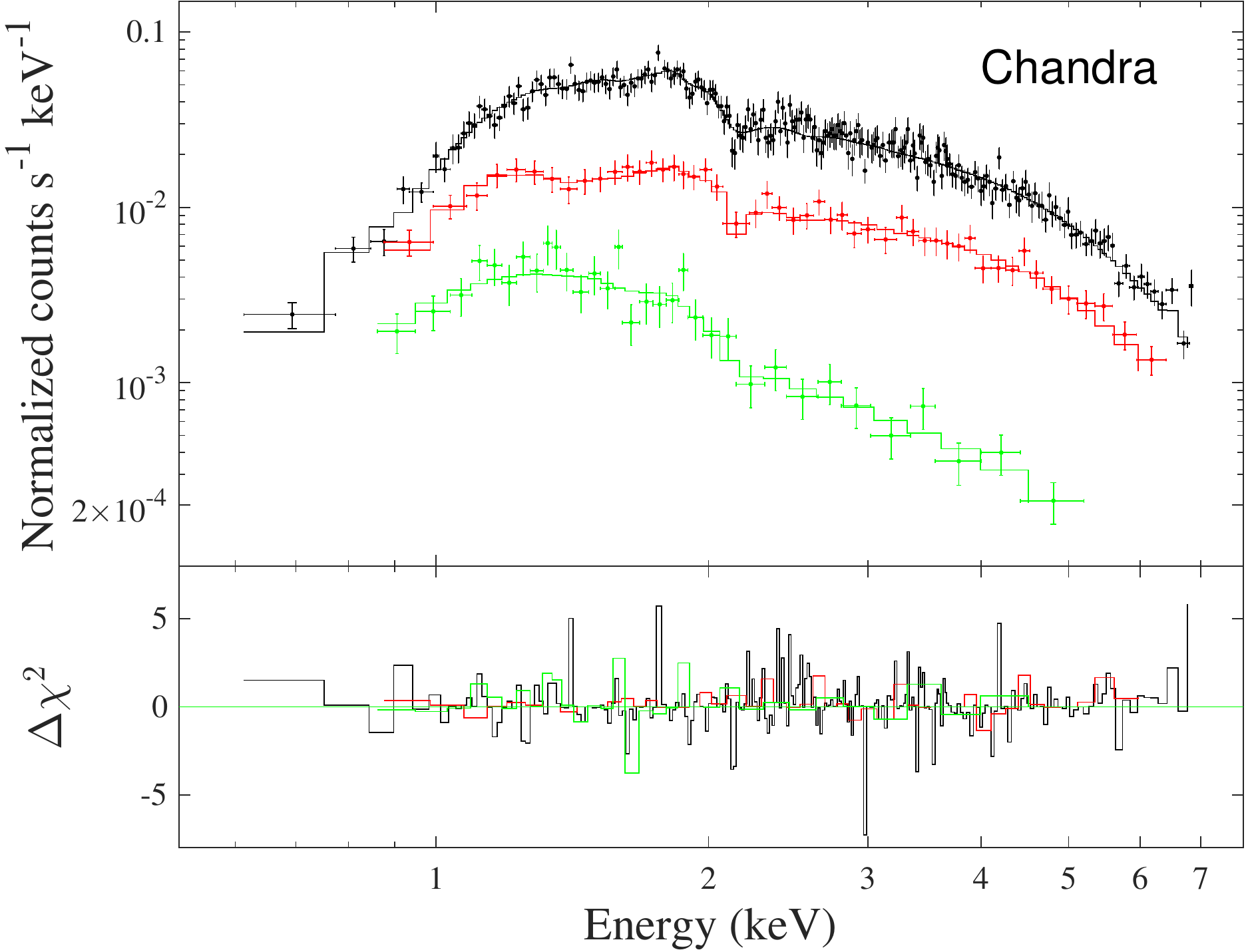}\hspace{2mm}
\includegraphics[width=8cm]{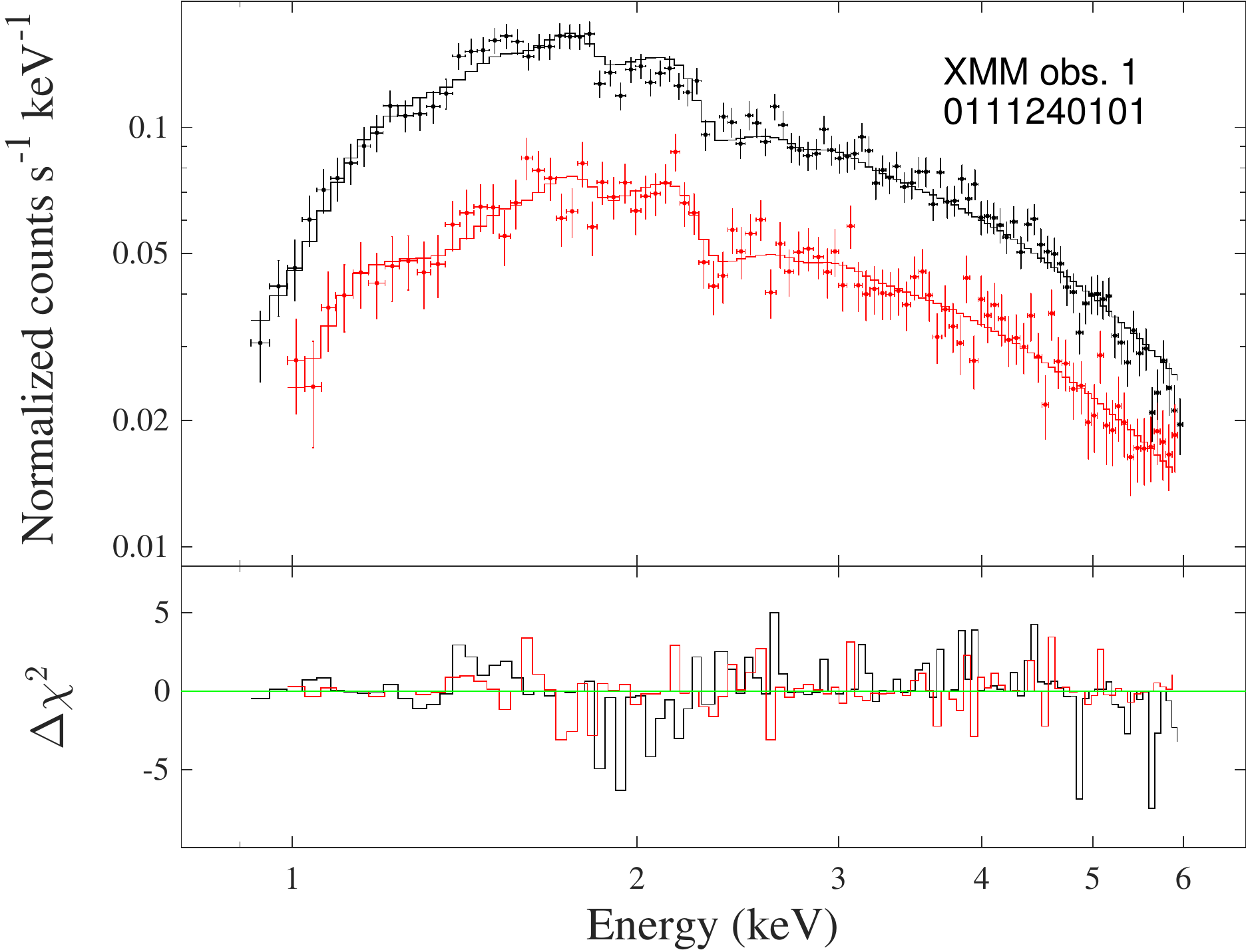}}\vspace{2mm}
\resizebox{ \hsize}{!}{
\includegraphics[width=8cm]{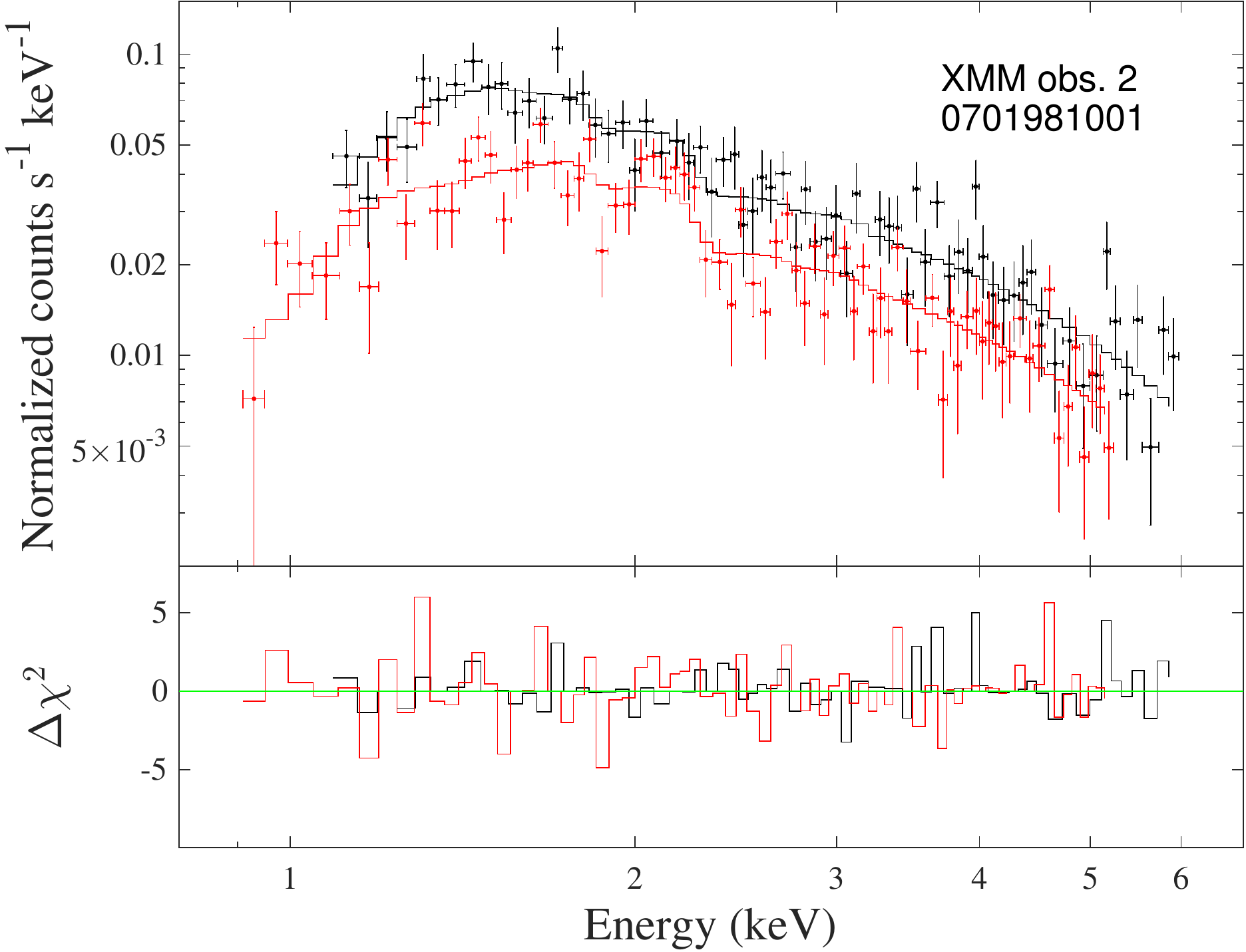}\hspace{2mm}
\includegraphics[width=8cm]{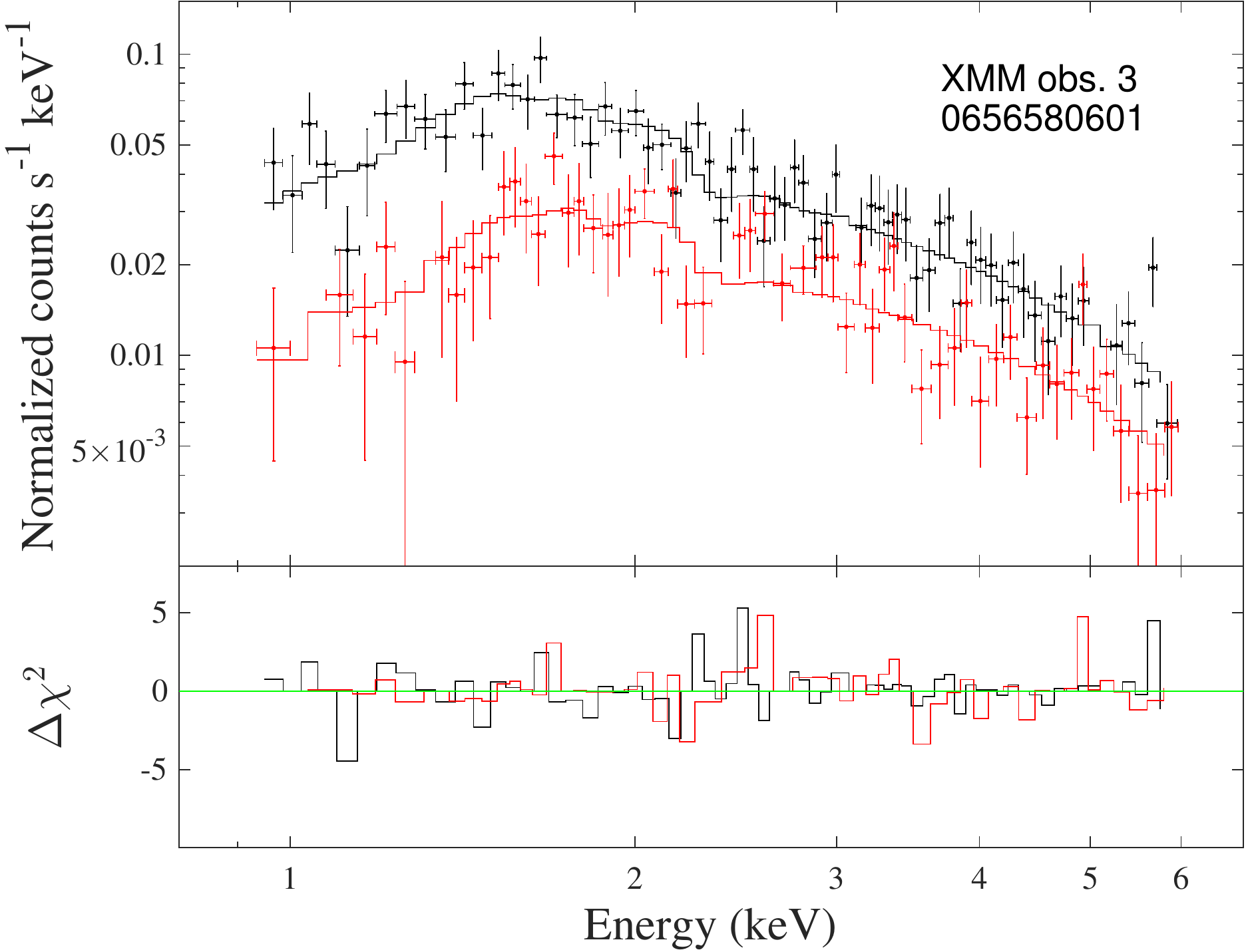}}\vspace{2mm}
\resizebox{\hsize}{!}{
\includegraphics{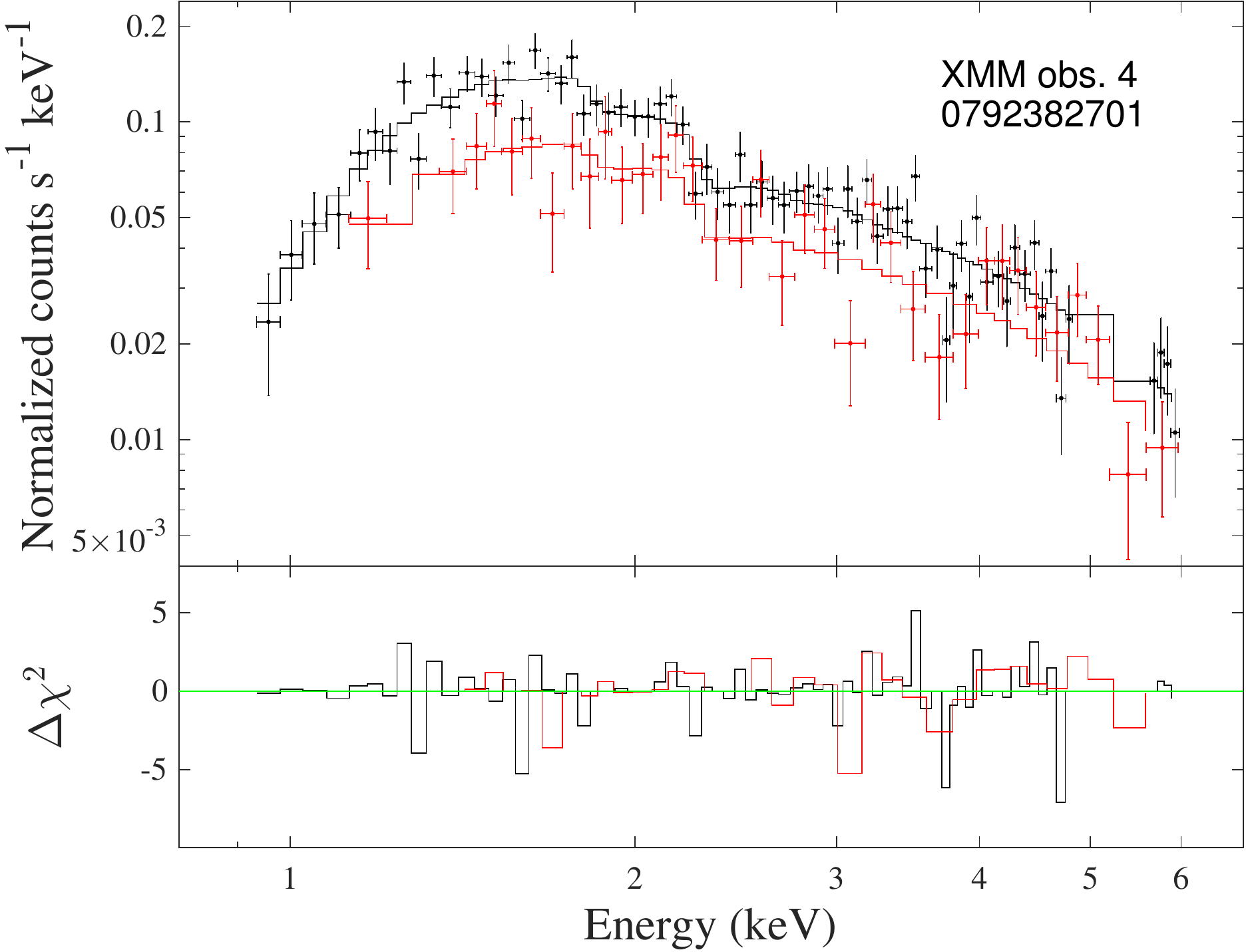}\hspace{2mm}
\includegraphics{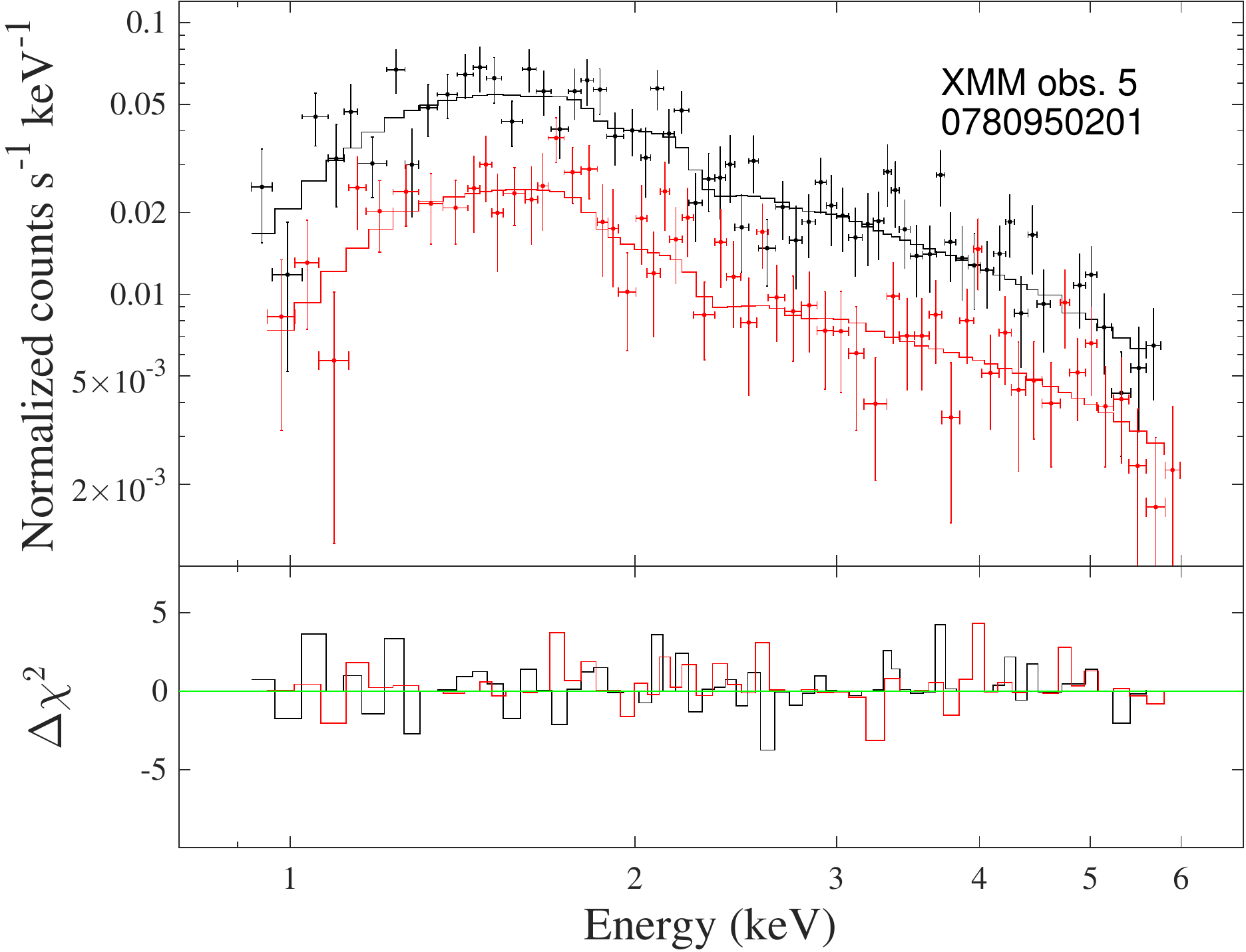}\hspace{2mm}}\vspace{2mm}
\vspace{1mm}
\caption{Datapoints, best-fitting models, and $\chi^2$ residuals for our sample of  \chandra\ and \xmm\  phase-resolved spectra; see Table \ref{tab:spec_fit} for the values of the best-fitting parameters. The top left panel shows the \chandra\ spectra (combined from ObsIDs 9140, 10937, 12823 and 12824); the other panels show the \xmm\ EPIC spectra (ObsIDs labelled in each panel). For the \xmm\ spectra, MOS and pn were combined with the {\sc sas} task {\it epicspeccombine}. Spectra and models corresponding to the bright, intermediate, and faint phases are plotted in black, red, and green, respectively. (Faint phases are not analyzed for \xmm\ spectra, because they are too contaminated by diffuse emission.)\\
\label{fig:fit_xmm} }
\end{figure*}

For the \xmm\ data, we only fitted the spectra of the bright and intermediate phases,
%in the energy band of 0.9--6.0 keV for \xmm\ data, 
because the faint-phase emission is highly contaminated by the diffuse emission in the inner region of the Circinus galaxy, and by the PSF wings from the active galactic nucleus. We also ignored all data above 6 keV, because the strong and inhomogeneous background emission (including Fe lines from the nuclear source) always dominates over the CG\,X-1 emission in that band.

We used the same combination of models discussed for the \chandra\ data: 
%were used to fit the \xmm\ phase-resolved spectra. The general model is 
{\it tbabs}$\times${\it absori}$\times${\it pcfabs}$\times${\it tbabs}$\times${\it  po}. The first {\it tbabs} component represents the line-of-sight absorption and is fixed at $N_{\rm H} = 5.6 \times 10^{21}$ cm$^{-2}$; the second {\it tbabs} and the {\it absori} components represent Compton-thin neutral and ionized absorption around CG\,X-1, respectively; the {\it pcfabs} component is a partial-covering, Compton-thick absorber responsible for the dips and occultations. As before, we assumed that in the bright phase, the covering fraction $f$ of {\it pcfabs} is $f=0$. We also assumed that the photon index and normalization of the power-law emission is the same in the bright and intermediate spectra, so that the difference between the two phases is entirely due to changes in the column densities of the Compton-thin absorbers and to an increased covering fraction of the Compton-thick medium. Although the best-fitting values change from observation to observation without a clear trend, most spectra are well fitted with intrinsic cold-absorber column densities $\sim$2--5 $\times 10^{21}$ cm$^{-2}$ and ionized-absorber column densities $\sim$2--4  $\times 10^{22}$ cm$^{-2}$ (Table \ref{tab:spec_fit}), consistent with the properties of the {\it Chandra} spectra. The covering fraction for the intermediate spectra varies between $\sim$0.3--0.6; this is not surprising, because the intermediate spectra were extracted (by definition) from time bins with count rates between 15\% and 70\% of the maximum count rate. 

In summary, as for the {\it Chandra} spectra, the main difference between the bright and intermediate spectra from each individual observation is the energy-independent covering fraction of the Compton-thick screen (Figure \ref{fig:fit_xmm}). The photon index of the power-law emission is consistent with $\Gamma \approx 2.1$ within the 90\% confidence limit of every observation (which is also consistent with the {\it Chandra} result). The intrinsic luminosity of the bright phase is itself variable from orbit to orbit, ranging from $\approx$5.5 $\times 10^{39}$ \ergs\ to $\approx$2.6 $\times10^{40}$ \ergs, in the 0.3--8 keV band. 

\begin{figure}%[ht!]
%\centering
\resizebox{ \hsize}{!}{
\vspace{2mm}
\includegraphics[width=9.0cm]{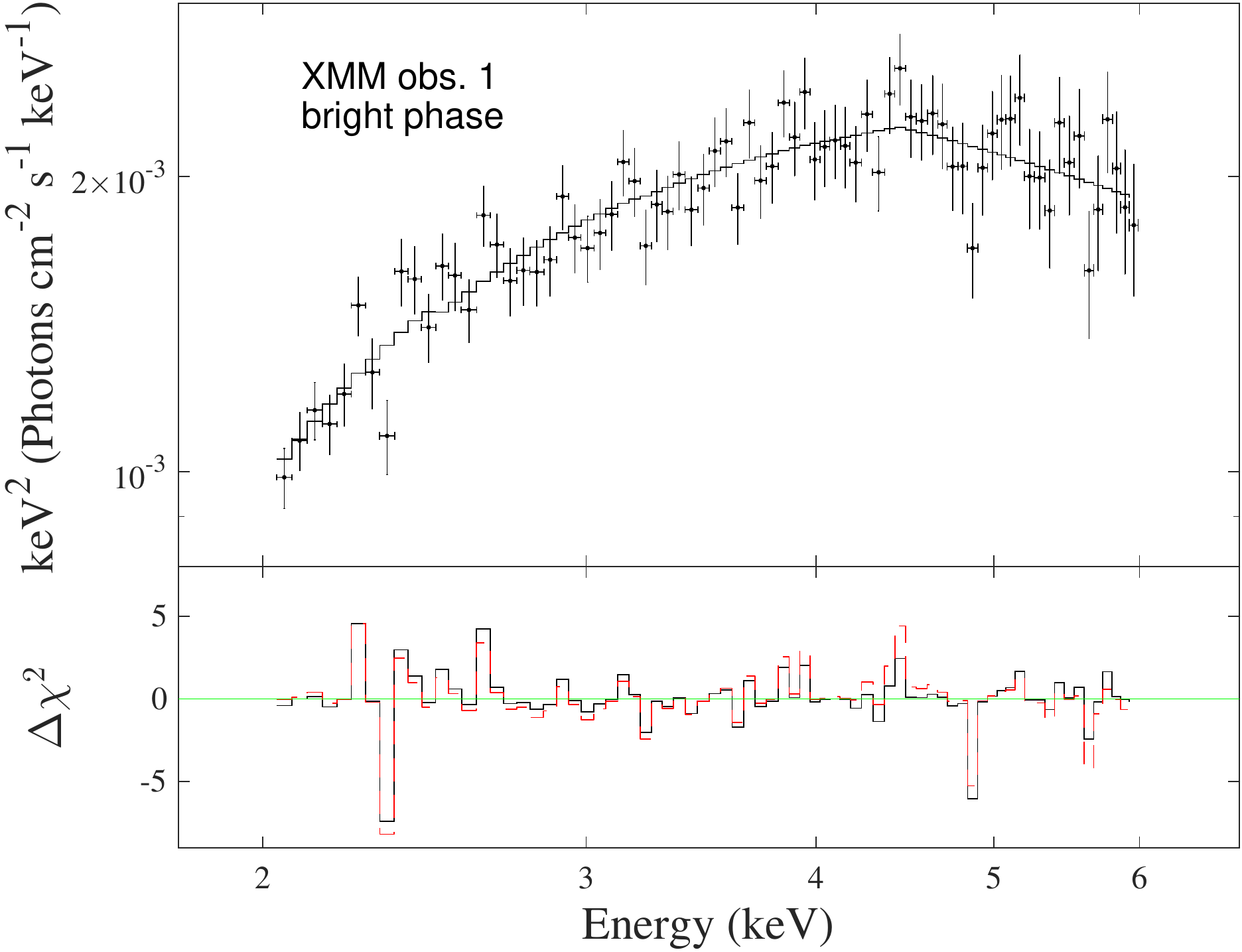}\hspace{2mm}}\vspace{2mm}
\caption{Unfolded spectrum and $\chi^2$ residuals for the bright phase of \xmm\ observation 1. Upper panel: the model used for the unfolding is an absorbed broken power-law ({\it tbabs}$_1 \times${\it tbabs}$_2 \times${\it bknpo}).  Lower penal: the black solid line is the $\Delta \chi^2$ of a  {\it bknpo}  model, and the dashed red line is for a {\it po} model. See Table 4 for the best-fitting parameters and for a comparison between power-law and broken power-law models.}
\label{fig:curvature_fit} 
\end{figure}

Finally, we used the bright-phase \xmm\ spectra to test for the presence of a high-energy downturn, one of the defining properties of ULXs \citep{Stobbart2006,Gladstone2009,Sutton2013,Walton2018}. 
To do so, we fitted the \xmm\ spectra above 2 keV with a simple power-law and a broken power-law ({\it bknpow} in {\sc xspec}), independently; we restricted our fit to energies $>$2 keV because the effect of neutral and ionized absorption is statistically negligible there. We find that the broken power-law model is statistically preferred over the simple power law, at the 98\% significance level, according to the F-test (Table \ref{tab:curvature_fit}). The best-fitting break energy is $(4.4 \pm 0.3)$ keV, and the continuum slope steepens from $\Gamma \approx 2.0$ to $\Gamma \approx 2.8$ (Figure \ref{fig:curvature_fit}) . 

% about the ULX state

%\textcolor{magenta}{May don't need this: The high X-ray luminosity (\Lx $\gtrsim 10^{40}$ \ergs) and the disk-like high energy curvature of CG\,X-1 is resemble to a few broadened disk ULXs, {\it{e.g.}}, NGC 1313 X-2 and Ho IX X-1 \citep{Gladstone2009, Sutton2013}. 
%But the $\sim 10^{40}$ \ergs\ luminosity is unusual for most of the broadened disk ULXs, which dominated the X-ray luminosity $\lesssim 3\times 10^{39}$ \ergs. As discussed by \citet{Sutton2013}, these sources could be misclassified, because it is difficult to distinguish between  broadened disk spectra and hard ultraluminous spectra with the strong spectral curvature.}

%\clearpage
\hspace{-0.5cm}
\begin{deluxetable*}{lccccccccc}
\centering
\tabletypesize{\footnotesize}
%\tabletypesize{\small}
\tablewidth{0pt}
\tablecolumns{10}
\tablecaption{ Best-fitting parameters for \chandra\ and \xmm\ phase-resolved spectra, for our fiducial model.  \label{tab:spec_fit}}
\tablehead{
 %& & &  &   &  \\%[1pt]
\colhead {Phase}  & \colhead{ $N_{\rm H}$}  & \colhead{$\xi$}  & \colhead{f} & \colhead{$N_{\rm H}$} & \colhead{$\Gamma$} & \colhead{Norm} & \colhead{$F_{\rm X}$}  & \colhead{ $L_{\rm X}$} & \colhead{$\chi^2_{\nu}$ (dof)} \\[5pt]
   & ($10^{22}$ cm$^{-2}$) & ($10^2$)  & &($10^{22}$ cm$^{-2}$) & & ($10^{-4}$)   & (10$^{-13}$ CGS)   & (10$^{38}$ CGS) &  \\[3pt]
  & {\it absori} & {\it absori} & {\it pcfabs}  & {\it tbabs}$_2$ & {\it po} & {\it po} & {\it cflux} &  & \\[3pt]
\colhead{(1)}&(2) & (3)& (4)& (5) & (6) &(7) & (8) & (9)  &(10)\\[-3pt]
}
\startdata
\\[-3pt]
\multicolumn{10}{l}{ \chandra \ combined ObsIDs 9140, 10937, 12823 and 12824}\\[3pt]
%faint$^a$ &-  & -  &- & $<0.25$ & 2.42$^{+0.38}_{-0.29}$ &  0.33$^{+0.16}_{-0.07}$ & 0.57$^{+0.06}_{-0.07}$ & 3.32$^{+1.69}_{-0.58}$  &  0.76 (30)  \\%tbabs*tbabs*po
Faint$^a$ & 2.34$^{+4.13}_{-2.34}$  & unconstr.  & - & 0.12$^{+0.09}_{-0.12}$ & 2.63$^{+0.25}_{-0.31}$ &  0.50$^{+0.07}_{-0.03}$ & 0.58$^{+0.07}_{-0.06}$ & 5.3$^{+1.1}_{-2.4}$  &  0.73 (28)  \\[3pt]%tbabs*absori*tbabs*po
Bright$^b$ &1.96$^{+1.01}_{-0.91}$  & 4.9$^{+3.8}_{-2.4}$  & $[0]$ & 0.57$^{+0.8}_{-0.8}$ & \multirow{2}{*}{2.10$^{+0.16}_{-0.16}$}  & \multirow{2}{*}{9.04$^{+2.48}_{-1.94}$}   & 16.5$^{+0.5}_{-0.5}$ & \multirow{2}{*}{89$^{+16}_{-12}$ }  &\multirow{2}{*}{ 0.82 (363)}  \\[3pt] % po.xcm
Interm.$^b$ & 5.11$^{+2.02}_{-1.68}$  & 6.0$^{+4.3}_{-2.5}$ & 0.60$^{+0.04}_{-0.04}$  & 0.40$^{+0.11}_{-0.11}$  & & & 5.9$^{+0.3}_{-0.3}$ & &  \\ 
     & & &  &  & & &  &  &  \\%[0.1pt]
 \hline
 & &  & &  & & &  &  &  \\[0.01pt]
 \multicolumn{10}{l}{ \xmm\ obs. 0111240101 (1)}\\[3pt]
 Bright$^b$  &2.04$^{+0.73}_{-0.59}$  & 1.7$^{+1.5}_{-0.9}$ & %$\left \langle0\right \rangle$ &0.38$^{+0.13}_{-0.16}$ & 
 $[0]$ &0.38$^{+0.13}_{-0.16}$ &
  \multirow{2}{*}{1.97$^{+0.11}_{-0.10}$} & \multirow{2}{*}{22.72$^{+4.35}_{-3.45}$} & 48.5$^{+1.0}_{-1.0}$ & \multirow{2}{*}{255$^{+36}_{-27}$} & \multirow{2}{*}{1.00 (192)} \\[3pt]% 1.simul po.xcm
 Interm.$^b$   &3.38$^{+0.84}_{-0.72}$  &1.4$^{+1.4}_{-0.4}$ & 0.42$^{+0.03}_{-0.03}$  &0.16$^{+0.28}_{-0.13}$&  &  & 26.5$^{+0.8}_{-0.8}$ &  &\\ % egress_po
 &  &  & &  & & &  &  &  \\[0.01pt]
 \hline
   & &  &  & & & &  &  &  \\[0.01pt]
 \multicolumn{10}{l}{ \xmm\ obs. 070198100 (2)}\\[3pt]
Bright$^b$   & 4.31$^{+1.41}_{-2.15}$ & unconstr.  & $[0]$ & 0.71$^{+0.11}_{-0.27}$ & \multirow{2}{*}{2.44$^{+0.16}_{-0.31}$} & \multirow{2}{*}{14.03$^{+1.41}_{-2.15}$}  & 15.7$^{+1.1}_{-2.4}$ & \multirow{2}{*}{131$^{+38}_{-45}$}   & \multirow{2}{*}{1.17 (134) } \\[3pt]% po.xcm
Interm.$^b$   & 2.39$^{+0.92}_{-1.42}$ &  8.4$^{+30.6}_{-0.8}$ & 0.38$^{+0.07}_{-0.11}$  &  0.47$^{+0.15}_{-0.25}$ & & & 10.0$^{+2.5}_{-2.3}$  &   & \\ %
 &  &  & &  & & &  &  &  \\[0.01pt]
 \hline
   & &  & &  & & &  &  &  \\[0.01pt]
 \multicolumn{10}{l}{ \xmm\ obs. 0656580601 (3)}\\[3pt]
Bright$^b$   &1.65$^{+2.49}_{-1.54}$   &2.4$^{+5.5}_{-2.4}$   & $[0]$ & $<0.59$ & \multirow{2}{*}{ 2.04$^{+0.37}_{-0.36}$ } & \multirow{2}{*}{7.60$^{+5.89}_{-3.52}$ } &17.2 $^{+2.4}_{-2.2}$  & \multirow{2}{*}{71.1$^{+34.4}_{-10.6}$ } & \multirow{2}{*}{0.95 (113) } \\[3pt]% po.xcm,
 Interm.$^b$  & 2.07$^{+1.38}_{-2.07}$  & unconstr.  &0.43$^{+0.09}_{-0.11}$  & $<1.0$ & & & 9.1 $^{+6.2}_{-1.5}$  &   &  \\
    & &  & &  & & &  &  &  \\[0.01pt]
 \hline
   & &  & & & & &  &  &  \\[0.01pt]
 \multicolumn{10}{l}{ \xmm\ obs. 0792382701 (4)}\\[3pt]
 Bright$^b$  & 3.21$^{+2.49}_{-0.88}$ & unconstr.  & $[0]$ & 0.59$^{+0.15}_{-0.14}$ & \multirow{2}{*}{2.10$^{+0.12}_{-0.18}$} & \multirow{2}{*}{10.81$^{+3.35}_{-1.62}$} &21.7 $^{+0.6}_{-1.2}$  & \multirow{2}{*}{112$^{+27}_{-12}$ } & \multirow{2}{*}{1.12 (98)} \\[3pt] %po.xcm
 Interm.$^b$ & 1.29$^{+3.04}_{-1.29}$ & 24.6$^{+1.4}_{-2.1}$ & 0.25$^{+0.13}_{-0.15}$ &$<0.98$ &  &   & 14.8$^{+1.6}_{-1.1}$  &   & \\ 
    & &  & & & & &  &  &  \\[0.01pt]
    \hline
     & &  & & & & &  &  &  \\[0.01pt]
 \multicolumn{10}{l}{ \xmm\ obs. 0780950201 (5) }\\[3pt]
 Bright$^b$  & 0.41$^{+2.08}_{-0.41}$ & $<9.2$  & $[0]$ &0.29$^{+0.16}_{-0.24}$ & \multirow{2}{*}{ 1.99$^{+0.28}_{-0.22}$ } & \multirow{2}{*}{ 4.85$^{+1.83}_{-0.09}$ } & 11.9$^{+0.9}_{-0.9}$  & \multirow{2}{*}{54.8$^{+24.4}_{-8.0}$} & \multirow{2}{*}{0.93 (110)} \\[3pt] % pps/qyl/po.xcm %bright_po.xcm
 Interm.$^b$ & 12.38$^{+11.41}_{-12.37}$ & unconstr. & 0.45$^{+0.22}_{-0.22}$  & 0.34$^{+0.22}_{-0.22}$ & &  & 5.59$^{+0.76}_{-0.58}$  &  &   \\
\enddata
\tablenotetext{}{\hskip-2.5pt {\it Columns}: 
(1) Identification of the orbital phase corresponding to the best-fitting spectral parameter listed on each row (see Section 4 for the definition of Faint, Intermediate and Bright phases); (2) equivalent hydrogen column density of the ionized absorber (modelled with {\it absori}); (3) {\it absori} ionization parameter $\xi = L/(N_eR^2)$, where $L$ is the integrated luminosity between 5 eV and 300 keV, $N_e$ is the electron number density, and $R$ is the distance from the source to the ionized material  \citep{Done1992}; (4) dimensionless covering fraction of the Compton-thick absorber modelled with {\it pcfabs} ($0\leqslant f \leqslant 1$); (5) Equivalent hydrogen column density of the intrinsic neutral absorber (in the host galaxy and around CG\,X-1); (6) photon index of the power-law component, assumed identical for Bright and Intermediate phases; (7) power-law normalization, in units of $10^4$ photons keV$^-1$ cm$^{-2}$ s$^{-1}$ at 1 keV; (8) absorbed flux in the 0.3--8 keV band; (9) unabsorbed luminosity in 0.3--8 keV band ($L_{\rm X} \equiv 4 \pi d^2 F_{\rm X}$, with $d = 4.2$ Mpc); (10) reduced $\chi^2$ and degrees of freedom.
\\ 
\hskip-2.5pt~ {\it Notes}: The models used for the phase-resolved spectra are as follows:
$^a$: {\it tbabs}$_{1} \times${\it absori}$\times${\it tbabs}$_{2}  \times${\it po}. 
$^b$: {\it tbabs}$_{1} \times${\it absori}$\times$ {\it pcfabs}$\times${\it tbabs}$_{2} \times${\it po}.
\\
\hskip-2.5pt~ Other parameters not shown in this table are fixed at default or assumed values. Specifically: 
(i) for {\it absori}, the temperature, Fe abundance and redshift  are fixed at $3\times10^4$ K, 1.00, and 0, respectively; 
(ii) for {\it pcfabs}, the column density of the Compton-thick absorber is fixed at $N_{\rm H} = 2 \times 10^{24}$ cm$^{-2}$, but identical fitting results are obtained for any other higher values;
(iii) the line-of-sight absorption was modelled with an additional {\it tbabs}$_1$ component, with column density fixed at $N_{\rm H} = 0.56 \times 10^{22}$ cm$^{-2}$. \\}
%\vspace{0.5cm}
\end{deluxetable*}
%\clearpage

% from /Users/qyl/work/cgx1/xmm/0111240101/pps/bright/bright_bknpo_2_6.xcm 
% matlab pl_spectra_xmm.m

\subsection{    Long-term  spectral and luminosity variations   }

After analyzing the differences between brighter and fainter phases within individual orbital cycles (Sections 4.3 and 4.4), we set out to estimate the range of luminosity variability for the bright phase from cycle to cycle, over the past two decades. We have already shown a range of luminosities based on the maximum observed count rates converted through {\sc pimms} (Section 3.3 and Figure \ref{fig:dphi}); now we want to estimate peak luminosities more accurately from spectral analysis. To do so, we used six \chandra\ and five \xmm\ observations from 2000 to 2018, as already mentioned in Section 4.1. This time, we extracted and fitted the average spectrum of each observation ({\it i.e.}, no longer split into faint, intermediate and bright intervals); MOS and pn data were fitted simultaneously for each \xmm\ observation. We applied our fiducial model with intrinsic ionized and neutral absorption, and foreground neutral absorption: {\it tbabs} $\times$ {\it absori} $\times$ {\it tbabs} $\times$ {\it powerlaw}. Here, we did not include a Compton-thick {\it pcfabs} term because we assume that it gives a grey correction to the spectrum, equivalent to a rescaling of its normalization. For each observation, we determined the average de-absorbed luminosity in the 0.3--8 keV band (Table \ref{tab:spec_fit_obs}); we also measured the average count rate and the maximum count rate (see Section 4.2 for the definition of maximum count rate). Then, we multiplied the average luminosity of each observation by the ratio of maximum over average count rate for that same observation.  This gave us the peak luminosity of each observation.

\begin{figure}%[t]
\center
\includegraphics[width=8.5cm]{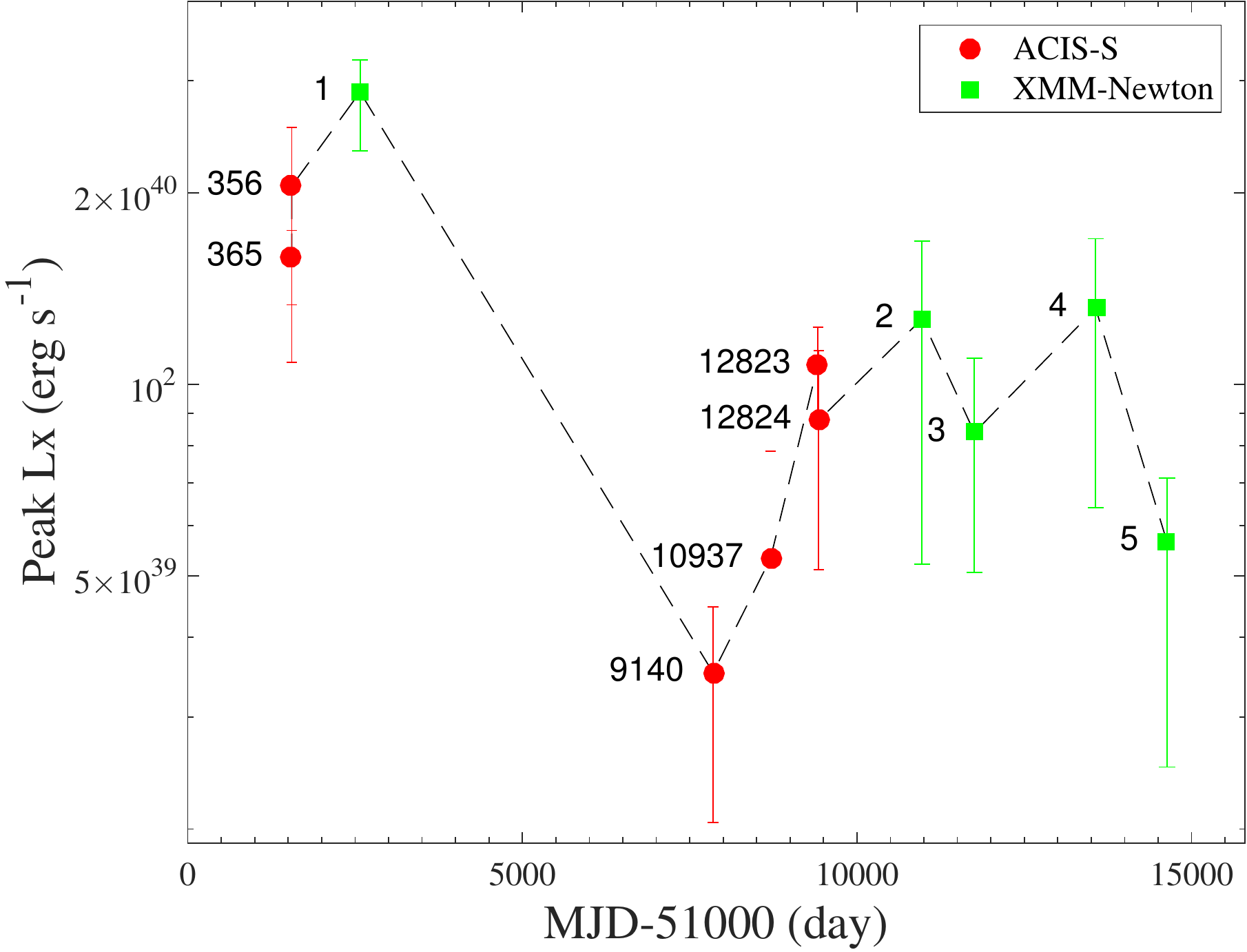}
\caption{ Long-term 0.3--8 keV lightcurve of CG\,X-1. The peak X-ray luminosity of each observation is the value inferred for the unocculted portion of that observation; luminosities were obtained from individual spectral modelling of each observations (see Table \ref{tab:spec_fit_obs} for the best-fitting parameter values).  Red dots represent \chandra/ACIS-S3 observations, and green squares are for \xmm/EPIC observations. Error bars are for the 90\% confidence levels. ObsIDs are labelled in the plot next to each datapoint (see Table 1 for details). The lightcurve spans between 2000 March 13 and 2018 February 7.\\}
\label{fig:long_lc}
\end{figure}

The highest peak luminosity was seen in the first \xmm\ observation (2001 August 6), at $L_{\rm X,peak} \approx 2.9 \times 10^{40}$ \ergs; the lowest luminosity was recorded in \chandra\ ObsID 9140 (2008 October 26), with $L_{\rm X,peak} \approx 3.5\times 10^{39}$ \ergs\ (Table \ref{tab:spec_fit_obs}). Thus, $L_{\rm X,peak}$  varies by a factor of $\approx$8 across our sub-sample of eleven observations (Figure \ref{fig:long_lc}), whereas the average luminosity of each observation varies by a factor of $\approx$13; the variability range of the average luminosities is higher because observations with lower peaks also have longer occultation phases (Figure \ref{fig:dphi}) and therefore even lower average luminosities. There is no trend in the hardness-luminosity plane, that is the best-fitting photon index $\Gamma$ is approximately the same for all observations. For 10 out of the 11 observations in our spectral analysis sub-sample, {\it i.e.}, all except \chandra\ ObsID 365, the mean photon index is $\left \langle \Gamma \right \rangle \approx 2.11$ with a scatter $\sigma \approx 0.15$. Only the spectrum of \chandra\ ObsID 365 is significantly different from all the others, with $\Gamma = 1.5 \pm 0.2$.
%, and steeper than $\approx 2$, 
%except for {\it Chandra} ObsID 9140 (the lowest luminosity one) where $\Gamma \approx 1.8$.

% from /Users/qyl/work/cgx1/matlab/pl_Figure_long_lc.m

%%%%%%%%%%%%%%%%%%%%%%%%%%%%%%%%%%%%%%%%%%%%%%%%%%%%%%%%%%%%%%%%%%%%%%%%%%%%%

\section{ The Nature of the donor star   }

\subsection{  A possible optical counterpart }

\begin{figure}%[hp!]
\center{
%\vspace{2mm}
\includegraphics[width=8.8cm]{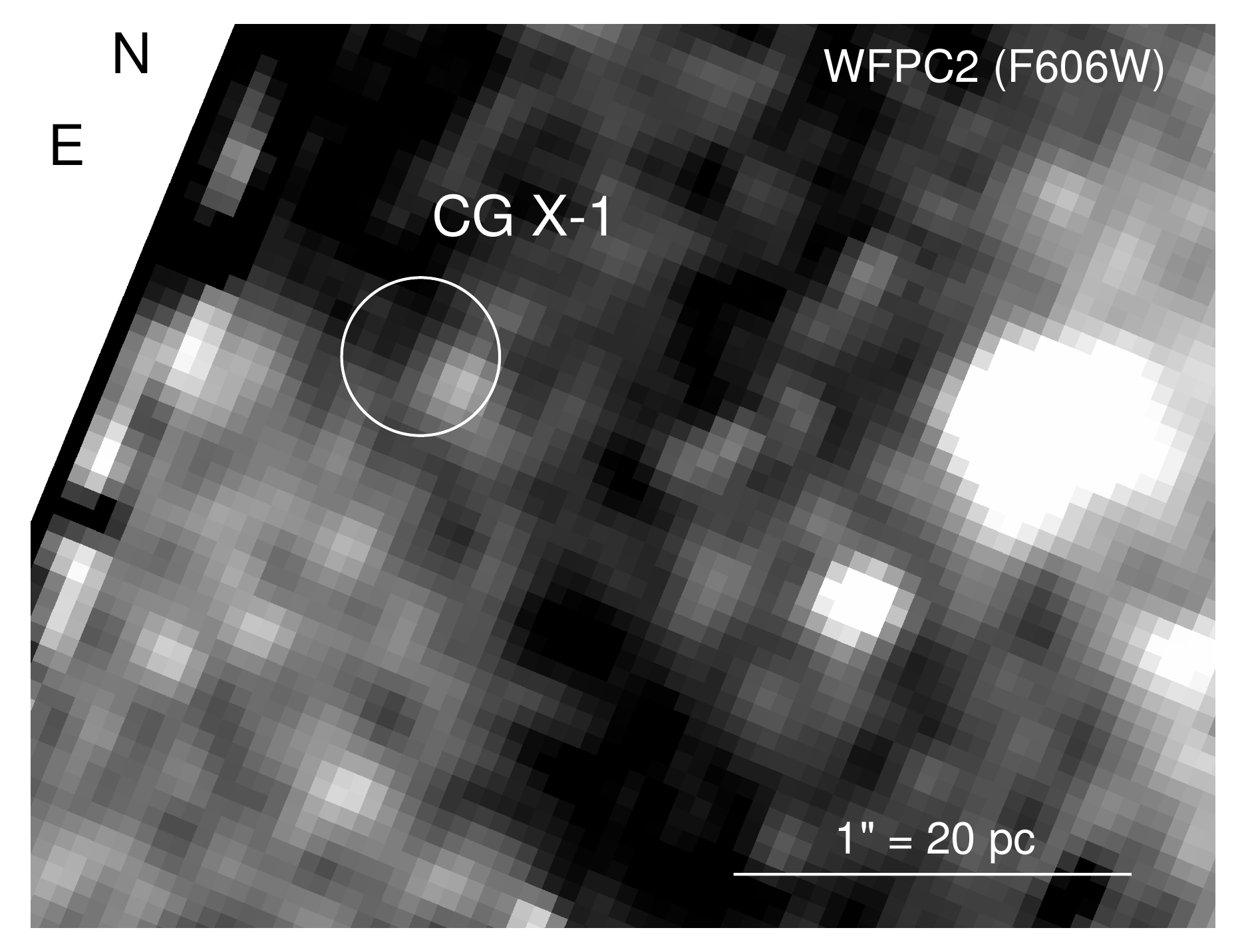}}
\caption{{\it {Hubble Space Telescope}} ({\it {HST}}) image of the CG\,X-1 field, taken with the PC chip of the Wide Field Planetary Camera 2 (WFPC2), in the F606W filter (600-s exposure). The white circle with a radius of 0\arcsec.2 represents the positional error of CG\,X-1, after we re-aligned the \chandra\ and {\it HST} astrometry (see Section 5.1). The image was smoothed with a Gaussian core of 3 pixels. The optical source inside the X-ray error circle has an observed brightness $m_{\rm {F606W}} \approx 24.3$ mag (Vegamag) and is our best candidate counterpart of CG\,X-1. This intrinsic reddening is very uncertain because of the irregular dust filaments (dark regions in the image) across the field. \\} 
\label{fig:hst}
\end{figure}

% how to do the asytromety
A faint, point-like optical source was found at the X-ray position \citep{Bauer2001, Weisskopf2004}, in a combined 600s exposure from \hst\ WFPC2, in the F606W filter (Figure \ref{fig:hst}). We re-processed the data to check or improve the astrometric alignment between X-rays and optical bands, and verify whether that source is indeed the most likely counterpart of CG\,X-1. We aligned the \chandra\ and \hst\ positions using five bright, point-like sources detected in both images. The residual random scatter between the best-fitting positions of the reference sources in the two bands gives us an estimate of the error radius for the position of CG\,X-1. We are able to narrow down the error circle to $\approx$0.\arcsec 2 (white circle in Figure \ref{fig:hst}), improving on the previous results. The error circle does indeed include the optical source previously suggested as the most likely counterpart. Applying standard techniques of {\it HST}/WFPC2 aperture photometry \citep{Sirianni2005} to the source, and using the most updated values of the WFC2 zeropoints\footnote{http://www.stsci.edu/hst/acs/analysis/zeropoints}, we confirm an apparent brightness m$_{\rm F606W} \approx V \approx (24.3\pm0.1)$ mag in the Vega system, in agreement with \cite{Weisskopf2004}.

Taking into account the line-of-sight extinction $A_{F606W} \approx 3.5$ mag \citep{Schlafly2011}, and a distance modulus of 28.1 mag for the Circinus galaxy, we estimate the intrinsic brightness of the source as $M_{V} \lesssim -7.3$ mag (depending on the amount of additional local extinction).
This is already at the high end of the luminosity range for a WR star ($M_{V}$ ranging approximately from $-7$ to $-2.5$ mag: \citealt{Crowther2007,Massey2003}). 
 Given that CG\,X-1 is located in a star-forming region of Circinus, near dust lanes and filaments, we expect also non-negligible local extinction, which is likely very high and uncertain. Thus, it is more likely that the optical counterpart is an unresolved young star cluster. 
The field of CG\,X-1 was also observed with {\it HST}/WFPC2 in the narrow-band H$\alpha$ filter F656N and \ion{O}{3} filter F502N, but no excess line emission is detected in those bands. In conclusion, with only one image in one optical filter, we do not have enough information to constrain the colour, spectral type and luminosity of the optical counterpart.

% from 0111240101/pps/bright/bright_po_2_6.xcm  bright_bknpo_2_6.xcm 
\begin{deluxetable*}{ccccccc}%\small
\centering
\tabletypesize{\footnotesize}
\tablewidth{0pt}
\tablecolumns{7}
\tablecaption{ Comparison of power-law and broken power-law models in the 2--6 keV band (\xmm\ observation 1)
%, Bright phase in the 2--6 keV band.  
\label{tab:curvature_fit}}
\tablehead{
 %& & &  &   &  \\%[1pt]
%\colhead{spectra} & 
\colhead{ $\Gamma$} & \colhead{ $\chi^2$/dof} & \colhead { $\Gamma_1$ }  & \colhead{ $E_{\rm break}$ }  & \colhead{ $\Gamma_2$  }  & \colhead{  $\chi^2$/dof  } 
%& \colhead{  $\Delta \chi^{2d}$ } 
& \colhead{ 1-P }  \\
%& \colhead{ 1-P({\it F}-test)$^e$ }  \\
 & &  &   &  &  &    \\[-3pt]
%     &  \multicolumn{2}{c}{{\it po}}                 & \multicolumn{4}{c}{ {\it bknpower}} &    &    \\
 {\it po} & {\it po} &{\it bknpo}&{\it bknpo}&{\it bknpo}&{\it bknpo}  &\\[3pt]
 \colhead{(1)} & (2) & (3) & (4) & (5) & (6) & (7) \\[-3pt]
}
\startdata\\
%XMM obs 1 bright &   
2.36$^{+0.17}_{-0.16}$  &  73.82/72   & 1.97$^{+0.29}_{-0.33}$   & 4.44$^{+0.27}_{-0.53}$  & 2.78$^{+0.43}_{-0.37}$  & 66.41/70 & 97.5  \\
\enddata
\tablenotetext{}{\hskip-2.5pt {\it Columns}: (1) Photon index of the single power-law model; (2) goodness-of-fit for the single power-law model; (3) photon index below the break for the {\it bknpo} model; (4) break energy (keV); (5) photon index above the break; (6) goodness-of-fit for the {\it bknpo} model; (7)  percentage probability that the broken power-law model is a better fit than the single power-law model, from the F-test. \\}
\end{deluxetable*}

\subsection{ Ruling out foreground Galactic sources    }

 % optical envirenment
We have already mentioned (Section 1) that the identification of CG\,X-1 as a possible foreground Galactic source (in particular, an mCV projected by chance in front of Circinus) has been a point of contention in previous studies \citep{Weisskopf2004}, but was subsequently rejected by \cite{Esposito2015}. Here, we confirm those arguments and suggest additional ones against the mCV interpretation, based on our X-ray and optical results.

The X-ray over optical flux ratio $F_{\rm X}/F_{\rm opt}$ is a useful criterion to identify and classify CVs, because it removes the uncertainty on the source distances. Although the precise definition of the energy bands used for the X-ray and optical fluxes may change from author to author, the general conclusions remain the same (within a factor of two). The ratio $F_{\rm X}$(0.1--4.0 keV)$/F_{\rm opt}$(5000--6000\AA) shows an upper boundary at around 4 \citep{Patterson1985}. Using the standard $V$ band, and converting it to a flux with the relation\footnote{This relation comes from the standard definition of the apparent $V$ magnitude as $V = -2.5 \log F_{\lambda} -21.100$ where $F_{\lambda}$ is the flux density at the top of the Earth's atmosphere, in units of erg cm$^{-2}$ s$^{-1}$ \AA$^{-1}$ \citep{Bessell1998}. Then, $F_{\rm opt} \equiv F_{\lambda} \times 1000$ \AA. The same relation $m_{\lambda} = -2.5 \log F_{\lambda} -21.100$ defines the ``STmag'' photometric system.} $\log F_{\rm opt} = -0.4 V -5.44$, we infer $F_{\rm X}$(2--10 keV)$/F_{\rm opt} \lesssim 7$ (Figure~3 in \citealt{Mukai2017}), for both magnetic and non-magnetic CVs\footnote{The only exception to this rule is a small number of ultracompact white dwarf - white dwarf systems \citep{Solheim2010}, such as RX J1914$+$24, with orbital periods $\lesssim$10 min, which can reach $F_{\rm X}/F_{\rm opt} \sim 1000$: \citep{Ramsay2008,Ramsay2005,Ramsay2002}. However, such (very rare) sources have a super-soft, thermal X-ray spectrum with almost no emission above 1 keV, inconsistent with the spectrum of CG\,X-1. Their orbital period is also much shorter than the one measured in CG\,X-1, so there is no possibility of mis-identification.}. 
A similar analysis by \cite{Revnivtsev2014} shows that $F_{\rm X}$(0.5--10 keV)$/F_{\rm opt} \lesssim 5$ for non-magnetic CV
(for this relation, we have used the conversion $\nu F_{\nu} \approx 5 F_{\lambda} \times 1000$ \AA $\equiv F_{\rm opt}$ at $\lambda \approx 5000$ \AA). All those empirical relations are based on observed (rather than de-absorbed) fluxes, but most of the sources are within a few 100 pc, and their line-of-sight absorbing column density is negligible.  For CG\,X-1, even if it were a CV in the Milky Way, its foreground line-of-sight optical extinction and X-ray absorption would be significant, because of its location in the direction of the Galactic plane. Given the uncertainty on its true distance, for CG\,X-1 we have calculated both the observed and the de-absorbed $F_{\rm X}/F_{\rm opt}$ ratios, and compared them with the upper limits found in the CV surveys cited earlier. We have not removed the additional intrinsic absorption component derived from X-ray fitting, because that component is not removed in those CV surveys, either.

First, we compare the observed fluxes. From the net count rate in the WFPC2 F606W band, and the tabulated value of {\it photflam}{\footnote{http://www.stsci.edu/hst/wfpc2/analysis/wfpc2\_photflam.html}}, we obtain $F_{\lambda} \approx 5.2 \times 10^{-19}$ erg cm$^{-2}$ s$^{-1}$ \AA$^{-1}$, and (by our definition, assuming a flat spectrum) $F_{\rm opt} \approx 5.2 \times 10^{-16}$ erg cm$^{-2}$ s$^{-1}$. For the X-ray flux, we take the unocculted phases of {\it Chandra}/ACIS ObsID 12823 (the longest in our series, Table 1 and Figure \ref{fig:ks}). The observed 2--10 keV flux is $F_{\rm X} \approx 1.9 \times 10^{-12}$ erg cm$^{-2}$ s$^{-1}$, giving an implausibly high X-ray over optical flux ratio of $\approx 3650$. For a more physical result, we remove the line-of-sight extinction ($A_V \approx 4$ mag) from the optical flux, and the line-of-sight absorption (corresponding to $N_{\rm H} \approx 6 \times 10^{21}$ cm$^{-2}$) from the X-ray flux. We obtain $F_{\rm opt} \approx 2.06 \times 10^{-14}$ erg cm$^{-2}$ s$^{-1}$ and $F_{\rm X} \approx 2.03 \times 10^{-12}$ erg cm$^{-2}$ s$^{-1}$, corresponding to $F_{\rm X}$(2--10 keV)$/F_{\rm opt} \approx 100$.  These values are well outside the range observed in CVs, which are then definitively ruled out. Instead, $F_{\rm X}/F_{\rm opt} \sim 100$--1000 is what we expect and observe in typical ULXs  \citep{Ambrosi2018,Gladstone2013,Tao2011}, because of an X-ray luminosity $\approx 10^{40}$ erg s$^{-1}$ (from the compact object) and an optical luminosity of $\approx 10^{38}$ erg s$^{-1}$ (combination of the contributions from the irradiated disk and the massive donor star).

Finally, as an additional test of the foreground source scenario, we point out that the 6.4-keV Fe line is usually strong in mCVs, and their absence is very rare \citep{Butters2011}. In order to check whether there is significant 6.4-keV line emission in CG\,X-1, we extracted {\it Chandra}/ACIS X-ray images in the 6.4--6.7 keV band, and then combined all the images from different observations. We found no concentration of photons in this band at the position of CG\,X-1; the few detected photons are clearly distributed as diffuse background. Also, no significant 6.4--6.7 keV lines were detected in stacked background-subtracted X-ray spectra (Section 4.4). 
%Having no Fe line for mCV \citep{Butters2011} is a rare thing.

In conclusion, we suggest that our new arguments about the flux ratio and the lack of Fe lines, together with those of \cite{Esposito2015}, put the final nail in the coffin of the Galactic mCV scenario.

\subsection{ Mass density of the secondary Roche lobe }
The binary separation in CG\,X-1 is 
\begin{eqnarray} 
&a& = \left[\frac{GP^2(M_1+M_2)}{4\pi^2}\right]^{1/3} \\ \nonumber
& &\simeq 5.8 ~ \left(\frac{P}{7.2~ {\rm hr}}\right)^{2/3} ~ \left(\frac{M_1+M_2}{30 ~{\rm M_\odot}}\right)^{1/3} ~ \rm R_\odot ,
 \end{eqnarray}
where G is the gravitational constant, $P$ the orbital period in unit of hr, $M_1$ and $M_2$  the masses of the compact object and donor star, respectively, in unit of \Msolar. 
The radius of the secondary Roche lobe is
\begin{equation}
    R_{\rm L,2} = a \, \frac{0.49q^{2/3}} { 0.6q^{2/3} + {\rm ln}(1+q^{1/3} ) }
  %\approx 2.6~{\rm R_\odot ~(for}~q = 2), 
\end{equation}
\citep{Eggleton1983}, where $q\equiv M_2/M_1$ is the mass ratio.
The average mass density inside the secondary Roche lobe (lower limit to the average density of the donor star) is 
\begin{equation}
\rho = 0.2101\frac{q}{1+q}
\left(\frac{R_{\rm L,2}}{a}\right)^{-3}
\left(\frac{P}{7.2\ {\rm hr}}\right)^{-2} \rm ~g~ cm^{-3}
\end{equation}
%It is only a function of the mass ratio $q$ 
\citep{Eggleton1983}.

In principle, the duration of the eclipse provides constraints on the mass ratio (coupled with the inclination angle); this was already well discussed by \cite{Weisskopf2004}. However, the problem for CG\,X-1 is that we are no longer sure what the duration of the true eclipse is (given the large cycle-to-cycle variability), and what part of the occultation is instead due to other thick structures between donor star and accretor. So, at this point we want to use only the most general constraint from Equation (7). For any plausible value of $q$ suitable to stellar-mass binaries ($0.01 \lesssim q \lesssim 100$), the average density in the donor Roche lobe must be $0.55 \lesssim \rho {\rm~ g~cm}^{-3} \lesssim 2.2$ ($0.4 \lesssim \rho/\rho_{\odot} \lesssim 1.6$). For a more likely range of $0.1 \lesssim q \lesssim 10$, the density is $1 \lesssim \rho ~{\rm g~cm}^{-3} \lesssim 2.2$ ($0.7 \lesssim \rho/\rho_{\odot} \lesssim 1.6$).
Such density range rules out massive early-type main-sequence stars, supergiants, red giants, asymptotic giants, and white dwarfs as the donor star. Instead, low-mass main-sequence stars, slightly evolved stars (such as low-mass helium stars) and massive WR stars are consistent with this tight constraint.

% spect fit: 0111240101/pps/m1_m2_pn_absori_po.xcm, for obs2 obs3: m1_m2_pn_po.xcm, obs4:m1_m2_m1u_m2u_pn_po.xcm
% obs5: m1_m2_pn_po.xcm
%\clearpage
\begin{deluxetable*}{ccccccccc}\small
%\rotate
\centering
\tablewidth{0pt}
\tablecolumns{9}
\tablecaption{ Best-fitting parameters of the phase-averaged spectra from a sample of individual observations.  \label{tab:spec_fit_obs}}
%\vspace{0.5cm}
\tablehead{
 %& & &  &   &  \\%[1pt]
 \colhead {Obs.}  &\colhead{$N_{\rm H}$}&\colhead{$N_{\rm H}$} & \colhead{$\Gamma$} & \colhead{Norm} & \colhead{$\left \langle F_{\rm X} \right \rangle$}  & \colhead{$\left \langle L_{\rm X} \right \rangle$} & \colhead{$L_{\rm {X,peak}}$} & \colhead{$\chi^2_{\nu}$ (dof)} \\[3pt]
   & $(10^{22}$ cm$^{-2})$ & $(10^{22}$ cm$^{-2})$  & & $(10^{-4})$   & (10$^{-13}$ \ergcms)   & (10$^{38}$ \ergs)  & (10$^{38}$ \ergs) &  \\[3pt]
  & {\it absori} & {\it tbabs} & {\it po} & {\it po} &  &  & \\[3pt]
\colhead{(1)}&(2) & (3)& (4)& (5) & (6) &(7) & (8) & (9)\\[-3pt]
}
\startdata
\\[-3pt]
%365$^a$  & 0.65$^{+0.63}_{-0.65}$ & 0.18$^{+0.61}_{-0.17}$ & 1.72$^{+0.16}_{-0.36}$ & 14.46$^{+5.62}_{-5.01}$ &  49.65$^{+3.81}_{-3.63}$ & 186.86$^{+43.24}_{-25.94}$ & 186.86$^{+43.24}_{-25.94}$ & 0.89 (85)\\[3pt] % with pileup
%
%158.27$^{+24.93}_{-16.23}$
%160.44$^{+75.67}_{-38.14}$ 204.83 $^{+90.61}_{-48.69}$
%179.02$^{+34.29}_{-21.91}$ & 287.92$^{+55.15}_{-35.24}$
%13.91$^{+5.67}_{-2.55}$   & 35.20$^{+14.74}_{-9.52}$
%26.84$^{+39.46}_{-12.77}$    & 53.20$^{+78.21}_{-25.31}$
%46.09$^{+9.11}_{-6.67}$    & 107.35$^{+21.22}_{-15.54}$
%43.84$^{+18.30}_{-12.60}$    & 87.76$^{+36.63}_{-25.22}$
%61.82$^{+36.38}_{-20.12}$    & 126.67$^{+74.54}_{-41.23}$
%54.93$^{+21.81}_{-16.91}$    & 83.96$^{+33.34}_{-25.85}$
%88.56$^{+45.51}_{-25.36}$   & 131.64$^{+67.65}_{-37.70}$
%29.62$^{+16.37}_{-7.62}$ & 56.56$^{+31.53}_{-14.53}$
%
365 & 0.48$^{+3.16}_{-0.48}$ & 0.16$^{+0.51}_{-0.16}$ & 1.51$^{+0.22}_{-0.20}$ & 10.32$^{+3.79}_{-2.49}$ &  46.2$^{+3.0}_{-3.0}$ & 158$^{+25}_{-16}$ & 158$^{+25}_{-16}$ & 0.89 (85)\\[3pt] % with no pileup
356$^a$ & 1.94$^{+1.55}_{-1.10}$ & 0.56$^{+0.19}_{-0.30}$ & 2.09$^{+0.28}_{-0.52}$ &  14.98$^{+9.03}_{-6.20}$ & 29.0$^{+1.8}_{-2.3}$ & 160$^{+76}_{-38}$ & 205$^{+91}_{-49}$ & 0.87 (166)\\[3pt]
1& 2.43$^{+1.36}_{-0.90}$ & 0.71$^{+0.16}_{-0.16}$  & 1.99$^{+0.13}_{-0.12}$  &  16.15$^{+4.02}_{-2.83}$  & 33.2$^{+0.6}_{-0.3}$& 179$^{+34}_{-22}$ & 288$^{+55}_{-35}$  & 1.00 (752) \\[3pt]%0111240101/pps/m1_m2_pn_absori_po.xcm
9140 & 6.47$^{+2.13}_{-4.59}$ & 0.21$^{+0.09}_{-0.17}$ & 1.79$^{+0.21}_{-0.15}$ & 1.12$^{+0.69}_{-0.47}$ & 2.98$^{+0.32}_{-0.16}$& 13.9$^{+5.7}_{-2.6}$   & 35.2$^{+14.7}_{-9.5}$ & 1.02 (63)\\[3pt]% chandra/evt/acis9140/repro/acis9140_po.xcm
10937 & 2.12$^{+7.04}_{-2.12}$ & 0.20$^{+0.50}_{-0.20}$ & 2.25$^{+0.66}_{-0.70}$ &  2.62$^{+3.84}_{-1.62}$ & 4.2$^{+0.6}_{-0.5}$& 26.8$^{+39.5}_{-12.8}$    & 53.2$^{+78.2}_{-25.3}$ & 0.88 (22) \\[3pt] %chandra/evt/acis10973/repro/acis10973_po.xcm
12823 & 1.57$^{+1.05}_{-0.89}$ & 0.49$^{+0.07}_{-0.07}$ & 2.02$^{+0.14}_{-0.14}$ &  4.18$^{+1.09}_{-0.83}$ & 8.82$^{+0.22}_{-0.21}$ & 46.1$^{+9.1}_{-6.7}$    & 107$^{+21}_{-16}$ & 0.97 (303) \\[3pt] %cgx1/chandra/evt/acis12823/repro/acis12823_po.xcm
12824 & 2.25$^{+2.01}_{-1.70}$ & 0.50$^{+0.15}_{-0.16}$ & 2.16$^{+0.29}_{-0.25}$ &  4.19$^{+2.03}_{-1.55}$ & 7.06$^{+0.37}_{-0.36}$ & 43.8$^{+18.3}_{-12.6}$    & 87.8$^{+36.6}_{-25.2}$ & 0.87 (120)\\[3pt] %cgx1/chandra/evt/acis12824/repro/acis12824_po.xcm
2 & 1.90$^{+2.28}_{-1.72}$ & 0.61$^{+0.07}_{-0.54}$ & 2.31$^{+0.17}_{-0.31}$ &  6.07$^{+4.21}_{-2.33}$ & 9.60$^{+0.47}_{-0.47}$ & 61.8$^{+36.4}_{-20.1}$    & 127$^{+75}_{-41}$ & 0.98 (300) \\[3pt] %xmm/0701981001/pps/m1_m2_pn_po.xcm
3 & 1.97$^{+1.93}_{-1.74}$ & 0.50$^{+0.21}_{-0.50}$ & 2.15$^{+0.27}_{-0.34}$ &  5.22$^{+2.56}_{-2.01}$ & 10.01$^{+0.60}_{-0.57}$ & 54.9$^{+21.8}_{-16.9}$    & 84.0$^{+33.3}_{-25.9}$ & 0.94 (212) \\[3pt]%xmm/0656580601/pps/m1_m2_pn_po.xcm
4 & 1.33$^{+1.47}_{-1.33}$ & 0.59$^{+0.22}_{-0.59}$ & 2.15$^{+0.26}_{-0.31}$ &  8.42$^{+5.34}_{-3.19}$ & 15.33$^{+0.95}_{-0.79}$& 88.6$^{+45.5}_{-25.4}$   & 132$^{+68}_{-38}$ & 0.88 (182) \\[3pt] %xmm/0792382701/pps/m1_m2_pn_po.xcm
5 & 2.60$^{+4.03}_{-2.60}$ & 0.58$^{+0.42}_{-0.22}$ & 2.28$^{+0.25}_{-0.34}$ & 3.21$^{+0.25}_{-0.34}$ & 5.34$^{+0.46}_{-0.43}$ & 29.6$^{+16.4}_{-7.6}$ & 56.6$^{+31.5}_{-14.5}$  &  1.01 (152)\\[3pt] % xmm/0780950201/pps/qyl/m1_m2_pn_po.xcm
\enddata
\tablenotetext{}{\hskip-2.5pt {\it Notes}: $^a$: \chandra\ ObsID 356 suffers from pile-up at the 40\% level (ACIS-S3 count rate $\approx$0.2 ct s$^{-1}$). We fitted its spectrum with {\it pileup}$\times${\it tbabs}$\times${\it absori}$\times${\it tbabs}$\times${\it po}. The model used for all the other observations is {\it tbabs}$\times${\it absori}$\times${\it tbabs}$\times${\it po}. $\left \langle F_{\rm X} \right \rangle$ is the average observed flux of each observation, and $\left \langle L_{\rm X} \right \rangle$ the average unabsorbed luminosity ($\left \langle L_{\rm X} \right \rangle \equiv 4 \pi d^2 \left \langle F_{\rm X} \right \rangle$), both in the 0.3--8 keV band. $L_{\rm {X,peak}}$ is the inferred luminosity of the unocculted portions of each observation (see Section 4.5).\\}
\end{deluxetable*}

\subsection{ Low-mass or high mass donor?}

For a peak \Lx\ $\approx 3\times10^{40}$ \ergs, the mass accretion rate must be a few 10$^{-6}$ $M_{\odot}$ yr$^{-1}$ even for radiatively efficient accretion. Considering also that in the super-critical regime, radiative efficiency decreases because of advection and outflows ({\it e.g.}, \citealt{Poutanen2007}), we suggest that the mass transfer rate into the Roche lobe of the accretor approaches 10$^{-5}$ \Msolar \ $yr^{-1}$. 
Can a low-mass donor (with the constraints mentioned in Section 5.3) provide such high mass transfer rate? A small number of ULXs have been found in elliptical galaxies \citep{vanHaaften2019,Plotkin2014,David2005}, and old globular clusters \citep{Dage2018, Roberts2012, Shih2010, Maccarone2007}; white dwarf donors in ultracompact systems are a plausible scenario for the ULX population in globular clusters \citep{Dage2018,Steele2014}. But ULXs in old stellar populations are much rarer than in young stellar environments (an order of magnitude rarer above $\approx$5 $\times 10^{39}$ erg s$^{-1}$: \citealt{Swartz2004,Plotkin2014}). Moreover, we have already ruled out the possibility of a white dwarf donor, and also ruled out a globular cluster as the optical counterpart of CG\,X-1. Other ULXs with a low- or intermediate-mass donor have been seen in intermediate-age environments of star-forming galaxies \citep{Soria2012}. Population synthesis models show the possibility of ULXs with a NS accretor and a low-mass companion stars: the donor may have started its life as an
%a $\approx$8--10 \Msolar star \citep{Fragos2015} or even as low as 
$\approx$6 \Msolar\ \citep{Wiktorowicz2015}. Those models show that, after a common-envelope phase, the companion star is stripped of its hydrogen envelope, the binary separation is reduced, and the system goes through a ULX phase with a helium-star donor mass of $\approx$1--2 \Msolar (either a Helium Hertzsprung gap or a Helium giant branch star).

While we cannot dismiss the possibility of a stripped, low-mass helium-star, we suggest that the most likely type of donor is a WR star, in agreement with \cite{Esposito2015}. We briefly summarize a few arguments that are consistent with this scenario. First, a WR donor can provide a sufficiently high mass transfer rate onto a BH accretor, producing persistent X-ray luminosities $L_{\rm X} \sim 10^{39}$--$10^{41}$ erg s$^{-1}$ \citep{Wiktorowicz2015,Bogomazov2014}, for timescales of a few 10$^5$ yr.

Second, WR stars have typical radii of only $\approx$1--2 $\times 10^{11}$ cm ($\approx$1.5--3 $R_{\odot}$) \citep{Crowther2007}, and can fit in a binary orbit with a period of a few hours. For a $M_1 = 10$ \Msolar\ BH  and  a  $M_2 = 20$ \Msolar\ donor, and a period of 7.2 hr,  $R_{L,2}\approx 2.6$ \Rsolar\ (large enough to fit a WR star), and $\rho\approx 1.6$ g cm$^{-3}\approx 1.2 ~\rho_{\odot}$ (also consistent with a typical WR). 
Third, the short orbital period and eclipsing/dipping X-ray lightcurve profile of CG\,X-1 resemble the observed properties of other X-ray binaries that have been interpreted as WR systems \citep[e.g.][]{Bauer2004, Carpano2007, Zdziarski2012, Laycock2015a, Ghosh2006, Maccarone2014, Esposito2013}. The reason for this interpretation is that a WR provides a dense wind that strongly affects the X-ray lightcurve via photoelectric absorption, especially when the compact object transits behind the donor star. Stronger stellar winds and smaller binary separations make this effect more important in WR HMXBs than for example in those with a supergiant donor. In Sections 6.1 and 6.2, we will discuss in more detail how the dense WR wind may explain the peculiar X-ray lightcurve of CG\,X-1; we will revisit the comparison with other candidate WR X-ray binaries in Section 6.5.

%%%%%%%%%%%%%%%%%%%%%%%%%%%%%%%%%%%%%%%%%%%%%%%%%%%%%%%%%%%%%%%%%%%%%%%%%%%%%
\section{   DISCUSSION }

%\subsection{  Explanations for the eclipsing behaviours   }

\subsection{     What causes the asymmetric occultations?   }

We have shown (Section 3.2 and Figure \ref{fig:flc}) that the folded X-ray lightcurve (averaged over dozens of cycles) has an apparent eclipse (lasting for $\approx 1/4$ of the period), with a sharp ingress and a slow egress. We have also shown (Figure \ref{fig:alc}) that this simple picture is complicated by irregular dips in each single cycle, and that the X-ray profile and the duration of the full occultation differ substantially from cycle to cycle. From spectral analysis, we showed (Sections 4.2 and 4.3) that the transition from lower to higher fluxes is energy-independent; hence, it is best explained as a decreasing level of partial covering by a totally opaque medium---rather than, for example, by a gradual decrease of the column density of the  photoelectric absorber. Any model of the system need to explain these unusual X-ray properties.

The sharp ingress and regular occurrence of the occultation around phase 1 suggest that there is a proper eclipse by the companion star, but the irregular duration of this phase also suggests a contribution from other optically thick material, such as clouds or clumps of dense gas (See Figure \ref{fig:cartoon} for a cartoon of the binary geometry). Irregular dips are well known in several LMXBs seen at moderately high inclination \citep{White1982, Frank1987, D'A&igrave;2014, D&iacute;az Trigo2006}. In those systems, the occulting material is either the accretion stream, or the geometrically thick bulge where the stream impacts the outer rim of the accretion disk. However, because of conservation of angular momentum along its ballistic trajectory, the stream is always trailing the compact object along the orbit. Therefore, dips caused by the accretion stream and impact bulge happen before eclipse ingress, contrary to what we see in CG\,X-1. 

Asymmetric eclipses are seen in some HMXBs \citep{Falanga2015}, such as the NS system Vela X-1. In that case, the asymmetric absorption has been attributed to a stream-like region of slower and denser wind trailing the NS \citep{Doroshenko2013}; photoelectric absorption gradually increases during the slow ingress and reaches the maximum at superior conjunction. Again, this type of enhanced wind absorption is not consistent with the energy-independent dips of CG\,X-1 and with their preferential location after the eclipse. However, energy-independent occultations were detected (``Type B dips'': \citealt{Feng2002}) in another well-known HMXB, the BH system Cyg X-1; a possible explanation for such dips is partial covering of an extended X-ray emitting region by an opaque screen \citep{Feng2002}.

WR X-ray binaries are the subclass of HMXBs with the thicker stellar wind. Typical mass-loss rates are $\dot{M} \sim 10^{-4}$--$10^{-5}$ \Msolar\ yr$^{-1}$, with terminal velocities of $\sim$1000--3000 km $s^{-1}$ \citep{Gr&auml;fener2017, Crowther2007}. Thus, we expect that WR X-ray binaries are most likely to show the effect of variable wind absorption on X-ray lightcurves, especially just before and just after superior conjunction of the compact object. Indeed, this is what is seen in the BH-WR systems NGC\,300 X-1 \citep{Carpano2007, Crowther2010, Binder2015, Carpano2018} and IC\,10 X-1 \citep{Silverman2008, Barnard2014, Laycock2015b, Steiner2016}. In particular, the X-ray lightcurve of NGC\,300 X-1 also shows dips during a slow egress, consistent with variable partial covering of an extended X-ray emitting region \citep{Binder2015}; unlike CG\,X-1, the variable absorbing clumps are not Compton-thick (column densities of only $\approx$10$^{23}$ cm$^{-2}$), so they have an energy-dependent effect on the 0.3--10 keV spectrum. A Compton-thick absorber was found in IC\,10 X-1, from {\it NuSTAR} observations \citep{Steiner2016}, when the BH passes behind the WR star. The X-ray lightcurve of IC 10 X-1 also has the same type of asymmetric profile as CG\,X-1, with a steep ingress and a slow egress; however, it was suggested \citep{Barnard2014} that in IC\,10 X-1, the X-ray source is never completely occulted by the star, and even its eclipse (more exactly, a dip in flux by a factor of 10) is caused by the thick wind.

Based on the analogy with the wind absorption in WR HMXBs, in the next Section we will try to assess where the Compton-thick clumps could be located in CG\,X-1, in order to explain the asymmetric lightcurve and dips. We also keep in mind another crucial piece of evidence: CG\,X-1 is a super-Eddington source, while NGC\,300 X-1 and IC\,10 X-1 are sub-Eddington BHs. For example, the peak luminosity of CG\,X-1 ($\approx$3 $\times 10^{40}$ erg s$^{-1}$) is 300 times higher than the luminosity of IC\,10 X-1 ($\approx$7 $\times 10^{37}$ \ergs, \citet{Laycock2015a}). Therefore, the compact object itself in CG\,X-1 is expected to launch a much stronger wind  \citep{Ohsuga2005, King2003, Pinto2016,Pinto2017, Walton2016, Kosec2018} than in the other systems.

\subsection{ Colliding winds and bow shock   }

 % from /Users/qyl/work/cgx1/cartoon/
 \begin{figure}
%\center{
\hspace{-0.7cm}
\includegraphics[width=10.0cm]{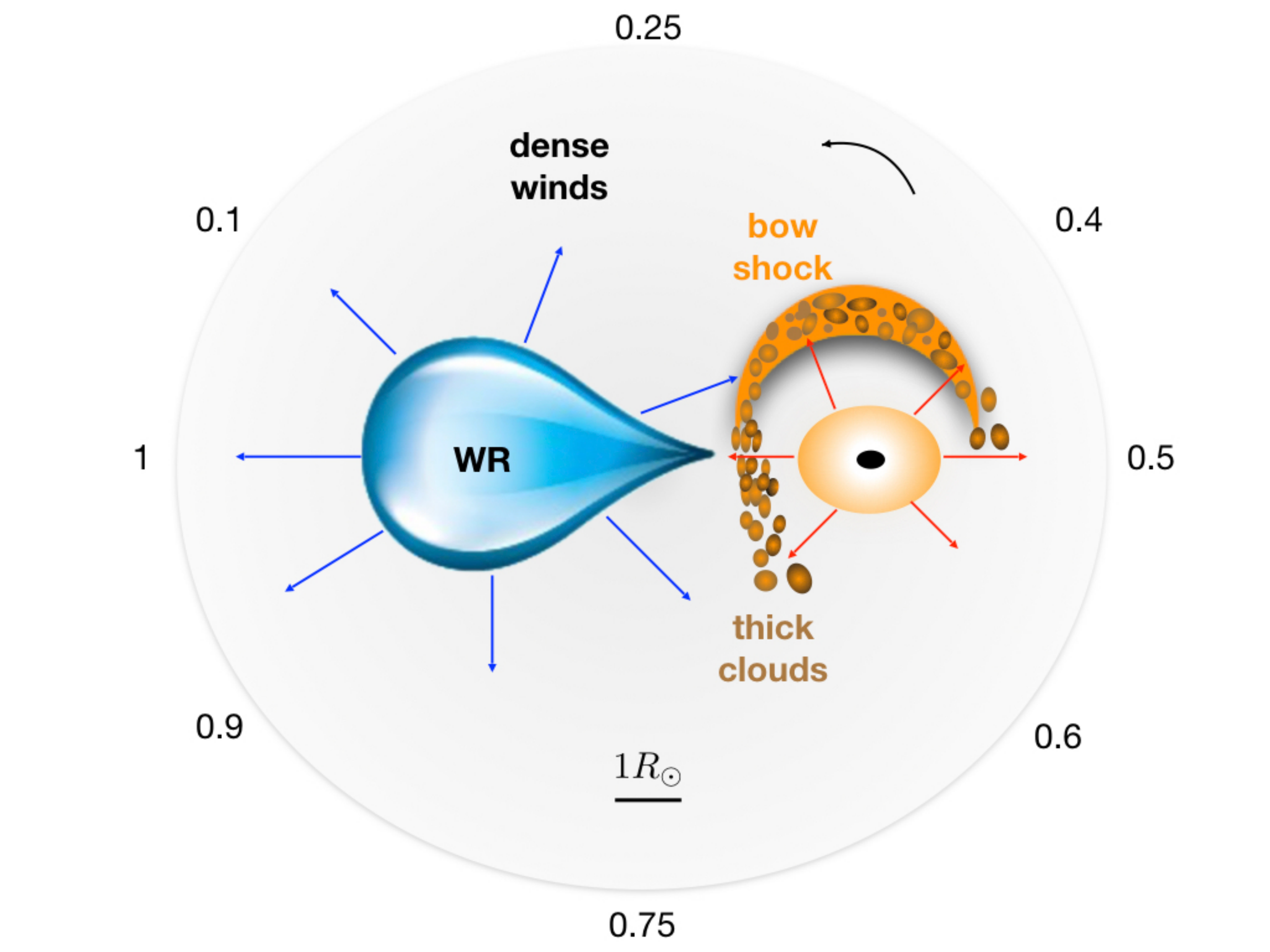}
\caption{A cartoon of our proposed geometry of CG\,X-1. The binary is rotating counter-clockwise. The numbers at the edge refer to the corresponding phase when we observe the system from that direction. $\phi=1$ is the phase when the compact object is directly behind the donor star. Because of the strong stellar winds and super-critical disk winds, the binary is surrounded by dense gas. Compton-thick clouds are produced in the colliding wind region between the two stars, and in the bow shock ahead of the (supersonically moving) compact object. The bow shock introduces an asymmetry in the lightcurve profile, causing the X-ray emission at phase $\approx$0.25 (compact object moving towards the observer) to be generally more occulted than at phase $\approx$0.75 (compact object receding). The bright (unocculted) phase is roughly from $\phi \approx $ 0.5--0.8. \\}
\label{fig:cartoon}
\end{figure}

There are two obvious regions where we expect a density enhancement (over the already high value inside a WR wind) with possible Compton-thick clump formation. The first location is along the line between WR and compact accretor, caused by wind-wind collision (WR wind and super-Eddington wind from the accretion disk). The second location is the bow shock in front of the compact object, caused by its fast orbital motion inside a dense medium. For simplicity, we shall also assume that the compact object is a stellar-mass BH (Figure \ref{fig:cartoon}).

Let us start from the wind collision. Shocks produced by colliding winds have been extensively studied (especially in the context of WR-O star binaries), both theoretically ({\it {e.g.}}, \citealt{Stevens1992, Usov1992, Kenny2005, Parkin2008, Lamberts2012}) and observationally ({\it {e.g.}}, \citealt{Pollock2005, Dougherty2005, Zhekov2012, Hill2018, Naz&eacute;2018}). Revisiting the detailed properties of the shocked gas is well beyond the scope of this paper: here we just want to check (with order-of-magnitude arguments) whether the shocked gas layer can act as a Compton-thick absorber.  Let us assume for simplicity that both the ULX disk wind (component 1) and the WR wind (component 2) have a similar projected speed in the orbital plane, $v_{\rm {w1}} \sim v_{\rm {w2}} \sim 1000$ km s$^{-1}$. For the ULX outer disk wind, this is comparable to the escape velocity from the outermost disk annulus. From the observed X-ray luminosity, we inferred an average mass accretion rate $\gtrsim$10$^{-6} M_{\odot}$ yr$^{-1}$. At such super-Eddington rates, the amount of mass lost in disk outflows is predicted to be larger than the mass accreted through the BH event horizon. Therefore, we estimate a mass loss rate of a few times $10^{-6} M_{\odot}$ yr$^{-1}$, in the super-Eddington disk outflow; this is consistent with our expectations that $\sim$10\% of the mass lost by the WR is accreted by the BH\footnote{A much lower fraction of stellar wind would be accreted by a NS, as the accretion cross section scales as $(M_{\rm NS}/M_{\rm BH})^2$.}. The intersection of the contact discontinuity between the two colliding winds with the line of centres will be located at a distance $d_1$ from the compact object and $d_2$ from the donor star, with $d_1/d_2 = (\dot{M}_1/\dot{M}_2)^{1/2}\, (v_{\rm w1}/v_{\rm w2})^{1/2}$ and $d_1 + d_2 = a$ \citep{Stevens1992}. For example, for $\dot{M}_1 \approx 0.1\dot{M}_2$, $d_2 \approx (3/4)a$. The density $\rho_{\rm w2}$ of a spherically symmetric wind from the WR star at a distance $d_2$ from its centre is 
%\begin{equation}
%    \begin{scriptsize}
\begin{eqnarray}
%&n_w& =\frac{\dot{M}}{4\pi a^{2} v_{\rm w} \mu {\rm m}_{\rm H}}\\
\rho_{\rm w2}& =&\frac{\dot{M}_2}{4\pi d_2^{2} \, v_{\rm w2}}
%&  &  \simeq 1.78\times10^{13} 
\approx 6.4\times10^{-12} \!
%\frac{1}{\mu}\!
\left(\frac{ \dot{M_2}}{10^{-5}~ \rm{M_{\odot} ~\! yr^{-1}}}\right)  \nonumber \\
&&\left(\frac{d_2}{ 4 ~ \!{\rm R_{\odot}} }\right)^{-2} \,
\left(\frac{v_{\rm w2}}{10^3~{\rm km~ s^{-1}}} \right)^{-1} \! \rm g~cm^{-3}\!.
\end{eqnarray}
%\end{scriptsize}
%\end{equation}
From standard shock theory, the density of the shock-heated WR wind is a factor of 4 higher, which corresponds to a number density $n_{\rm s2} \sim 10^{13}$ cm$^{-3}$. The temperature of the shock-heated gas is $kT_{\rm s2} \approx (3/16) \, \mu \, m_{\rm H}\, v_{\rm w2}^2 \approx 2$ keV for the assumed wind speed of $\approx$1000 km s$^{-1}$, where $m_{\rm H} \approx 1.7 \times 10^{-24}$ g is the mass of the hydrogen atom, and $\mu$ is the mean  mass per particle of gas measured in unit of $m_{\rm H}$. On the other side of the contact discontinuity, there will be another layer of shocked gas from the disk wind, with a similar temperature and density $n_{\rm s1} \approx (v_{\rm w2}/v_{\rm w1})^2\, n_{\rm s2} \sim 10^{13}$ cm$^{-3}$.

The cooling timescale for the shocked WR wind is 
\begin{equation}
    t_{\rm cool,2}=\frac{ 3 (n_e + n)_{\rm s2} \, kT_{\rm s2}}{2n^2_{\rm s2}\, \Lambda(T_{\rm s2})} 
    \approx \frac{ 3 kT_{\rm s2}}{n_{\rm s2}\, \Lambda(T_{\rm s2})}
\end{equation}
\citep{Dopita2003}, where $\Lambda$ is the cooling function \citep{Sutherland1993}. An analogous expression holds for the cooling timescale $t_{\rm cool,1}$ of the ULX wind. Thus, for a WR mass-loss rate of a few $10^{-5} M_{\odot}$ yr$^{-1}$, we expect a cooling timescale of a few seconds for the shocked wind, and a few 10s of seconds for the shocked ULX wind, in a system with the characteristic size of CG\,X-1. The cooling timescales can be compared with the timescales for the hot shocked gas on either side of the contact discontinuity to leave the interaction region; that is, the escape timescales $t_{\rm esc,1} = d_1/c_{s1}$ and $t_{\rm esc,2} = d_2/c_{s2}$, where $c_{\rm s1}$ and $c_{\rm s2}$ are the sound speeds in the two shocked layers. If $\chi \equiv t_{\rm cool}/t_{\rm esc} < 1$, a shocked shell is radiative; otherwise, it is adiabatic \citep{Stevens1992}. For the physical parameters assumed for our toy model of CG\,X-1, $\chi_2 \sim 10^{-3}$ for the shocked WR wind and $\chi_1 \sim 10^{-2}$ for the shocked ULX wind. Thus, both shells are radiative and will collapse into geometrically thin, cold shells \footnote{As an aside, this is a crucial difference between the colliding winds in a compact system such as CG\,X-1 and those in WR-O star binaries. In the latter class of systems, because of their typically much larger binary separation, the wind density at the shock is lower, and both shells are usually adiabatic \citep{Parkin2008}.}.

Next, we want to estimate the surface density of the radiative shells, to estimate whether they can provide the column density required to absorb essentially all X-ray emission below 10 keV.  For this purpose, we adapt the analytical solution of \cite{Kenny2005}. In addition to our previous assumptions on wind speed and mass loss rate, we assumed initial wind temperatures of a few $10^4$ K, both for the WR and the ULX wind, and a temperature of $\approx$10$^4$ K for the radiatively-cooled shocked shells. The density in a radiative shell is enhanced by a compression factor approximately equal to $\mathcal{M}^2$, where $\mathcal{M}$ is the Mach number \citep{Weaver1977}. For typical sound speeds $\sim$10 km s$^{-1}$ and wind speeds $\sim$1000 km s$^{-1}$, the compression factor is $\sim$10$^4$. Hence, we expect the two layers of radiatively cooled gas on both sides of the contact discontinuity to have densities of a few $10^{-8}$ g cm$^{-3}$. From the solutions of \cite{Kenny2005}, inserting typical parameters for this system, we obtain that each of the two shells has a thickness $\delta \sim 10^{-4}$ times the binary separation, that is a total thickness $\delta_1 + \delta_2 \sim 10^8$ cm. Thus, the surface density of the cold screen is $\sim$ a few g cm$^{-2}$, corresponding to a Compton-thick column density of a few $10^{24}$ cm$^{-2}$: this is enough to block all X-ray photons in the {\it Chandra} and {\it XMM-Newton} bands. 

Crucially, the opaque screen is subject to a dynamical process of fragmentation known as thin-shell instability \citep{Stevens1992}. We expect the cold gas to get shredded into clumps and filaments, with characteristic sizes $\sim$10$^8$ cm. The X-ray emitting region in a ULX is approximately the region of the geometrically thick accretion flow inside the spherization radius, which is a function of the mass accretion rate. A plausible value of the spherization radius for the observed X-ray luminosity of CG\,X-1 is a few 1000 km $=$ a few 10$^8$ cm, comparable with the size of the occulting clumps. Thus, we conclude that the shocked wind between the two system components is dense enough to produce Compton-thick absorption, and it can fragment into clumps with the right size to produce partial covering of the X-ray emission. If the obscuring clumps were much smaller in size, we would not see sharp, individual dips in the lightcurve, but only smooth flux changes; conversely, if the clumps were much bigger than the X-ray emitting region, we would either see total eclipses of the full flux, rather than variable partial covering. An analogous scenario of partial covering of the X-ray source by Compton-thick clouds was used to explain occultations in the Seyfert galaxy NGC\,1365 \citep{Risaliti2009a,Risaliti2009b}.

As we mentioned earlier, the wind collision region is not the only place where a thin shell of shocked gas can be formed. Under the plausible approximation of a circular orbit, the Keplerian orbital velocity $v_{\rm K1}$ of the compact object in CG\,X-1 is 
\begin{equation}
    v_{\rm K1} = \left[\frac{GM_2^2}{a\left(M_1 + M_2\right)}\right]^{1/2}.
\end{equation}
For example, for $M_1 = 10 M_{\odot}$ and $M_2 = 20 M_{\odot}$, we obtain $v_{\rm K1} \approx 660$ km s$^{-1}$; for $M_1 = 5 M_{\odot}$ and $M_2 = 30 M_{\odot}$, $v_{\rm K1} \approx 900$ km s$^{-1}$; for $M_1 = M_2 = 20 M_{\odot}$, $v_{\rm K1} \approx 550$ km s$^{-1}$. An object with a strong wind, moving into dense ambient medium at such high speed, will necessarily produce a strong bow shock. The physics of bow shocks in front of fast-moving OB stars is also a well-studied problem ({\it e.g.}, \citealt{Wilkin1996, Comeron1998, Meyer2017}), based on the same physical elements as the colliding wind problem (forward shock into the WR wind; contact discontinuity; reverse shock into the ULX wind). Near the apex of the bow shock, the WR wind is moving perpendicularly to the shock, but the forward shock is driven by the supersonic orbital motion of the BH, which, in the case of CG\,X-1, is only slightly lower than the WR wind speed. The distance between the BH and the contact discontinuity at the apex is 
\begin{equation}
    R_{\rm 0} = \left(\frac{\dot{M}_1 \, v_{\rm w1}}{4\pi \, \rho_{\rm a} \, v_{\rm K1}^2}\right)^{1/2}
\end{equation}
\citep{Comeron1998}, where $\rho_{\rm a}$ is the density of the unshocked WR wind near the apex, at a distance of $\approx$\,$a$ from the stellar centre. Using Equation (8), we can recast this distance as 
\begin{equation}
    R_{\rm 0} = \left(\frac{\dot{M}_1}{\dot{M}_2}\right)^{1/2} \, 
    \left(\frac{v_{\rm w1} \, v_{\rm w2}}{v_{\rm K1}^2}\right)^{1/2}\, a.
\end{equation}
For our assumptions of $\dot{M}_1 \sim 0.1 \dot{M}_2$, and wind speeds of $\approx$1000 km s$^{-1}$, this imples $R_{\rm 0} \approx 0.5a \approx 3R_{\odot}$. In terms of angular distance, the apex of the bow shock is $\approx$30$^{\circ}$ in front of the BH. 

As for the case of colliding winds, it is easy to verify that both the shocked ambient gas and the shocked ULX wind are fully radiative and collapse to two thin, cold shells on either side of the contact discontinuity. The surface density $\sigma$ of the cold gas can be approximated \citep{Comeron1998} as $\sigma \approx 1.25 \rho_{\rm a} R_{\rm 0} \sim 1$ g cm$^{-2}$ for $\dot{M}_2 \sim 10^{-5} M_{\odot}$ yr$^{-1}$. This corresponds to an absorbing column density $\sim$10$^{24}$ cm$^{-2}$, again sufficient to block most of the X-ray emission below 10 keV. The bow shock will be seen directly in front of the BH at the orbital phase $\phi \approx 1.2$. The combined effect of true stellar eclipse, colliding winds and bow shock may keep the X-ray emission continuously occulted from phase $\phi \approx 0.90$ to $\phi \approx 1.1$, and partially covered from phase $\phi \approx 1.1$ to $\phi \approx 1.5$, as seen in some orbital cycles (Figure \ref{fig:alc}). Instead, we expect to see unocculted emission roughly from $\phi \approx 1.5$ to $\phi \approx 1.8$ when the bow shock is behind the X-ray source (Figures \ref{fig:flc} and \ref{fig:cartoon}). As for the colliding wind shell, the cold bow-shock shell is also subject to instabilities and fragmentation \citep{Blondin1998}, which cause irregular dips during each cycle, variable partial covering, and profile variability from cycle to cycle (including cycles where the X-ray emission is already close to its maximum value around phase 1.2). 

In addition to the slower equatorial disk winds considered for our model, ULXs also launch relativistic or semi-relativistic outflows at lower inclination angles from the polar axis \citep{Walton2016,Pinto2016,Kosec2018,Pinto2017}. Such outflows may create larger, parsec-scale shells of shocked circumstellar gas around the binary system. However, such shells will be Compton-thin and are not directly relevant to the phase-dependent occulting behaviour we are studying in this work; therefore, a study of the circumbinary ULX bubble is beyond the scope of this paper. 
%\qyl{For this section, our first assumption is that the speed of the ULX disk wind is similar as the WR stellar wind, $\sim 1000$ km s$^-1$. However, faster ULX outflows, with speeds $\gtrsim$0.1$c$, are detected in several ULXs \citep{Walton2016,Pinto2016,Kosec2018,Pinto2017} . 
%The ultrafast outflows are launched from the innermost region of the non-standard disk, mostly at lower inclination angles from the polar axis ( but we are not concerned here with this polar outflow). The supersoft source NGC 55 ULX, which is observed at higher inclination, also presents high speed (0.01--0.2c) powerful wind \citep{Pinto2017}. For an extreme case, if CG\,X-1 has  $\sim$0.1c disk wind as NGC 55 ULX, we need recalculate our toy model using equation (8), (9), (11), (10). The number density of the shocked WR winds and ULX winds will be $n_{\rm s2} \sim 10^{14}$ cm$^{-3}$ and $n_{\rm s1} \sim 10^{11}$ cm$^{-3}$, respectively; $\chi_{\rm 2}  \sim 10^{-4}$,  
%$\chi_{\rm 1} \sim 10^{3}$, $R_{\rm 0}\approx$ 2.6a $\approx 15$ \Rsolar.  Then only the WR shells are radiative, and the ULX shells are adiabatic.  The bow shock will be a bit outside the binary system, and no longer Compton-thick to X-ray below 10 keV. Therefore, the ultrafast outflows will lead to a thinner layer of opaque screen between the two stars, and the bow shock in front of the compact object will be transparent to the X-ray, instead of partially covering it.}

\subsection{  Anti-correlation between egress width and full luminosity}
We showed (Figure \ref{fig:dphi}) a trend of faster egress times in cycles with higher X-ray luminosities in the bright phase. We also argued from spectral analysis (Section 4) that the transition between bright and occulted phases of a cycle is due to an increase of partial covering by an opaque screen. Thus, we suggest that the egress width is correlated with the area and location of the opaque screen with respect to the X-ray emission region: higher X-ray luminosities seem to reduce the phase interval affected by partial covering. 

One simple explanation for the anti-correlation is that at higher (intrinsic) luminosities, larger parts of the opaque screen are ionized by the strong X-ray flux and become optically thin to X-rays; this leads to shorter occultations and faster egress times. 
%An alternative explanation is that the higher luminosity corresponds also to stronger outflows from the compact object, and this may push the opaque screen to larger distances from the X-ray source, or even blow it away, thus reducing the fraction of the X-ray emission region covered along our line of sight.
An alternative explanation is that the higher luminosity corresponds also to stronger outflows from the compact object, and this may push the opaque screen and the bow shock to larger distances from the X-ray source (as discussed in Section 6.2), or even blow it away, thus reducing the fraction of the X-ray emission region covered along our line of sight.
A third, more speculative, scenario is that changes in the observed X-ray luminosity are due to disk precession: when the X-ray-emitting inner disk is seen more face-on, we expect to see a higher X-ray luminosity in the unocculted phases, and a smaller degree of occultation by opaque clouds in the orbital plane \citep{Hu2013}. Further investigations on these possibilities are beyond the scope of this paper.

It is significant that the duration of the full or partial occultation phase (as defined in Section 3.3) saturates at a minimum value $\Delta \phi \approx 0.2$ at X-ray luminosities $\gtrsim$10$^{40}$ \ergs (Figure \ref{fig:dphi}). We speculate that this minimum value is evidence of a true eclipse by the donor star, which can never be eliminated even when the partial-covering clouds are ionized or blown away. However, even during the full occultation/eclipse phase, there is still residual emission, which we will discuss in the next section.

\subsection{ Residual emission in the eclipse} \label{sect:residual}
One of the results of our spectral analysis of phase-resolved stacked spectra is that even during the true eclipse (or deepest occultation) at $\phi \approx 0.9$--1.1, we still detect residual flux from CG\,X-1, at a level of $\approx$1\% of the bright-phase flux (\Lx$\approx 2.5\times 10^{38}$ \ergs). The residual flux is softer and less absorbed than the flux in the bright phase (Section 4). {\it Chandra}'s image resolution shows that the residual emission is associated with the point-like source, and is not, for example, due to incomplete background subtraction of diffuse X-ray emission in the inner region of the Circinus galaxy. Here, we assess possible origins for this residual emission.

The residual emission cannot be from the WR star itself: the X-ray luminosity of an isolated WR star does not exceed a few times $10^{32}$ \ergs\ \citep{Skinner2012}. Wind-wind collisions might offer a more promising explanation, analogous to the X-ray emission seen from WR + O star binaries ({\it e.g.,} \citealt{Guerrero2008}). Expected wind speeds $\gtrsim$1000 km s$^{-1}$ for both the WR wind and the super-Eddington disk wind guarantee that the shocked gas reaches temperatures $\gtrsim$1 keV.
In the simplest case of identical winds, each with mass loss rate $\dot{M}$ and speed $v_{\rm w}$, in the fully radiative regime, the predicted luminosity is $L_{\rm X} \approx f\dot{M} v_{\rm w}^2 \approx 10^{36} \dot{M}_{-5} \, v_{\rm w,3}^2$ erg s$^{-1}$ \citep{Stevens1992}, where the geometrical factor $f \approx 1/6$ parameterizes the fraction of wind kinetic luminosity perpendicular to the wind shock, $\dot{M}_{-5}$ is in units of $10^{-5}$\Msolar\ yr$^{-1}$, and $v_{\rm w,3}$ is in units of 1000 km s$^{-1}$.
For unequal winds, in the isothermal regime, the ratio of the luminosities from the two shocked winds is $L_1/L_2 \approx (\dot{M}_1/\dot{M}_2)^{1/2} (v_{\rm w1}/v_{\rm w2})^{3/2}$ \citep{Stevens1992}. In the case of CG\,X-1, we can plausibly assume $\dot{M}_1 \lesssim 0.1 \dot{M}_2$, as already discussed in Section 6.2. For the X-ray luminosity of the shocked winds to exceed $10^{38}$ erg s$^{-1}$, values of $\dot{M}_1 \gtrsim 10^{-6}$ \Msolar\ yr$^{-1}$, and $v_{\rm w1} \sim 0.1c$ would be required. The luminosity would be completely dominated by the disk wind component. We cannot rule out this scenario for a ULX, because we know that the kinetic power carried by the outflows is comparable to the radiative power \citep{Siwek2017}. So far, we have not identified any other ULXs where such kinetic power is dissipated as X-rays in a strong shock at small distances from the compact object, as opposed to large ionized bubbles (ULX bubbles: \citealt{Pakull2002,Pakull2006}) at $\sim$100 pc scale. On the other hand, before CG\,X-1, we had not identified any WR ULX, immersed in a circumstellar medium with densities $\sim$10$^{12}$ cm$^{-3}$. So, the theoretical properties of a compact WR ULX bubble are an intriguing topic for further work.

Another scenario is that the residual soft X-ray emission comes from the photosphere of the optically thick outflows driven by the super-Eddington source. Such thermal component has been invoked to explain the spectra of ultraluminous supersoft sources and the ``soft excess'' in ULX spectra \citep{Tao2019,Soria2016,Urquhart2016,Middleton2015}. It is generally consistent with a blackbody spectrum at temperatures $\approx$50--120 eV for supersoft sources, and $\approx$0.1--0.4 keV for ULXs, with a soft X-ray luminosity from $\sim$10$^{38}$ erg s$^{-1}$ up to a few 10$^{39}$ erg s$^{-1}$ \citep{Zhou2019,Urquhart2016}. If we fit the residual emission of CG\,X-1 with a blackbody model, we obtain a temperature of $\approx$0.5 keV (higher than seen in the soft excess of any other ULX) and an X-ray luminosity of $\approx$10$^{38}$ erg s$^{-1}$ (Table 2). In the outflow model of \cite{Zhou2019}, this solution corresponds to a NS accreting at almost 100 times Eddington. The physical size of the outflow photosphere ($r_{\rm {ph}}$) is related to the apparent blackbody radius ($r_{\rm {bb}}$) obtained from spectral fitting as $r_{\rm {ph}} \approx \sqrt{\tau_{\rm {es}}} \, r_{\rm {bb}}$ \citep{Meier1982,Zhou2019}, where $\tau_{\rm {es}}$ is the scattering optical depth at the photosphere ($\tau_{\rm {es}} \sim $ a few 100). The photons emitted from the photosphere will be scattered multiple times before they can 
%because of the high scattering optical depth, and will then 
propagate freely to us from the last scattering sphere, where $\tau_{\rm {es}}$ decreases to unity. The radius of this surface may be $\sim$100 times larger than the radius of the photosphere. In our case, $r_{\rm {bb}} \approx 130$ km, plausible values for $r_{\rm {ph}}$ are a few $10^8$ cm, and the radius of the last scattering sphere may be $\sim$10$^{11}$ cm, which is comparable to the size of the occulting star and the bow shock.  We note that a blackbody model is a poor fit to our eclipse spectrum, with $\chi^2_{\nu} > 2$ (Table 2). However, the spectrum emerging from the last scattering sphere may be significantly different from the input blackbody emitted at the photosphere of the optically thick outflows.

Alternatively, we suggest that the most likely source for the residual eclipse emission is a fraction of X-ray photons emitted along the polar funnel and downscattered by the wind into our line of sight, at scales larger than the companion star. Residual emission in eclipse is seen in various other Galactic HMXBs, and is usually interpreted as scattered light \citep{Sako1999,Schulz2002,Wojdowski2003,Lopez2006}. It is usually a few percent of the direct luminosity out of eclipse, consistent with our flux ratio. A WR system such as CG\,X-1, with strong outflows from both binary components, is precisely the type of system where we expect that some radiation emitted perpendicular to the binary plane will be scattered into our line of sight even when the direct emission is obscured.
 
%========================================================
\subsection{ Binary evolution }

To-date, we know of only seven compact ({\it i.e.}, with a binary period $\lesssim 1$ d) HMXBs with a candidate WR donor: in addition to CG\,X-1, they are 
NGC\,4214 X-1 ($P=3.6$ hr, \citealt{Ghosh2006}); 
Cyg X-3 in the Milky Way ($P=4.8$ hr, \citealt{Zdziarski2012});  
NGC\,4490 CXOU J123030.3+413853 (X-1) ($P=6.4$ hr,  \citealt{Esposito2013}); 
%CG\,X-1 ($P=7.2$ hr, \citealt{Esposito2015}); 
NGC\,253 CXOU J004732.0-251722.1 ($P=14.5$ hr \citealt{Maccarone2014}); 
NGC\,300 X-1 ($P=32.8$ hr, \citet{Carpano2007}); 
IC\,10 X-1 ($P=34.8$ hr, \citealt{Laycock2015a}). 
One more candidate WR X-ray binary is M\,101 ULX-1 \citep{Liu2013}, but it has a tentative period of 8.2 d, which suggests a different evolutionary history than the compact class. Within this group, CG\,X-1 is the most luminous one (peak X-ray luminosity $L_{\rm X} \approx 3\times 10^{40}$ erg s$^{-1}$), and the only ULX, clearly exceeding the Eddington limit for a $10$\Msolar\ BH.

We focus on compact WR binaries because they help us understand two key evolutionary features of binary systems, one in their past history and one in their future one. In the past, they must have come through a phase of dramatic shrinking of the binary separation from initial values of $\sim$100--1000 \Rsolar\ to $\lesssim$10 \Rsolar. The most likely mechanism for such shrinking is a common envelope phase after the formation of the compact object \citep[{\it {e.g.}},][]{Dominik2012,Bogomazov2014,Belczynski2016,vandenHeuvel2017,Bogomazov2018,Giacobbo2018}. In some cases, theoretical calculations suggest \citep{Pavlovskii2017,vandenHeuvel2017,Bogomazov2018} that a substantial but gradual shrinking of a compact binary system (a ``spiral-in'' during stable Roche-lobe overflow mass transfer) may occur without a common envelope phase, if the donor star has a radiative envelope and the mass ratio is $\lesssim$3; however, a common envelope phase seems unavoidable \citep{vandenHeuvel2017} if we want to obtain binary periods as short as a few hours (like in CG\,X-1 and Cyg X-3).
%In the past, they must have come through a common envelope phase after the formation of the compact object, that shrank the binary separation from initial values of $\sim$100--1000 \Rsolar\ to $\lesssim$10 \Rsolar\ \citep[{\it {e.g.}},][]{Dominik2012,Bogomazov2014,Belczynski2016,vandenHeuvel2017,Bogomazov2018,Giacobbo2018}. 
In the future, if they remain bound and their orbit does not widen too much after core collapse of the WR star, systems like CG\,X-1 and Cyg X-3 will form double compact objects (DCOs) that can spiral in and merge via gravitational wave emission on timescales much shorter than the Hubble time \citep[{\it {e.g.}},][]{Bulik2011,Belczynski2013,Esposito2015,Belczynski2016}. Other types of DCOs originating for example from a compact object orbiting a supergiant donor star ({\it i.e.}, evolved from supergiant HMXBs) are too widely separated to do so. Thus, population studies of short-period WR X-ray binaries, and especially ULX WR systems (detectable with X-ray telescopes in a larger volume of space), provide crucial observational constraints to the predicted rate of LIGO/Virgo detections \citep{Abbott2016a, Abbott2016b, Abbott2017a, Abbott2017b, Abbott2017c}. An analysis of the physical scenarios and possible steps between the X-ray binary stage and the final merger is beyond the scope of this work. Instead, we will briefly discuss the past and present evolutionary stages of these systems.

The common envelope phase required for the existence of systems such as CG\,X-1 is the one that occurs after the formation of the first compact object; this results in the loss of the hydrogen-rich envelope of the companion star and the dramatic shrinking of the binary separation. After the end of common envelope, the donor star is helium-rich and can be classified as a WR star\footnote{As the historical definition of a WR star is based on spectroscopic features rather than a precise range of physical parameters, the distinction between stripped-envelope helium stars and classical WR stars has always been somewhat blurred in the literature; helium stars more massive than $\approx$10 \Msolar\ with WR-like spectra are usually simply called WR stars. See \cite{G&ouml;tberg2018} for a comprehensive discussion of the relation between stripped helium-rich stars and WR stars across the mass range.}.
From this point, further evolution of the system (which is now an X-ray binary) depends on the mass ratio, on how mass and angular momentum are transferred from the donor to the compact object, and on how they are lost from the system \citep{Huang1963,Lommen2005,vandenHeuvel2017}. For example, mass transfer from a more massive WR donor to a less massive compact component, and outflows from the accretion disk wind, are two factors that lead to orbital shrinking, while mass loss in the wind from the more massive component results in orbital widening. In particular, for WR donors with wind mass loss rates of $\sim$10$^{-5}$ \Msolar\ yr$^{-1}$, we expect a secular increase of binary separation and orbital period \citep{Tutukov2013}. This is already observed in Cyg X-3 ($\dot{P}/P \approx$ 1.2--4 $\times 10^{-6}$ yr$^{-1}$: \citealt{Lommen2005}). In this work (Section 3.2), we have shown a period increase at the 90\% confidence limit also in CG\,X-1 ($\dot{P}/P = (10.2\pm4.6) \times 10^{-7}$ yr$^{-1}$).

Orbital widening inevitably leads to the donor star becoming detached from its Roche lobe, as WR stars do not expand in their final stages of evolution (unlike supergiants). For a fixed mass loss rate $\dot{M}_2$ in the WR wind, the mass transfer rate $\dot{M}_1$ into the Roche lobe of the compact object scales as $\dot{M}_1/-\dot{M}_2 \approx G^2 M_1^2/(a^2 \, v_{\rm w2}^4)$ \citep{Frank2002}. As the WR donor becomes more detached, we expect a decrease in the mass accretion rate and luminosity, both because of the increase in the binary separation $a$, and because the WR wind reaches a higher velocity $v_{\rm w2}$ before being intercepted by the compact object; the latter effect makes it harder to form an accretion disk around the accretor \citep{Ergma1998}.

We have already noted that CG\,X-1 is the only ULX in the known sample of seven close WR X-ray binaries. The higher luminosity of CG\,X-1 compared with NGC\,300 X-1 and IC\,10 X-1 can be explained by a larger orbital separation in those two systems (period longer than a day). However, Cyg X-3 has an even shorter period than CG\,X-1, and yet its X-ray luminosity is almost two orders of magnitude lower, $L_{\rm X} \sim$  a few $10^{38}$ erg s$^{-1}$ \citep{Koljonen2018,Koljonen2010,Hjalmarsdotter2008}. Clearly, the binary period alone cannot be used to predict the X-ray luminosity of compact WR binaries; the mass, evolutionary stage, metal abundance of the WR star certainly plays an equally important role. Moreover, the accretion cross section scales as $M_1^2$; the mass of the compact object is not known, either in Cyg X-3 or in CG\,X-1. For example, for Cyg X-3, mass estimates for the compact object based on accretion-state behaviour range from $\approx$2 \Msolar\ \citep{Zdziarski2013}, consistent with either a BH or a NS, to a run-of-the-mill stellar-mass BH around 10 \Msolar\ \citep{Shrader2010}, to a heavy BH with a mass $\gtrsim$20 \Msolar\ \citep{Hjalmarsdotter2008}.

This uncertainty leads us to the last issue we want briefly to mention in this work: whether the compact object in CG\,X-1 is more likely a BH or a NS. Based on X-ray luminosity alone, until a few years ago it would have been straightforward to classify it as a BH in the ultraluminous state. However, we now know from observational studies that NS ULXs do exist and can also reach apparent luminosities of $\sim$10$^{40}$ erg s$^{-1}$ \citep{Bachetti2014, Israel2017a, Israel2017b, Furst2016}, although so far no NS ULX has been found with a period as short as a few hours. Theoretical models have also caught up with the observations, to explain the existence and abundance of NS ULXs \citep{ Koliopanos2017,Mushtukov2019,Pintore2017, Walton2018, Wiktorowicz2017, Wiktorowicz2019}. 

If CG\,X-1 is a wind-accreting rather than Roche lobe overflow system, one argument in favour of the BH scenario is that a BH can intercept a larger fraction of stellar wind, and reach higher luminosities. For a typical WR wind with $-\dot{M}_2 \sim $ a few $\times 10^{-5}$ \Msolar\ yr$^{-1}$ \citep[{\it {e.g.}},][]{Hillier2003,Vink2005,Belczynski2010,Smith2014}, the fraction intercepted by the compact object must be at least $\sim$10\% of that rate, to explain the luminosity of CG\,X-1. The intercepted wind fraction can be estimated as $\dot{M}_1/-\dot{M}_2 \approx (1/4) \, (M_1/M_2)^2 \, (R_2/a)^2$ \citep{Frank2002}. Only with a BH accretor ($M_1 \sim M_2$) can this fraction approach $\sim$10\%, perhaps with the aid of focused winds if the donor is closed to filling its Roche lobe. A WR + NS system is more likely to have $\dot{M}_1/-\dot{M}_2 \lesssim 10^{-3}$, insufficient to explain CG\,X-1. In addition to the smaller accretion cross section, there is another (possibly even more important) reason why a NS orbiting in the wind of a WR donor is not expected to become a ULX \citep{Lipunov1982}. Most of the WR wind is expected to be deflected by the magnetosphere of the NS ({\it i.e.}, the magnetospheric radius can be larger than the Bondi capture radius). Moreover, any phase of super-critical accretion will quickly spin-up the NS above the threshold for the propeller or ejector regimes \citep{Lipunov1982,Ergma1998}. The characteristic spin-down timescale is then longer than the WR lifetime, so that the NS would not have time to become a ULX again.

Another argument in favour of the BH scenario in short-period WR X-ray binaries \citep{vandenHeuvel2017} is that a NS would be unlikely to survive the common envelope stage and would likely merge with the WR star. In particular, \citet{vandenHeuvel2017} showed (their Figs 1,2) that for a mass ratio $M_2/M_1 \gtrsim 3.5$, a 1.5-\Msolar\ NS may survive common envelope only for a very limited range of initial binary periods and donor masses ($M_2 \approx 40$--70 \Msolar); a 5-$\rm M_{ \odot}$ BH has the highest chance to survive and produce a short-period WR system such as Cyg\,X-3 and CG\,X-1, for initial donor masses $\gtrsim$18 \Msolar; more massive BHs ($M_1 \gtrsim 15$ \Msolar) may avoid common envelope altogether for any mass of the donor star, and go through a slower spiral-in phase during stable Roche lobe overflow, with periods always $\gtrsim$1d \citep{Pavlovskii2017,vandenHeuvel2017,Bogomazov2018}.

%%%%%%%%%%%%%%%%%%%%%%%%%%%%%%%%%%%%%%%%%%%%%%%%%%%%%%
\section{   SUMMARY  }

We have studied the X-ray timing and spectral properties of the brightest X-ray source in the Circinus galaxy, confirming that it is a ULX with an X-ray luminosity $L_{\rm X} \sim$10$^{40}$ erg s$^{-1}$. The possibility of a foreground mCV is ruled out because the X-ray to optical flux ratio is $\gtrsim 100$, inconsistent with a CV.

We used a set of observations (two from \rosat, 24 from \chandra, and five from \xmm) that spans over 20 years. We phase-connected all the observations, thanks to the recurrent presence of an eclipse in each orbital cycle. This gave us a binary period $P = (25970.0 \pm 0.1)$ s $\approx$7.2 hr (the most precise period measured in a ULX to date), and a period derivative $\dot{P}/P = (10.2\pm4.6) \times 10^{-7}$ yr$^{-1}$ (error range at the 90\% confidence level); $\dot{P}/P$ is larger than 0 at the 10$\sigma$ level. We showed that the X-ray profiles of each orbital cycle share the same general structure (fast ingress, eclipse, slow egress with dips), but each cycle has a unique shape, in terms of duration of the occultation phase, and dip structure. This is more complicated than a clean stellar eclipse, and suggests that the X-ray source is occulted by a complex structure of absorbers, located at approximately the same position with respect to the two binary components, but varying randomly over timescales shorter than a few hours.

As a first-order approximation, the X-ray spectrum is a power-law with photon index $\Gamma \approx 2$, seen through a cold local absorber (column density $\approx$ a few times $10^{21}$ cm$^{-2}$). More detailed modelling of the spectra with highest signal to noise reveals two additional features. First, there is a spectral downturn above $\approx$ 4--5 keV. This curvature is one of the defining properties of ULXs, as opposed to sub-Eddington stellar-mass BHs. Second, there is a significant, additional absorption component from ionized gas, with column density $\sim$10$^{22}$, more prominent during the egress phase than during the unocculted bright phase. This is typical of HMXBs, where we see the X-ray source through a thicker (and ionized) stellar wind when it passes behind the donor star.

The gradual transition from full occultation to fully bright phase does not correspond to a gradual decrease of the absorbing column density from Compton-thick to Compton-thin values. To a first approximation, we see an increase of the flux normalization during egress, without significant changes in the spectral shape; the increase is not monotonic, but is instead punctuated by irregular dips. This is consistent with a decrease of the covering fraction by an additional, Compton-thick partial-covering absorber. From the varying duration and structure of the occultation phases and dipping behaviour in each orbit, we suggest that the occulting material is made of opaque clouds with a characteristic size comparable to the size of the X-ray emitting region. From the orbital phase in which such occultations occur, we also infer that the absorbing material is mostly located between the two binary components and ahead of the accretor in its orbital trajectory. The duration of the occultation and dipping phases in each cycle is anti-correlated with the luminosity observed in the bright phase of that cycle. The peak luminosity varied between $\approx$4 $\times 10^{39}$ \ergs\ and $\approx$3 $\times 10^{40}$ \ergs\ over our 20-year coverage. Finally, we noticed residual X-ray emission in eclipse, at a luminosity $\sim$10$^{38}$ erg s$^{-1}$, with a softer spectrum than out of eclipse.

%Both model-independent photon energy distribution and phase-resolved spectral fittings demonstrate that the spectra during the eclipse are much softer and less absorbed than the spectra outside the eclipse. 
%The continuum spectra of the non-eclipse phase were well fitted by a power-law model with $\Gamma\sim 2$; adding a hot plasma absorption component {\it absori} significantly improved the fitting results.
%More detailed phase-resolved spectral analysis reveals that the spectra of the egress phase and bright phase share the similar spectral slope but are only different in the fluxes. The flux difference can be represented by a partial-covering component {\it pcfabs}, which is Compton-thick to X-ray emission. 

We proposed a simple model that can account for all these observational timing and spectral properties. We agree with the previous suggestions in the literature that CG\,X-1 is most likely a WR X-ray binary, as a WR star is the only type of massive star that can fit inside a Roche lobe radius with an inferred size $\lesssim$ 3\Rsolar, and at the same time provide enough mass transfer onto the compact object to generate X-ray luminosities $\sim$10$^{40}$ erg s$^{-1}$. In addition to the stellar wind, the super-Eddington accretor itself is expected to generate a powerful radiatively driven wind (this makes this system different from ordinary sub-Eddington HMXBs). Thus, we argued that the system should contain two regions of shocked wind: one between compact object and donor star (wind-wind collisions), and the other in front of the compact object along its orbital motion (bow shock into the dense medium created by the WR wind). Simple order-of-magnitude calculations show that the shocks are fully radiative; the shocked gas is so dense that it cools on timescales of a few seconds, collapses into a cold, Compton thick shell, and is expected to fragment into small clumps. Such clumps may be responsible for the partial covering and the irregular dips. The presence of a bow shock is the reason for the asymmetric location of the dips (mostly after rather than before the eclipse). The mass loss rate from the binary system is consistent with the observed increase in the period: $\dot{M}/M \sim \dot{P}/P \approx 10^{-6}$ yr$^{-1}$ (similar also to what is measured in the Galactic WR system Cyg X-3). The residual, softer emission in eclipse may come from the shocked wind, or, more likely, from X-ray photons scattered into our line of sight by the dense circumbinary medium.

We do not have direct observational evidence to determine whether the compact object is a BH or a NS. (BH identifications based only on super-Eddington luminosity arguments have been proven spectacularly unreliable in ULXs). {However, we} do favour the BH interpretation for two indirect arguments. First, the accretion cross section in wind-fed binaries scales as the square of the accretor mass, and a 1.5-$\rm M_{ \odot}$  accretor is unlikely to intercept enough wind to generate steady luminosities $\gtrsim$10$^{40}$ erg s$^{-1}$ for any plausible stellar wind model. Second, the short orbital period suggests that the system underwent a common envelope phase, which stripped the hydrogen envelope of the donor star and shrank the binary to a separation of a few solar radii; binary evolution models suggest that NSs are much less likely than stellar-mass BHs to survive common envelope.
%The Compton-thick medium that partially cover the egress phase, could be produced by powerful colliding winds. A massive WR star, which has strong stellar winds and is compact enough to reside in the secondary Roche lobe, is a bona fide candidate for the donor star of CG\,X-1. The Compton-thick medium produced in front of the moving direction of the BH can also interpret the asymmetric eclipse profile. The immense mass loss rate of a WR star may result in the expansion of the orbit.

%The peak X-ray luminosity in the 0.3--8 keV band can reach $2.9\times 10^{40}$ \ergs, which is $\sim$ 20 times higher than the Eddington luminosity of a 10\Msolar\ BH. In two of spectra with the most high quality, the high-energy spectra deviate the power-law continuum and curve down at $\sim$ 4-5 keV. This downturn curvature is ubiquitous in other ULXs, but not seen in sub-Eddington Galactic BHs. Given such high  X-ray luminosity and a hard-band curvature, CG\,X-1 is most likely a super-critical X-ray source. A stellar-mass black hole with a WR donor could be a better explanation for the timing and spectral properties of CG\,X-1, although a NS accretor can not be ruled out.

% importance of WR+BH
The general significance of our study is that compact WR X-ray binaries (period $\lesssim$1 d) are a rare (only seven identified so far) and intriguing type of system, providing clues for crucial steps of binary evolution (see also the review of \citealt{vandenHeuvel2019}). They are also a step towards the formation of DCOs that can decay and merge via gravitational wave emission; constraining the formation rate of compact WR X-ray binaries helps predicting the LIGO/Virgo detection rate. On top of that, CG\,X-1 is the most luminous (in fact, the only persistent ULX) in this subclass, at least an order of magnitude above the others. Thus, its behaviour can be used for various tests of wind collisions, shocked shells, and super-critical accretion inflow/outflow models, in more extreme conditions than other HMXBs. Thirdly, CG\,X-1 is now the ULX with the most precise and accurate binary period, and even a tentative detection of binary period evolution. Further observational campaigns should be aimed at detecting or placing stronger upper limits on its point-like optical counterpart, ionized ULX bubble, and radio flux.

%In recent years, seven WR-BH or WR-NS binaries are proposed, including four of them with orbital period less than one day. These compact WR XRBs are the best candidates for the progenitors of Gravitational wave events. If those WR XRBs can survive through the second SN explosion,  DCO systems will be produced, and will merge within Hubble time. CG\,X-1 is the brightest and the only one with peak $L_{\rm X} > 10^{40}$ \ergs\ among the seven WR XRBs. It is a valuable and promising ULX, and more importantly, is a kind of system which may end up with a BH-BH binary.  A bright ULX with eclipses and a well determined binary period is itself a very rare system, and makes CG\,X-1 an exceptional target for studying super-critical accretion inflows/outflows and constraining the evolutionary tracks from WR ULXs to DCOs.

%%%%%%%%%%%%%%%%%%%%%%%%%%%%%%%%%%%%%%%%%%%%%%%%%%%%%%
\acknowledgments
\begin{acknowledgements}
This work was supported by  National Program on Key Research and Development Project (Grant No. 2016YFA0400800), and the National Natural Science Foundation of China (NSFC) through grants NSFC-11425313/11603035/11603038.
We thank Ilaria Caiazzo, Paolo Esposito, JaeSub Hong, Chichuan Jin, Jiren Liu, Michela Mapelli, Manfred Pakull, Gavin Ramsay, Axel Schwobe, Lei Zhang for helpful discussions. Y.L. Q acknowledges the Harvard-Smithsonian Center for Astrophysics, and
RS thanks Curtin University and The University of Sydney, for hospitality during part of this research. DJW acknowledges support from an STFC Ernest Rutherford fellowship.
This work has made use of data obtained from the \chandra\ Data Archive, and software provided by the \chandra\ X-ray Center (CXC) in the application packages {\sc ciao}. 
This work has made use of software obtained from the High Energy Astrophysics Science Archive Research Center ({\sc heasarc}), a service of the Astrophysics Science Division at NASA/GSFC and of the Smithsonian Astrophysical Observatory's High Energy Astrophysics Division.
This work has also made use of the data from the \xmm, an ESA science mission funded by ESA Member States and USA (NASA). 
\end{acknowledgements}

%%%%%%%%%%%%%%%%%%%%%%%%%%%%%%%%%%%%%%%%%%%%%%%%%%%%%%
%\includegraphics[]{cgx-1-tmp.png}

\bibliography{bib}

\end{document}